\documentclass[draftcls,twoside,onecolumn,letter,12pt]{IEEEtran}
\usepackage{graphicx}
\usepackage{amssymb}
\usepackage{amsmath}
\usepackage{cite}

\newtheorem{definition}{Definition}
\newtheorem{theorem}{Theorem}
\newtheorem{lemma}{Lemma}
\newtheorem{corollary}{Corollary}
\newtheorem{example}{Example}
\newtheorem{property}{Property}

\DeclareMathOperator{\vecOp}{\mathrm{vec}}
\DeclareMathOperator{\tr}{\mathrm{tr}}
\DeclareMathOperator{\E}{\mathbb{E}}
\DeclareMathOperator{\var}{\mathbb{V}\mathrm{ar}}

\DeclareMathOperator{\etr}{\mathrm{etr}}

\newcommand{\EX}[1]{\E\left\{{#1}\right\}}
\newcommand{\EXs}[2]{\E_{{#1}}\left\{{#2}\right\}}
\newcommand{\PDF}[2]{p_{{#1}}\left({#2}\right)}

\newcommand{\CF}[2]{\Phi_{#1}\left({#2}\right)}
\newcommand{\MGF}[2]{\phi_{#1}\left({#2}\right)}

\newcommand{\Var}[1]{\var\left\{{#1}\right\}}

\newcommand{\B}[1]{{\pmb{#1}}}

\newcommand{\C}{\mathbb{C}}
\newcommand{\R}{\mathbb{R}}

\newcommand{\CGM}[5]{\tilde{\mathcal{N}}_{{#1},{#2}}\left({#3},{#4},{#5}\right)}
\newcommand{\CWM}[3]{\tilde{\mathcal{W}}_{#1}\left({#2},{#3}\right)}
\newcommand{\CQM}[5]{\tilde{\mathcal{Q}}_{{#1},{#2}}\left({#3},{#4},{#5}\right)}
\newcommand{\nTx}{{n_{\mathrm{T}}}}
\newcommand{\nRx}{{n_{\mathrm{R}}}}
\newcommand{\nSx}{{n_{\mathrm{S}}}}
\newcommand{\TxCM}{\B{\Phi}_{\mathrm{T}}}
\newcommand{\RxCM}{\B{\Phi}_{\mathrm{R}}}
\newcommand{\SxCM}{\B{\Phi}_{\mathrm{S}}}

\newcommand{\TxEV}[1]{{\lambda_{#1}^\mathrm{T}}}
\newcommand{\RxEV}[1]{{\lambda_{#1}^\mathrm{R}}}
\newcommand{\SxEV}[1]{{\lambda_{#1}^\mathrm{S}}}

\newcommand{\EVn}[1]{{\varrho\left(#1\right)}}
\newcommand{\EVm}[2]{{\tau_{#1}\left(#2\right)}}
\newcommand{\oEV}[3]{{{#1}_{#2}^{#3}}}
\newcommand{\odEV}[3]{{{#1}_{\langle #2 \rangle}^{#3}}}

\newcommand{\odTxEV}[1]{{\lambda_{\langle #1 \rangle}^\mathrm{T}}}
\newcommand{\odRxEV}[1]{{\lambda_{\langle#1 \rangle}^\mathrm{R}}}
\newcommand{\odSxEV}[1]{{\lambda_{\langle #1 \rangle}^\mathrm{S}}}

\newcommand{\snr}{\bar{\gamma}}
\newcommand{\ImUnit}{\jmath}
\newcommand{\PS}[2]{\left(#1\right)_{{#2}}}
\newcommand{\MPS}[2]{\left[#1\right]_{{#2}}}
\newcommand{\ZP}[2]{\tilde{C}_{#1}\left(#2\right)}
\newcommand{\HyperPFQ}[3]{{}_{#1}F_{#2}\left(#3\right)}
\newcommand{\matHyperPFQI}[3]{{}_{#1}\tilde{F}_{#2}\left(#3\right)}
\newcommand{\matHyperPFQII}[4]{{}_{#1}\tilde{F}_{#2}^{\left(#3\right)}\left(#4\right)}
\newcommand{\FN}[1]{\left\|{#1}\right\|_\mathrm{F}}
\newcommand{\DSM}[1]{\mathcal{L}\left(#1\right)}
\newcommand{\AF}{\mathsf{EFF}_\text{STBC}}
\newcommand{\CC}[2]{\B{\Phi}_{#1}^{(\text{c})}\left(#2\right)}
\newcommand{\EC}[2]{\B{\Phi}_{#1}^{(\text{e})}\left(#2\right)}
\newcommand{\TC}[2]{\B{\Phi}_{#1}^{(\text{t})}\left(#2\right)}
\newcommand{\EV}[1]{\pmb{\lambda}\left(#1\right)}
\newcommand{\CN}[1]{\zeta\left(#1\right)}
\newcommand{\CNs}[1]{\zeta\bigl(#1\bigr)}
\newcommand{\GF}{{\B{\mathcal{G}}_4}}
\newcommand{\PFC}[3]{\mathcal{X}_{#1,#2}\left(#3\right)}
\newcommand{\DsMIMO}[3]{{\left(#1,#3,#2\right)}}
\newcommand{\Tc}{{N_{\text{c}}}}
\newcommand{\tSNRmin}{\frac{E_\text{b}^\text{r}}{N_0}_\text{min}}
\newcommand{\SNRmin}{\frac{E_\text{b}}{N_0}_\text{min}}

\newcommand{\EFF}[1]{{\mathsf{EFF}_{#1}}}
\newcommand{\Kurt}[1]{\kappa\left(#1\right)}

\newcommand{\mySep}{\vspace{5pt}}
\newcommand{\EqF}{\nonumber\\[-0.7cm]}
\newcommand{\EqE}{\\[-0.8cm]\nonumber}

\makeatletter
\def\@setsize#1#2#3#4{
    \@nomath#1
    \let\@currsize#1
    \baselineskip #2
    \baselineskip \baselinestretch\baselineskip
    \parskip \baselinestretch\parskip
    \setbox\strutbox \hbox{
        \vrule height.7\baselineskip
            depth.3\baselineskip
            width\z@}
    \skip\footins \baselinestretch\skip\footins
    \normalbaselineskip\baselineskip#3#4}
\makeatother

\makeatletter
\newcommand{\setstretch}[1]{
    \def\baselinestretch{#1}%
    \@currsize
    }
\makeatother

\makeatletter
\newenvironment{salign}{
    \vskip 0.0\baselineskip
    \setstretch{1}
    \start@align\@ne\st@rredfalse\m@ne
    }{
    \math@cr
    \black@\totwidth@
    \egroup
    \ifingather@
        \restorealignstate@
        \egroup
        \nonumber
        \ifnum0=`{\fi\iffalse}\fi
        \else
        $$%
    \fi
    \ignorespacesafterend
    \vskip 0.6\baselineskip
    \par
    \noindent
    }
\makeatother

\date{July 5, 2006}

\begin{document}
\title{
        \hspace{4cm}\\[2.5cm]
        MIMO Diversity in the Presence of\\ Double Scattering
      }
\author{
        \hspace{4cm}\\
        {
        Hyundong Shin$^\dag$, \IEEEmembership{Member, IEEE}
        and
        Moe Z. Win, \IEEEmembership{Fellow, IEEE}
        }
\thanks{$^\dag$ Corresponding author}
\thanks{This research was supported in part by
            the Korea Research Foundation Grant funded by the
            Korean Government (KRF-2004-214-D00337),
            the Charles Stark Draper Endowment,
            the Office of Naval Research Young Investigator Award N00014-03-1-0489,
            and
            the National Science Foundation under Grant ANI-0335256.
            }
\thanks{H. Shin was with
            the Laboratory for Information and Decision Systems (LIDS),
            Massachusetts Institute of Technology,
            Cambridge, MA 02139 USA. He is now with the School of Electronics and Information, Kyung Hee
            University, 1 Seocheon, Kihung, Yongin, Kyungki 446-701, Korea
            (e-mail: {\tt hshin@khu.ac.kr}).}
\thanks{M. Win is with
            the Laboratory for Information and Decision Systems (LIDS),
            Massachusetts Institute of Technology,
            77 Massachusetts Avenue,
            Cambridge, MA 02139 USA
            (e-mail: {\tt moewin@mit.edu}).}
}

\markboth{Revised for Publication in the IEEE Transactions on
            Information Theory}
         {Shin and Win:
          MIMO Diversity in the Presence of Double Scattering}

\maketitle

\clearpage

\begin{abstract}
The potential benefits of multiple-antenna systems may be limited
by two types of channel degradations \hspace{-0.3em}---\emph{rank
deficiency} and \emph{spatial fading correlation} of the channel.
In this paper, we assess the effects of these degradations on the
diversity performance of  multiple-input multiple-output (MIMO)
systems, with an emphasis on orthogonal space--time block codes,
in terms of the symbol error probability, the effective fading
figure (EFF), and the capacity at low signal-to-noise ratio (SNR).
In particular, we consider a general family of MIMO channels known
as \emph{double-scattering} channels, which encompasses a variety
of propagation environments from independent and identically
distributed Rayleigh to degenerate keyhole or pinhole cases by
embracing both rank-deficient and spatial correlation effects. It
is shown that a MIMO system with $\nTx$ transmit and $\nRx$
receive antennas achieves the diversity of order
$\frac{\nTx\nSx\nRx}{\max\left(\nTx,\nSx,\nRx\right)}$ in a
double-scattering channel with $\nSx$ effective scatterers. We
also quantify the combined effect of the spatial correlation and
the lack of scattering richness on the EFF and the low-SNR
capacity in terms of the \emph{correlation figures} of transmit,
receive, and scatterer correlation matrices. We further show the
monotonicity properties of these performance measures with respect
to the strength of spatial correlation, characterized by the
eigenvalue majorization relations of the correlation matrices.
\end{abstract}


\section{Introduction}  \label{sec:Sec:1}

Recent rapid advances in multiple-input multiple-output (MIMO)
communication theory and growing cognizance of the tremendous
performance gains achieved by MIMO techniques
\cite{Win:87:JSAC,Fos:96:BLTJ,FG:98:WPC,Tel:99:ETT,WZH:05:IT,TSC:98:IT,Ala:98:JSAC,TJC:99:IT,TJC:99:JSAC}
have spurred efforts to integrate this technology into future
wireless systems such as wireless local area networks (WLANs) and
4G cellular systems. 
One of the approaches to exploiting diversity capability of MIMO
channels is the use of orthogonal space--time block codes
(OSTBCs), which have drawn considerable attention because they
attain full diversity with scalar maximum-likelihood (ML) decoding
\cite{Ala:98:JSAC,TJC:99:IT,TJC:99:JSAC}.\footnote{
        However, OSTBCs with arbitrary complex constellation cannot provide the
        full diversity and full transmission rate simultaneously for more
        than two transmit antennas \cite[Theorem~5.4.2]{TJC:99:IT} (see
        also \cite{TH:02:IT,SX:03:IT,WX:03:IT,LX:03:IT}). A new class of
        quasi-orthogonal codes has been proposed in
        \cite{Jaf:01:COM,SetRS:03:IT,SX:04:IT} with the tradeoff between
        the decoding complexity, transmission rate and/or diversity.
        }

In general, the potential benefits of multiple-antenna systems may
be limited by rank deficiency of the channel due to double
scattering or the keyhole effect, for example, as well as spatial
fading correlation due, for instance, to insufficient spacing
between antenna elements
\cite{SFGK:00:COM,CTKV:02:IT,KSPMF:02:JSAC,IUN:03:JSAC,SL:03:IT,CWZ:03:IT,SWLC:06:WCOM,GBGP:02:COM,CFGV:02:WCOM,MSTBT:02:JSAC,Mol:04:SP,LK:02:COM,ATM:03:CL,CF:04:CL}.
Some mechanism rendering a MIMO channel rank deficient cannot be
explained by the archetypal model based on single-scattering
processes \cite{MSTBT:02:JSAC,Mol:04:SP}. To address this issue, a
double-scattering MIMO model has been proposed recently in
\cite{GBGP:02:COM} wherein the channel matrix is characterized by
a product of two statistically independent complex Gaussian
matrices, in contrast to the common single complex Gaussian matrix
characterization for wireless MIMO channels.\footnote{
        In \cite{GBGP:02:COM}, the model was validated by
        simulations using ray tracing techniques.
        }
This double-scattering model can capture both rank-deficient and
spatial correlation effects of MIMO channels and encompass a
variety of propagation environments, bridging the gap between an
independent and identically distributed (i.i.d.) Rayleigh case and
a degenerate one-rank channel known as a keyhole or pinhole
channel. There are other recent attempts to modeling MIMO channels
for more realistic scattering environments (e.g., double or
multibounce diffuse scattering) beyond single scattering
\cite{Bur:03:JSAC,DM:05:IT,PBT:05:IT,PTB:06:IT}.

The effects of rank deficiency and spatial correlation on the
capacity of MIMO channels are relatively well understood (see,
e.g.,
\cite{SFGK:00:COM,KSPMF:02:JSAC,IUN:03:JSAC,GBGP:02:COM,CFGV:02:WCOM,MSTBT:02:JSAC,Mol:04:SP,CTKV:02:IT,LK:02:COM,ATM:03:CL,SL:03:IT,CWZ:03:IT,SWLC:06:WCOM,CF:04:CL}).
From a capacity point of view, it has been known that at high
signal-to-noise ratio (SNR), the spatial fading correlation
reduces the diversity advantage---a parallel shift of the capacity
curve over SNR in decibels (dB)---offered by multiple antennas,
whereas the rank deficiency decreases the spatial multiplexing
benefit---a slope of the capacity curve over SNR---of
multiple-antenna channels \cite{SL:03:IT}. 
%
%
Previously, the performance of space--time coding in the presence
of spatial fading correlation has been extensively studied for the
most popular Rayleigh, Rician, and Nakagami-$m$ fading
\cite{JS:04:CL,WSFY:04:COM,FMSL:04:IT,NBP:05:WCOM,MA:05:GLOBECOM,MA:05:WCOM}.
Also, the effect of rank deficiency has been investigated in
\cite{SL:03:CL,SL:04:VT,NBJ:04:EL,SHN:04:GLOBECOM} for a special
case of the keyhole channel.

The objective of this paper is to assess the effects of double
scattering on the diversity performance of MIMO systmes in a
communication link with $\nTx$ transmit antennas, $\nRx$ receive
antennas, and $\nSx$ effective scatterers on each of the transmit
and receive sides, which is referred to as a ``double-scattering
$\DsMIMO{\nTx}{\nRx}{\nSx}$-MIMO channel." Due to the channel
decoupling property, the OSTBC converts a MIMO fading channel into
identical single-input single-output (SISO) subchannels, each for
a different transmitted symbol, with a path gain given by the
Frobenius norm\footnote{
        The Frobenius norm of an $m \times n$ matrix
        $\B{A}=\left(A_{ij}\right)$ is defined as
        \begin{align*}
            \FN{\B{A}}
            \triangleq
                \sqrt{
                    \tr\bigl(
                            \B{AA}^\dag
                    \bigr)
                }
            =
                \left(
                    \sum_{i=1}^m
                    \sum_{j=1}^n
                        |A_{ij}|^2
                \right)^{1/2}
        \end{align*}
        %
        %
        where $\tr\left(\cdot\right)$ and $\dag$ denote the trace
        operator and the transpose conjugate of a matrix,
        respectively.
}
of the channel matrix $\B{H}$
\cite{SL:03:CL,SL:04:VT,NBP:05:WCOM,MA:05:GLOBECOM,MA:05:WCOM}. As
a result, the maximum achievable diversity performance of MIMO
systems can be characterized by the statistical property of
$\FN{\B{H}}$. Therefore, using the OSTBC as a pivotal MIMO
diversity technique\footnote{
        If the transmitter has channel knowledge, the maximum MIMO
        diversity can be achieved by \emph{transmit
        beamforming} (often called maximum ratio transmission (MRT) or MIMO maximal-ratio
        combining) in the eigenspace of the
        largest eigenvalue of the Gramian matrix $\B{H}^\dag\B{H}$ \cite{Lo:99:COM,KA:03:JSAC,DMJ:03:COM,Gra:05:COM}.
        }
(particularly, in the absence of channel knowledge at the
transmitter), we analyze the relevant performance measures in
double-scattering $\DsMIMO{\nTx}{\nRx}{\nSx}$-MIMO channels,
namely: i) the symbol error probability (SEP) \cite{SA:00:Book},
ii) the effective fading figure (EFF)
\cite{WW:99:COM,WK:99:JSAC,Win:05:WCOM}, and iii) the capacity in
a low-SNR regime \cite{Ver:02:IT,LTV:03:IT}.

Diversity in communication can ameliorate system performances in
behalf of error probability, information rate, and signal
fluctuation due to fading. From a error probability viewpoint, the
diversity attacks a high-SNR slope of the SEP curve, i.e.,
diversity order. In contrast, the diversity (from a capacity point
of view) affects a low-SNR slope of the capacity curve rather than
a high-SNR slope. For example, the high- and low-SNR slopes
(bits/s/Hz per $3$ dB) of the capacity for i.i.d.\ Rayleigh-fading
MIMO channels are given by
\begin{align*}
    S_{\infty}
    &=
        \min\left(\nTx,\nRx\right)
    \\
    S_{0}
    &=
        \frac{2\nTx\nRx}{\nTx+\nRx}
\end{align*}
respectively \cite{Ver:02:IT}. While the high-SNR capacity slope
$S_{\infty}$ is limited by the spatial multiplexing gain
$\min\left(\nTx,\nRx\right)$, the low-SNR capacity slope $S_{0}$
is limited by the diversity gain amounting to the harmonic mean of
$\nTx$ and $\nRx$. Therefore, the capacity is multiplexing-limited
in the high-SNR regime, but is diversity-limited in the low-SNR
regime. At high SNR, the diversity advantage serves only to
provide the power offset (i.e., the parallel shift of the capacity
curve) \cite{SL:03:IT}. These lessons stimulate a shift of focus
to the low-SNR regime in analyzing the diversity effect on the
capacity behavior. More inherently, diversity systems aim to
reduce signal fluctuations due to the nature of fading. The EFF
measure is defined as a \emph{variance-to-mean-square ratio
(VMSR)} of the instantaneous SNR (see Definition~\ref{def:EFF}).
This quantity can be used to assess the severity of fading and the
effectiveness of diversity systems on reducing signal
fluctuations. 
The main results of this paper can be summarized  as follows.
\begin{itemize}

\item

We show that the achievable diversity is of order
\begin{align*}
    \frac{\nTx\nSx\nRx}
         {\max\left(\nTx,\nSx,\nRx\right)}
    \, .
\end{align*}
Hence, if the channel is ``rich-enough," that is, the number of
effective scatterers is greater than or equal to the numbers of
transmit and receive antennas, the full spatial diversity order of
$\nTx\nRx$ can be achieved even in the presence of double
scattering.

\item

We derive exact analytical expressions for the SEP 
in three cases of
particular interest:

\begin{enumerate}

\item

spatially uncorrelated double scattering (includes i.i.d.\ and
keyhole channels as special cases);

\item

doubly correlated double scattering (includes a spatially
correlated MIMO channel where spatial correlation is present at
both the transmitter and the receiver);

\item

multiple-input single-output (MISO) double scattering (corresponds
to a pure transmit diversity system wherein a burden of diversity
reception at the receive terminal is moved to the
transmitter---original motivation of space--time coding
\cite{TSC:98:IT,Ala:98:JSAC,TJC:99:IT}).

\end{enumerate}

\item

We derive the EFF and the low-SNR capacity of double-scattering
$\DsMIMO{\nTx}{\nRx}{\nSx}$-MIMO channels. The results show that
these performance measures are completely characterized by the
\emph{correlation figures} of transmit, receive, and scatterer
correlation matrices.\footnote{
        The correlation figure is
        defined as a ratio of the second-order statistic of the spectra of
        correlation matrices to that of the fully correlated
        matrix (see Definition~\ref{def:CN}).
        }

\item

The EFF as a functional of the eigenvalues of correlation matrices
is \emph{monotonically increasing in a sense of Schur
(MIS)}.\footnote{
        See Appendix~\ref{sec:Appendix:A} for the notions of \emph{Schur
        monotonicity} and
        \emph{majorization}.
}
We show that the maximum possible increase in the EFF due to
double scattering is a sum of correlation figures of the transmit
and receive correlation matrices, which eventuates when the
scatterers tend to be fully correlated or the keyhole propagation
takes place, that is, when only a single degree of freedom is
available in the channel for communications.

\item

%
%
The low-SNR capacity slope as a functional of the eigenvalues of
correlation matrices is \emph{monotonically decreasing in a sense
of Schur (MDS)}. We also obtain the low-SNR capacity of a
double-scattering MIMO channel without the constraint of
orthogonal input signaling. This enables us to assess the penalty
of the use of OSTBCs (for achieving full diversity with simple
decoding) on spectral efficiency in the low-SNR regime. 

\end{itemize}


%
%
We note in passing that all the mathematical and statistical
results (on the monotonicity in a sense of \emph{Schur} and random
matrices) obtained in the appendices are applicable to many other
problems related to multiple-antenna communications---for example,
capacity analysis of MIMO relay channels \cite{WZH:05:IT} and
spatially correlated MIMO channels
\cite{SL:03:IT,CWZ:03:IT,SWLC:06:WCOM}, and error probability
analysis of multiple-antenna systems with cochannel interference
\cite{CWZ:03:COM,CWZ:05:COM}.

This paper is organized as follows. In Section~\ref{sec:Sec:2},
the system model considered in the paper is presented.
Section~\ref{sec:Sec:3} analyzes the achievable diversity and the
SEP in the presence of double scattering. Section~\ref{sec:Sec:4}
analyzes the EFF and the low-SNR capacity (with and without the
use of OSTBCs) of double-scattering
$\DsMIMO{\nTx}{\nRx}{\nSx}$-MIMO channels. Section~\ref{sec:Sec:5}
concludes the paper. Apropos of our study, the notions of
majorization and Schur monotonicity are briefly discussed in
Appendix~\ref{sec:Appendix:A}. In Appendix~\ref{sec:Appendix:B},
we provide supplementary useful results on some statistics derived
from complex Gaussian matrices.

\emph{Notation:} Throughout the paper, we shall use the following
notation. $\mathbb{N}$, $\R$, and $\C$ denote the natural numbers
and the fields of real and complex numbers, respectively. The
superscripts $\ast$, $T$, and $\dag$ stand for the complex
conjugate, transpose, and transpose conjugate, respectively.
$\B{I}_n$ and $\B{0}_{m \times n}$ represent the $n \times n$
identity matrix and the $m \times n$ all-zero matrix,
respectively. $\left(A_{ij}\right)$ denotes the matrix with the
$\left(i,j\right)$th entry $A_{ij}$ and $\det_{1 \leq i,j \leq n}
\left(A_{ij}\right)$ is the determinant of the $n \times n$ matrix
$\left(A_{ij}\right)$. $\tr\left(\B{A}\right)$,
$\etr\left(\B{A}\right)=e^{\tr\left(\B{A}\right)}$, and
$\FN{\B{A}}$ denote the trace, exponential of the trace, and
Frobenius norm of the matrix $\B{A}$, respectively. $\otimes$ and
$\oplus$ denote the Kronecker (direct) product and direct sum of
matrices and $\vecOp\left(\B{A}\right)$ denotes the vector formed
by stacking all the columns of $\B{A}$ into a column vector. Also,
we denote $\B{A}_1 \otimes \B{A}_2 \otimes \cdots \otimes \B{A}_n$
by $\bigotimes_{i=1}^n \B{A}_i$ and $\B{A}_1 \oplus \B{A}_2 \oplus
\cdots \oplus \B{A}_n$ by $\bigoplus_{i=1}^n \B{A}_i$. With a
slight abuse of notation, a positive-semidefinite matrix $\B{A}$
is denoted by $\B{A}\geq 0$ and a positive-definite matrix $\B{A}$
is denoted by $\B{A} > 0$. Finally, for a Hermitian matrix $\B{A}
\in \C^{n \times n}$ with the eigenvalues
$\lambda_1,\lambda_2,\ldots,\lambda_n$ in any order, $\EVn{\B{A}}$
denotes the number of distinct eigenvalues of $\B{A}$. Also,
$\odEV{\lambda}{k}{}$ and $\EVm{k}{\B{A}}$,
$k=1,2,\ldots,\EVn{\B{A}}$, denote the distinct eigenvalues of
$\B{A}$ in decreasing order and its multiplicity, respectively,
that is,
$\odEV{\lambda}{1}{}
> \odEV{\lambda}{2}{}
> \ldots > \odEV{\lambda}{\EVn{\B{A}}}{}$ and $\sum_{k=1}^\EVn{\B{A}} \EVm{k}{\B{A}}
=n$.

\section{System Model}   \label{sec:Sec:2}

We consider a MIMO wireless communication system with $\nTx$
transmit and $\nRx$ receive antennas, where the channel remains
constant for an integer multiple of $\Tc$ ($\geq \nTx$) symbol
periods and changes independently to a new value for each
coherence time. We assume that the channel is perfectly known at
the receiver but unknown at the transmitter.

\subsection{Orthogonal Space--Time Block Codes}

\begin{figure}[t]
    \centerline{\includegraphics[width=0.95\textwidth]{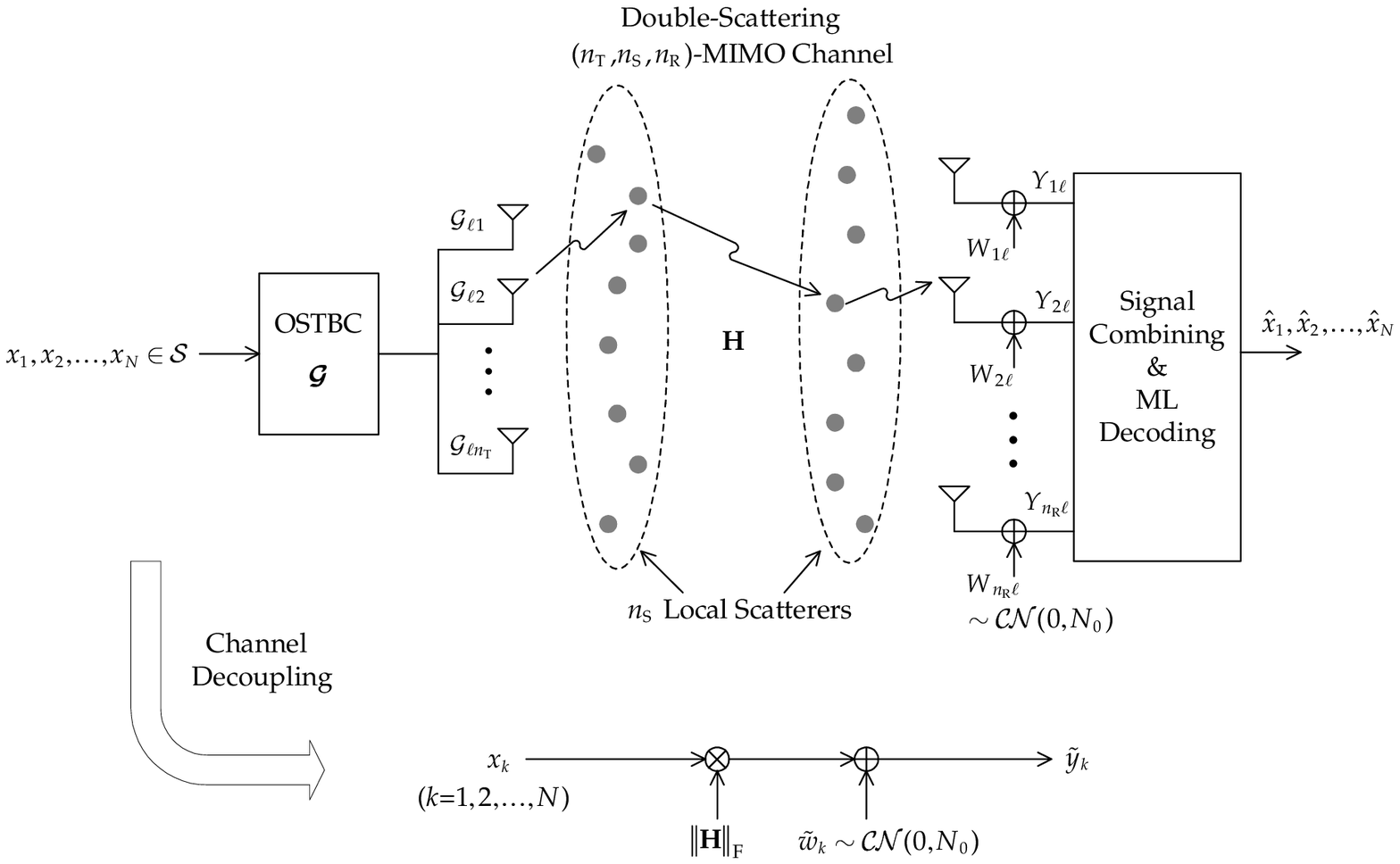}}
    \caption{
        Block diagram of a space--time block coded system in double-scattering $\DsMIMO{\nTx}{\nRx}{\nSx}$-MIMO channels and induced SISO subchannels.
    }
    \label{fig:Fig:1}
\end{figure}

A space--time block coded MIMO system in double-scattering
channels is illustrated in Fig.~\ref{fig:Fig:1}. During an
$\Tc$-symbol interval, symbols $x_i \in \mathcal{S}$,
$i=1,2,\ldots,N$, are encoded by an OSTBC defined by an $\Tc
\times \nTx$ transmission matrix $\B{\mathcal{G}}$, where
$\mathcal{S}$ is two-dimensional signaling constellation
\cite{TJC:99:IT,TJC:99:JSAC}. A general construction of complex
OSTBCs with the minimal delay and maximal achievable rate was
presented in \cite[Proposition~2]{TH:02:IT}. 
%
%
This construction of the OSTBC for $\nTx$ transmit antennas gives
the maximal achievable rate \cite[Theorem~1]{TH:02:IT}
\begin{align}    \label{eq:rate}
    \EqF
    \mathcal{R}=
        \frac{\lceil \log_2 \nTx \rceil+1}
             {2^{\lceil \log_2 \nTx \rceil}}
    \EqE
\end{align}
where $\lceil x \rceil$ denotes the smallest integer greater than
or equal to $x$. For example, Alamouti's code
$\left[\begin{smallmatrix} x_1 & x_2 \\ -x_2^\ast & x_1^\ast
\end{smallmatrix}\right]$ is a one-rate OSTBC
employing two transmit antennas \cite{Ala:98:JSAC} and
\begin{salign}
    \B{\mathcal{G}}_4
    =
        \begin{bmatrix}
            x_1 &   x_2 &   x_3 &   0
            \\
            -x_2^\ast &   x_1^\ast &   0    &   -x_3
            \\
            -x_3^\ast &   0 &   x_1^\ast    &   x_2
            \\
            0   &   x_3^\ast &  -x_2^\ast   &   x_1
        \end{bmatrix}
\end{salign}
is a $3/4$-rate OSTBC for four transmit antennas \cite{TH:02:IT}.

\subsection{Signal and Channel Models}

For a frequency-flat block-fading channel, the $\nRx \times \Tc$
received signal can be expressed in matrix notation as
\begin{align}    \label{eq:RS}
    \B{Y}
    =
        \B{H}
        \B{\mathcal{G}}^T
        +
        \B{W}
\end{align}
where $\B{H}\in \C^{\nRx \times \nTx}$ is the random channel
matrix whose $\left(i,j\right)$th entries $H_{ij}$, $i=1,2,\ldots
,\nRx$, $j=1,2,\ldots ,\nTx$, are complex propagation coefficients
between the $j$th transmit antenna and the $i$th receive antenna
with $\EX{|H_{ij}|^2}=1$, and $\B{W} \sim
\CGM{\nRx}{\Tc}{\B{0}_{\nRx\times\Tc}}{N_0\B{I}_{\nRx}}{\B{I}_{\Tc}}$
is the complex additive white Gaussian noise (AWGN) matrix (see
\cite[Definition~II.1]{SL:03:IT} and \cite[(1)]{SL:03:IT} for the
definition and distribution of complex Gaussian
matrices).\footnote{
        There exist minor typos in
        \cite[Definition~II.1]{SL:03:IT}; the covariance matrix
        $\B{\Sigma}\otimes\B{\Psi}$ should be read as
        $\B{\Sigma}^T \otimes\B{\Psi}$.
        }
The total power transmitted through $\nTx$ antennas is assumed to
be $\mathcal{P}$ and hence, the average SNR per receive antenna is
equal to $\snr \triangleq \mathcal{P}/N_0$.

For double-scattering $\DsMIMO{\nTx}{\nRx}{\nSx}$-MIMO channels
(see Fig. \ref{fig:Fig:1}), the channel matrix $\B{H}$ can be
written as \cite{GBGP:02:COM,SL:03:IT}
\begin{align}    \label{eq:H}
    \B{H}
    =
        \frac{1}{\sqrt{\nSx}} \,
        \RxCM^{1/2}
        \B{H}_1
        \SxCM^{1/2}
        \B{H}_2
        \TxCM^{1/2}
\end{align}
where $\nSx$ is the number of effective scatterers on each of the
transmit and receive sides, $\B{H}_1$ and $\B{H}_2$ are
statistically independent, $\B{H}_1 \sim
\CGM{\nRx}{\nSx}{\B{0}_{\nRx \times
\nSx}}{\B{I}_\nRx}{\B{I}_\nSx}$, $\B{H}_2 \sim
\CGM{\nSx}{\nTx}{\B{0}_{\nSx \times
\nTx}}{\B{I}_\nSx}{\B{I}_\nTx}$, and Hermitian positive-definite
matrices $\TxCM$, $\SxCM$, and $\RxCM$ are $\nTx \times \nTx$
transmit, $\nSx \times \nSx$ scatterer, and $\nRx \times \nRx$
receive correlation matrices with all diagonal entries $1$,
respectively.\footnote{
        In general, a correlation matrix is positive semidefinite
        with all diagonal entries $1$. 
        }
This model can include the rank-deficient effect of MIMO channels
as well as spatial fading correlation by controlling $\nSx$ and
the correlation matrices $\TxCM$, $\SxCM$, and $\RxCM$. Therefore,
\eqref{eq:H} is a general family of MIMO channels spanning from
the i.i.d.\ Rayleigh case ($\nSx \to \infty$ with
$\TxCM=\B{I}_\nTx$, $\SxCM=\B{I}_\nSx$, $\RxCM=\B{I}_\nRx$) to the
degenerate keyhole or pinhole case ($\nSx=1$ with
$\TxCM=\B{I}_\nTx$, $\RxCM=\B{I}_\nRx$) \cite{GBGP:02:COM}. Note
that the separability of correlation in \eqref{eq:H} is a
generalization of the well-known `Kronecker model'
\cite{SFGK:00:COM,CTKV:02:IT}. Although there are some attempts to
reporting discrepancy between this separable correlation model and
physical measurements (see, e.g.,
\cite{OHWWB:03:EL,WHOB:06:WCOM}), the Kronecker correlation model
has been accepted widely due to its experimental validation from
European Project \cite{KSPMF:02:JSAC} and analytical
tractability.

In \cite{IUN:03:JSAC}, so-called \emph{stochastic} rank
deficiency---meaning that the channel is rank deficient due to
fading correlation, i.e., the correlation matrices have zero
eigenvalues---was deemed as an important feature when dealing with
fading correlation. However, this form of channel degeneracy
cannot cover the case where the channel exhibits rank deficiency
even when fading is uncorrelated. In contrast, we shall restrict
$\TxCM$, $\SxCM$, and $\RxCM$ to positive-definite (i.e., full
rank) matrices in the paper. This implies that the rank of $\B{H}$
is equal to $\min\left(\nTx,\nSx,\nRx\right)$ with probability
one. Therefore, rank deficiency can be distinguished from the
fading correlation effect and may occur only due to the lack of
scattering richness with $\nSx$ less than
$\min\left(\nTx,\nRx\right)$. This also enables us to discriminate
a one-rank \emph{fully} correlated scenario from a degenerate
keyhole MIMO channel \cite{ATM:03:CL}, and grants the channel to
exhibit rank deficiency with uncorrelated fading (e.g., $\nSx <
\min\left(\nTx,\nRx\right)$ with $\TxCM=\B{I}_\nTx$,
$\SxCM=\B{I}_\nSx$, $\RxCM=\B{I}_\nRx$).

Let
$\B{\Xi}_1=\RxCM^{1/2}\B{H}_1$ and
$\B{\Xi}_2=\SxCM^{1/2}\B{H}_2 \TxCM^{1/2}$,
%
%
then we have
\begin{align}    \label{eq:H:1}
    \B{H}
    =
        \frac{1}{\sqrt{\nSx}} \,
        \B{\Xi}_1
        \B{\Xi}_2
\end{align}
where $\B{\Xi}_1 \sim \CGM{\nRx}{\nSx}{\B{0}_{\nRx \times
\nSx}}{\RxCM}{\B{I}_\nSx}$ and $\B{\Xi}_2 \sim
\CGM{\nSx}{\nTx}{\B{0}_{\nSx \times \nTx}}{\SxCM}{\TxCM}$ are
statistically independent complex Gaussian matrices.

\section{Symbol Error Probability}  \label{sec:Sec:3}

With perfect channel knowledge at the receiver, orthogonal
space--time block encoding and decoding convert a MIMO fading
channel into $N$ equivalent SISO subchannels, each for a different
symbol, with a path gain $\FN{\B{H}}$
\cite{SL:03:CL,SL:04:VT,NBP:05:WCOM,MA:05:GLOBECOM,MA:05:WCOM} (as
shown in Fig. \ref{fig:Fig:1}). Consequently, the performance of
OSTBCs is completely characterized by the statistical behavior of
$\FN{\B{H}}$ and the instantaneous SNR for each of the SISO
subchannels, denoted by $\gamma_\text{STBC}$, is given by
\cite{SL:03:CL,SL:04:VT}
\begin{align}    \label{eq:STBC:SNR}
    \gamma_\text{STBC}
    &=
        \frac{
                \snr\FN{\B{H}}^2
             }
             {
                \nTx
                \mathcal{R}
             }.
\end{align}

To evaluate the SEP, we need the probability density function
(pdf) or the moment generating function (MGF) of
$\gamma_\text{STBC}$. For double-scattering
$\DsMIMO{\nTx}{\nRx}{\nSx}$-MIMO channels, the MGF of
$\gamma_\text{STBC}$  can be written as
%
%
%
%
\begin{align}
\EqF
    \MGF{\gamma_\text{STBC}}{s; \snr}
    &\triangleq
        \EX{
            \etr\left(
                        -\frac{s\snr}{\nTx\mathcal{R}} \,
                        \B{HH}^\dag
                \right)
           }
    \nonumber \\
    &=
        \EXs{\B{\Xi}_1,\B{\Xi}_2}
            {
                \etr\left(
                        -\frac{s\snr}{\nSx\nTx\mathcal{R}} \,
                         \B{\Xi}_1
                         \B{\Xi}_2
                         \B{\Xi}_2^\dag
                         \B{\Xi}_1^\dag
                    \right)
            }
    \nonumber \\
    \label{eq:MGF:SNR:1}
    &=
        \EXs{\B{\Xi}_1}
            {
                \det\left(
                        \B{I}_{\nSx \nTx}
                        +\frac{s\snr}{\nSx\nTx\mathcal{R}} \,
                         \B{\Xi}_1^\dag
                         \B{\Xi}_1
                         \SxCM
                         \otimes
                         \TxCM
                    \right)^{-1}
            }
    \\
    \label{eq:MGF:SNR:2}
    &=
        \EXs{\B{\Xi}_2}
            {
                \det\left(
                        \B{I}_{\nRx \nSx}
                        +\frac{s\snr}{\nSx\nTx\mathcal{R}} \,
                         \RxCM
                         \otimes
                         \B{\Xi}_2
                         \B{\Xi}_2^\dag
                    \right)^{-1}
            }
\EqE
\end{align}
where \eqref{eq:MGF:SNR:1} and \eqref{eq:MGF:SNR:2} follow from
Lemma~\ref{le:CGM:etr} in Appendix~\ref{sec:Appendix:B}.

\subsection{Achievable Diversity}

Before devoting to deriving the SEP expressions, we discuss the
diversity order achieved by the OSTBC. In general, the achievable
diversity order can be defined as
\begin{align}
    d
    \triangleq
        \lim_{\snr \to \infty}
            \frac{-\log P_\text{e}}
                 {\log \snr}
\end{align}
where $P_\text{e}$ denotes the SEP for two-dimensional signaling
constellation with polygonal decision boundaries. In the absence
of double scattering, the OSTBC provides the maximum achievable
diversity order of $\nTx\nRx$. The corresponding diversity order
in double-scattering $\DsMIMO{\nTx}{\nRx}{\nSx}$-MIMO channels is
given by the following result.


\begin{theorem}     \label{thm:DO}

The diversity order achieved by the OSTBC over double-scattering
$\DsMIMO{\nTx}{\nRx}{\nSx}$-MIMO channels is
\begin{align}        \label{eq:DO:STBC}
    d_\text{STBC}
    =
        \frac{\nTx\nSx\nRx}
             {\max\left(\nTx,\nSx,\nRx\right)}.
\end{align}

\begin{proof}
See Appendix~\ref{sec:proof:thm:DO}.
\end{proof}

\end{theorem}

\mySep

Theorem~\ref{thm:DO} states that if the number of effective
scatterers is greater than or equal to the numbers of transmit and
receive antennas, the OSTBC provides the full diversity order of
$\nTx\nRx$ even in the presence of double scattering.

We now present analytical expressions for the SEP of the OSTBC for
three cases of particular interest---spatially uncorrelated double
scattering, doubly correlated double scattering, and MISO double
scattering. In what follows, a spatial correlation environment of
double-scattering channels is denoted by $\mathbb{T}=\left(\TxCM,
\SxCM, \RxCM\right)$ for given $\nTx$, $\nSx$, and $\nRx$.
%
%
%
%
%
%
%
%
%
%
%
%

\subsection{Spatially Uncorrelated Double Scattering}

Consider a spatial correlation environment
$\mathbb{T}_\text{uc}=\left(\B{I}_\nTx,\B{I}_\nSx,\B{I}_\nRx\right)$.
This spatially uncorrelated double-scattering scenario includes
i.i.d.\ and keyhole MIMO channels as special cases.

Let $n_1=\min\left(\nTx,\nSx\right)$,
$n_2=\max\left(\nTx,\nSx\right)$, and the $n_1 \times n_1$ random
matrix $\B{\Upsilon}$ be
\begin{salign}
    \B{\Upsilon}
    =
        \begin{cases}
            \,
            \B{\Xi}_2
            \B{\Xi}_2^\dag,
            &
            \text{if $\nSx \leq \nTx$}
            \\[5pt]
            \,
            \B{\Xi}_2^\dag
            \B{\Xi}_2,
            &
            \text{if $\nSx > \nTx$},
        \end{cases}
\end{salign}
which is a matrix quadratic form in complex Gaussian matrices
\cite[Definition~II.3]{SL:03:IT}. Then, from \eqref{eq:MGF:SNR:2}
and \eqref{eq:SEP:MPSK} in Appendix~\ref{sec:Appendix:C}, the SEP
of the OSTBC with $M$-PSK signaling in double-scattering
$\DsMIMO{\nTx}{\nRx}{\nSx}$-MIMO channels can be readily written
as
\begin{align}    \label{eq:SEP:MPSK:1}
\EqF
    P_{\text{e,\,MPSK}}
    =
        \frac{1}{\pi}
        \int_0^{\Theta}
        \EX
            {
                \det\left(
                        \B{I}_{n_1 \nRx}
                        +\frac{g \snr}{\nSx\nTx\mathcal{R} \sin^2\theta} \,
                         \RxCM
                         \otimes
                         \B{\Upsilon}
                    \right)^{-1}
            }
            d\theta
\EqE
\end{align}
where we have used the fact that $\B{\Xi}_2\B{\Xi}_2^\dag$ and
$\B{\Xi}_2^\dag\B{\Xi}_2$ have the same nonzero
eigenvalues.\footnote{
        As mentioned in the proof of Theorem~\ref{thm:DO},
        The SEP for the general case
        of arbitrary two-dimensional
        signaling constellation with polygonal decision
        boundaries can be written as a convex combination of terms
        akin to \eqref{eq:SEP:MPSK}.
        Thus, our results can be easily extended to any two-dimensional
        signaling constellation.
        }

In the absence of spatial correlation, the random matrix
$\B{\Upsilon}$ has the Wishart distribution
$\CWM{n_1}{n_2}{\B{I}_{n_1}}$
\cite[Definition~II.2]{SL:03:IT}. 
%
%
Applying Corollary \ref{cor:CGM:detVec:UC} in
Appendix~\ref{sec:Appendix:B} to \eqref{eq:SEP:MPSK:1}, we obtain
the SEP for this spatially uncorrelated environment
$\mathbb{T}_\text{uc}$ as
\begin{align}    \label{eq:SEP:MPSK:UC-DS}
\EqF
    P_{\text{e,\,MPSK}}^\text{uc-ds}
    =
        \frac{1}{\pi \mathcal{A}^\text{uc-ds}}
        \int_0^{\Theta}
            \det\left\{
                \B{\mathsf{G}}^\text{uc-ds}\left(\theta\right)
            \right\}
            d\theta
\end{align}
where
\begin{align}
    \mathcal{A}^\text{uc-ds}
    =
        \prod_{k=1}^{n_1}
            \left(n_2-k\right)!
            \left(k-1\right)!
\EqE
\end{align}
and $\B{\mathsf{G}}^\text{uc-ds}\left(\theta\right)
=\left(\mathsf{G}_{ij}^\text{uc-ds}\left(\theta\right)\right)$ is
the $n_1 \times n_1$ Hankel matrix whose $\left(i,j\right)$th
entry is given by
\begin{align}
\EqF
    \mathsf{G}_{ij}^\text{uc-ds}\left(\theta\right)
    =
        \left(
                n_2-n_1+i+j-2
        \right)!
        \
        \HyperPFQ{2}{0}
                 {
                    n_2-n_1+i+j-1,
                    \nRx;
                    -
                    \frac{g \snr}{\nSx\nTx\mathcal{R} \sin^2\theta}
                 }.
\EqE
\end{align}

\mySep

\begin{example}[Uncorrelated Extremes---Keyhole and I.I.D.]  \label{ex:extremes:UC-DS}
The i.i.d. and keyhole MIMO channels  are two extreme cases of
spatially uncorrelated double scattering (i.e., $\nSx=\infty$ and
$\nSx=1$, respectively). If $\nSx=1$, then $n_1=1$ and $n_2=\nTx$.
Hence, \eqref{eq:SEP:MPSK:UC-DS} reduces to
\cite[eq.~(11)]{SL:03:CL} for keyhole MIMO channels. As $\nSx \to
\infty$, \eqref{eq:SEP:MPSK:UC-DS} becomes
\cite[eq.~(26)]{SL:04:VT} (with a Nakagami parameter $m=1$) for
i.i.d.\ Rayleigh-fading MIMO channels.

\end{example}

\mySep

\begin{figure}[t]
    \centerline{\includegraphics[width=0.7\textwidth]{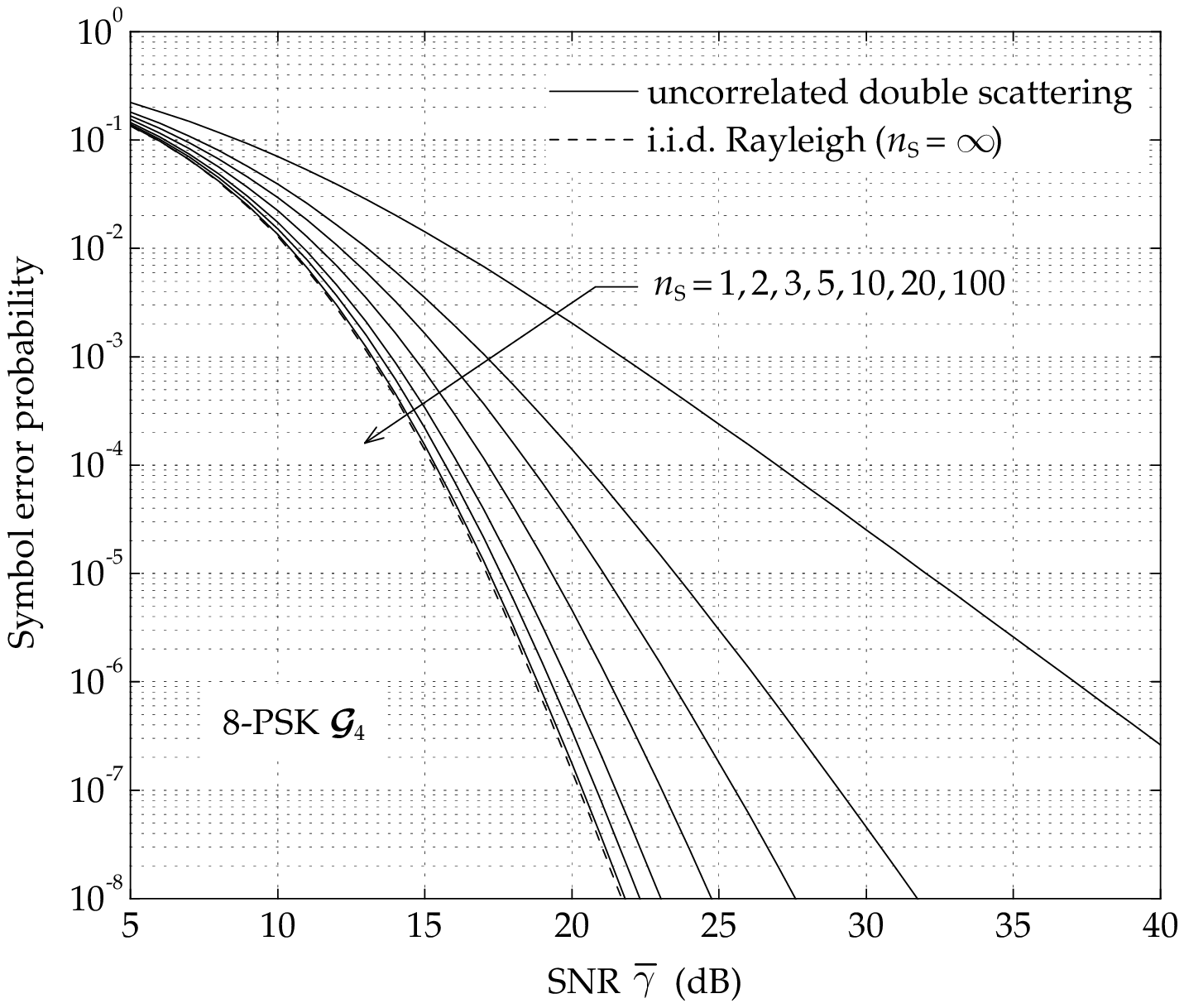}}
    \caption{
        SEP of $8$-PSK $\GF$ ($2.25$ bits/s/Hz) versus $\snr$ in spatially
        uncorrelated double-scattering
        $\DsMIMO{4}{2}{\nSx}$-MIMO channels.
        $\nSx=1$, $2$, $3$, $5$, $10$, $20$, $50$,
        $100$, $\infty$ (i.i.d.\ Rayleigh).
    }
    \label{fig:Fig:4}
\end{figure}

\begin{figure}[t]
    \centerline{\includegraphics[width=0.7\textwidth]{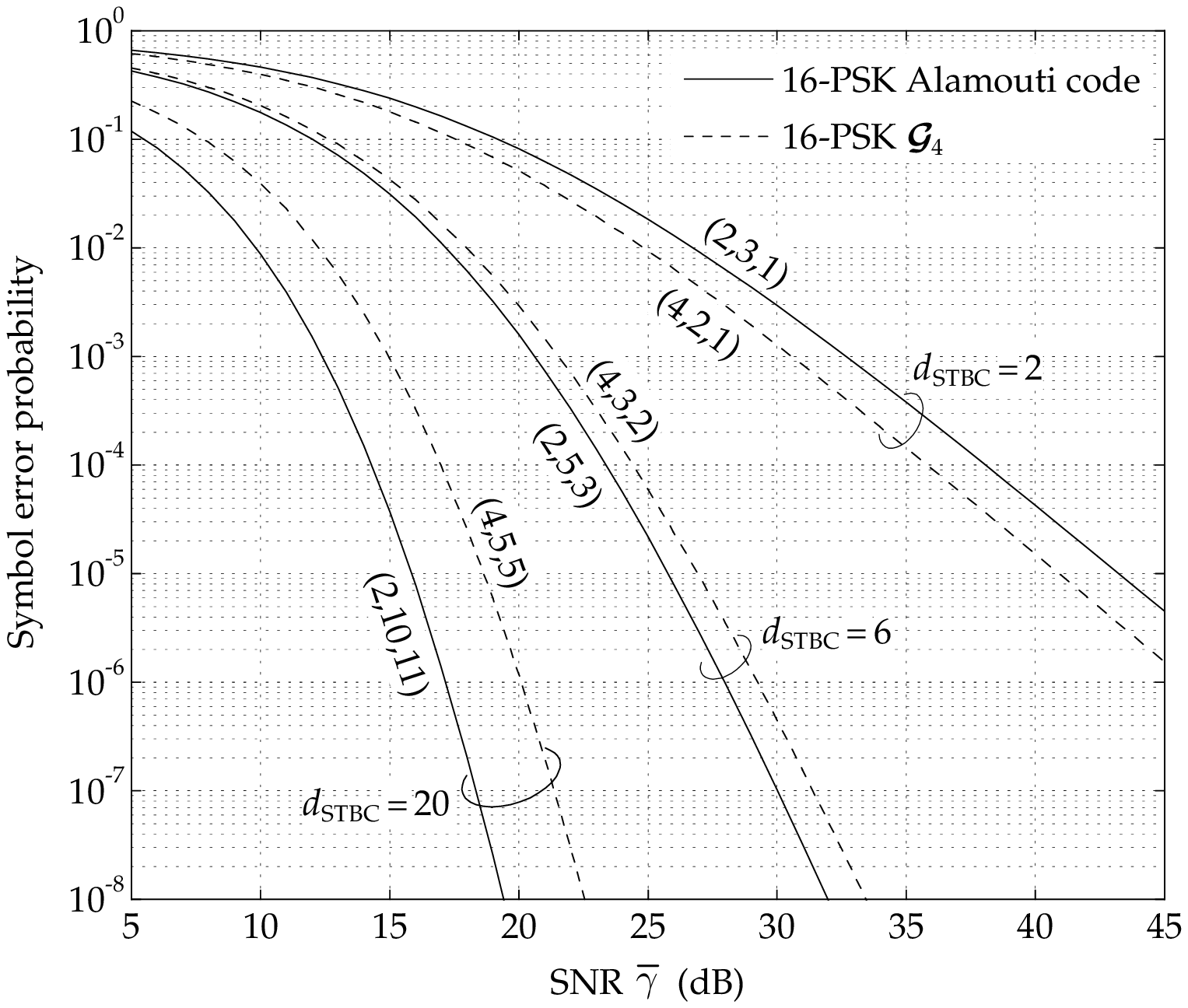}}
    \caption{
        SEP of $16$-PSK Alamouti ($4$ bits/s/Hz) and $\GF$ ($3$ bits/s/Hz) OSTBCs versus $\snr$
        in spatially uncorrelated
        double-scattering $\DsMIMO{\nTx}{\nRx}{\nSx}$-MIMO
        channels. The Alamouti and $\GF$ codes achieve the diversity order of
        $d_\text{STBC}=2$ in $\DsMIMO{2}{1}{3}$ and
        $\DsMIMO{4}{1}{2}$ links, respectively. The
        $d_\text{STBC}$'s for $\DsMIMO{2}{3}{5}$, $\DsMIMO{4}{2}{3}$
        and $\DsMIMO{2}{11}{10}$, $\DsMIMO{4}{5}{5}$ pairs are $6$ and $20$, respectively.
    }
    \label{fig:Fig:5}
\end{figure}

Fig.~\ref{fig:Fig:4} shows the SEP of $8$-PSK $\GF$ ($2.25$
bits/s/Hz) versus the SNR $\snr$ in spatially uncorrelated
double-scattering $\DsMIMO{4}{2}{\nSx}$-MIMO channels when $\nSx$
varies from $1$ (keyhole) to infinity (i.i.d.\ Rayleigh). We can
see that as $\nSx$ increases, the SEP approaches that of i.i.d.\
Rayleigh-fading MIMO channels in the absence of double scattering.
This resembles the behavior in Rayleigh-fading channels with
diversity reception, that is, the channel behaves like an AWGN
channel (diversity order of $\infty$) as the number of receive
antennas increases. Observe that when $\nSx \geq 4$, the slope of
the SEP curve at high SNR is identical to that of the i.i.d.\
case. This example confirms the result of Theorem~\ref{thm:DO}:
the diversity orders are equal to $d_\text{STBC}=2$, $4$, and $6$
for $\nSx=1$, $2$, and $3$, respectively, whereas
$d_\text{STBC}=8$ for $\nSx=5$, $10$, $20$, $100$, and $\infty$
(i.i.d.). A clearer understanding about the diversity behavior is
obtained by referring to Fig.~\ref{fig:Fig:5}, where the SEPs of
$16$-PSK Alamouti ($4$ bits/s/Hz) and $\GF$ ($3$ bits/s/Hz) OSTBCs
versus the SNR $\snr$ in spatially uncorrelated double-scattering
$\DsMIMO{\nTx}{\nRx}{\nSx}$-MIMO channels are shown. Using
\eqref{eq:DO:STBC}, we can easily show that the Alamouti and $\GF$
codes achieve the diversity order of $d_\text{STBC}=2$ for
$\DsMIMO{2}{1}{3}$ and $\DsMIMO{4}{1}{2}$ channels;
$d_\text{STBC}=6$ for $\DsMIMO{2}{3}{5}$ and $\DsMIMO{4}{2}{3}$
channels; and $d_\text{STBC}=20$ for $\DsMIMO{2}{11}{10}$ and
$\DsMIMO{4}{5}{5}$ channels. As can be seen, we obtain a close
agreement in the slopes of the SEP curves, corresponding to the
same value of $d_\text{STBC}$, at high SNR.

\subsection{Doubly Correlated Double Scattering}

Consider a spatial correlation environment
$\mathbb{T}_\text{dc}=\left(\TxCM, \B{I}_\nSx, \RxCM\right)$,
where spatial correlation exists only on the transmit and receive
ends. Note that this scenario includes a spatially correlated MIMO
channel in the absence of double scattering ($\nSx=\infty$) as a
special case. Let $\TxEV{i}$ and $\RxEV{j}$, $i=1,2,\ldots,\nTx$,
$j=1,2,\ldots,\nRx$, be the eigenvalues of $\TxCM$ and $\RxCM$ in
any order, respectively.
Suppose that $\nSx \geq \nTx$. Then, $\B{\Upsilon} \sim
\CWM{\nTx}{\nSx}{\TxCM}$. Applying Theorem~\ref{thm:CGM:detVec} in
Appendix~\ref{sec:Appendix:B} to \eqref{eq:SEP:MPSK:1}, we obtain
the SEP in the environment $\mathbb{T}_\text{dc}$ as
\begin{salign}    \label{eq:SEP:MPSK:DC-DS}
    P_{\text{s,\,MPSK}}^\text{dc-ds}
    =
        \frac{1}{\pi \mathcal{A}^\text{dc-ds}}
        \int_0^{\Theta}
            \det\left(
                        \begin{bmatrix}
                            \B{\mathsf{G}}_1^\text{dc-ds}\left(\theta\right)
                            &
                            \B{\mathsf{G}}_2^\text{dc-ds}\left(\theta\right)
                            &
                            \cdots
                            &
                            \B{\mathsf{G}}_\EVn{\TxCM}^\text{dc-ds}\left(\theta\right)
                        \end{bmatrix}
                \right)
            \,
            d\theta
\end{salign}
with
\begin{salign}
    \mathcal{A}^\text{dc-ds}
    =
        \det\left(
                    \begin{bmatrix}
                        \B{\mathsf{B}}_1^\text{dc-ds}
                        &
                        \B{\mathsf{B}}_2^\text{dc-ds}
                        &
                        \cdots
                        &
                        \B{\mathsf{B}}_\EVn{\TxCM}^\text{dc-ds}
                    \end{bmatrix}
             \right)
        \cdot
        \prod_{i=1}^{\nTx}
            \left(\nSx-i\right)!
\end{salign}
where
$\B{\mathsf{B}}_k^\text{dc-ds}=\bigl(\mathsf{B}_{k,ij}^\text{dc-ds}\bigr)$
and $\B{\mathsf{G}}_k^\text{dc-ds}\left(\theta\right)
=\left(\mathsf{G}_{k,ij}^\text{dc-ds}\left(\theta\right)\right)$,
$k=1,2,\ldots,\EVn{\TxCM}$, are $\nTx \times \EVm{k}{\TxCM}$
matrices whose $\left(i,j\right)$th entries are given respectively
by
\begin{align}
\EqF
    \mathsf{B}_{k,ij}^\text{dc-ds}
    =
        \left(-1\right)^{i-j} \,
        \PS{i-j+1}{j-1} \,
        \odTxEV{k}^{\nSx-i+j}
\end{align}
and
\begin{align}       \label{eq:G:DC-DS}
    \mathsf{G}_{k,ij}^\text{dc-ds}\left(\theta\right)
    =
        \sum_{p=1}^\EVn{\RxCM}
        \sum_{q=1}^{\EVm{p}{\RxCM}}
            \Biggl\{
            &
                \PFC{p}{q}{\RxCM}
                \cdot
                \odTxEV{k}^{\nSx-\nTx+i+j-1}
                \left(
                    \nSx-\nTx+i+j-2
                \right)!
    \nonumber \\
    & \times
                \HyperPFQ{2}{0}{
                                \nSx-\nTx+i+j-1,
                                q \, ;
                                -\frac{
                                        g
                                        \snr
                                        \odRxEV{p}
                                        \odTxEV{k}
                                      }
                                      {
                                        \nSx
                                        \nTx
                                        \mathcal{R}
                                        \sin^2\theta}
                                }
            \Biggr\}.
\EqE
\end{align}
In \eqref{eq:G:DC-DS}, $\PFC{p}{q}{\RxCM}$ is the
$\left(p,q\right)$th characteristic coefficient of $\RxCM$ (see
Definition~\ref{def:PFC} in Appendix~\ref{sec:Appendix:B}).

\begin{figure}[t]
    \centerline{\includegraphics[width=0.7\textwidth]{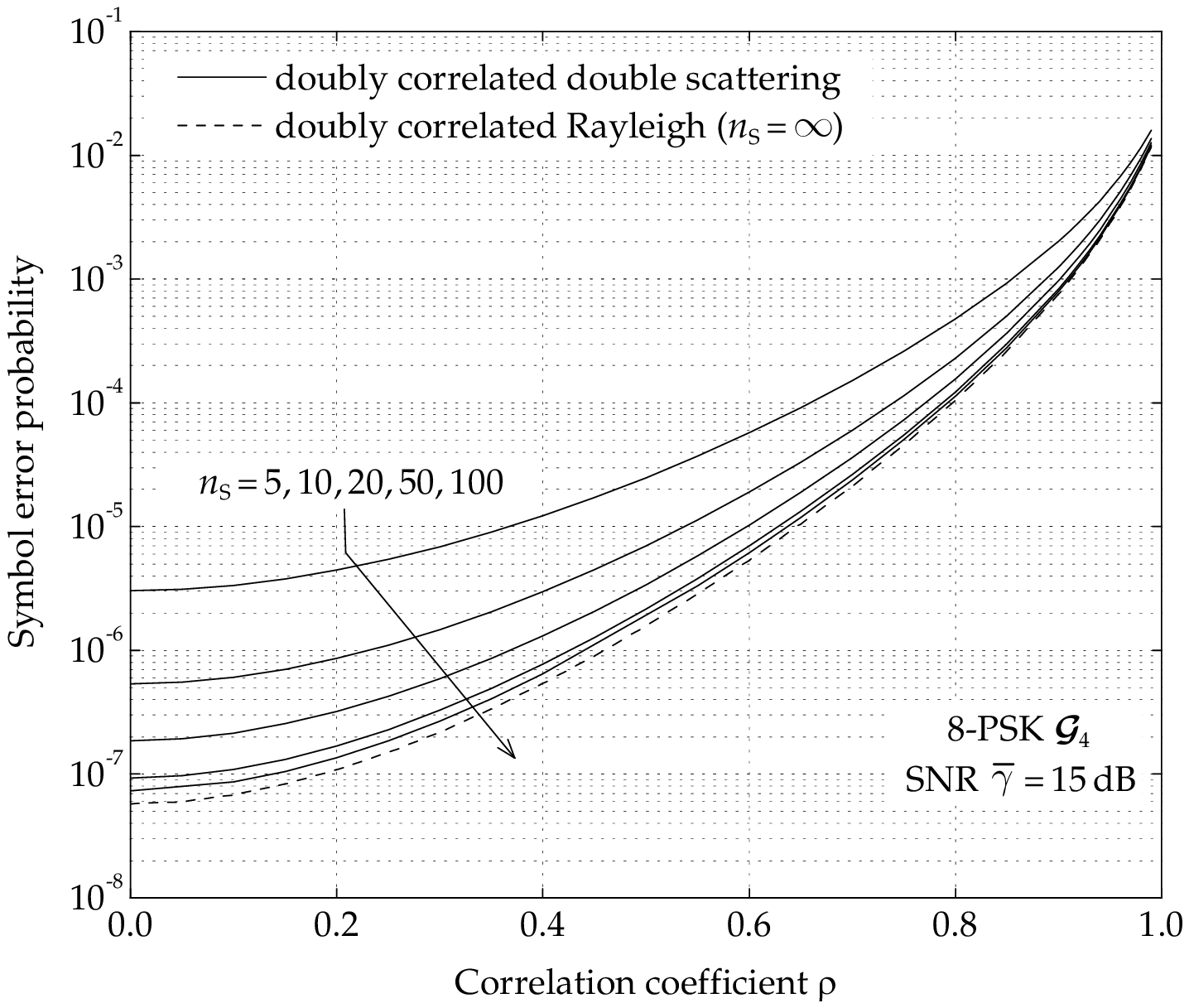}}
    \caption{
        SEP of $8$-PSK $\GF$ ($2.25$ bits/s/Hz) as a function of correlation coefficient
        $\rho$ in doubly correlated
        double-scattering $\DsMIMO{4}{4}{\nSx}$-MIMO
        channels with constant correlation
        $\TxCM=\RxCM=\CC{4}{\rho}$. $\nSx=5$, $10$, $20$, $50$,
        $100$, $\infty$ (doubly correlated Rayleigh) and $\snr=15$ dB.
    }
    \label{fig:Fig:7}
\end{figure}

Fig.~\ref{fig:Fig:6} shows the SEP of $8$-PSK $\GF$ versus the SNR
$\snr$ in doubly correlated double-scattering
$\DsMIMO{4}{4}{10}$-MIMO channels. In this figure, the transmit
and receive correlations follow the constant correlation
$\TxCM=\RxCM=\CC{4}{\rho}$, defined by \eqref{eq:CC} in
Appendix~\ref{sec:Appendix:A}, and the correlation coefficient
$\rho$ ranges from $0$ (spatially uncorrelated double scattering)
to $0.9$. The characteristic coefficients of the constant
correlation matrix are given by \eqref{eq:CC:CC:1} and
\eqref{eq:CC:CC:2} (see Example~\ref{ex:CC:PFC} in
Appendix~\ref{sec:Appendix:B}). For comparison, we also plot the
SEP of i.i.d.\ Rayleigh-fading MIMO channels. In
Figure~\ref{fig:Fig:6}, we can see that 
the SNR penalty due to double scattering with $\nSx=10$ (in the
absence of spatial correlation) is about $1$ dB at the SEP of
$10^{-6}$ and it becomes larger than $2.5$ dB for $\rho \geq 0.5$.
In Fig.~\ref{fig:Fig:7}, the SEP of $8$-PSK $\GF$ at $\snr=15$ dB
is depicted as a function of a correlation coefficient $\rho$ for
doubly correlated double-scattering $\DsMIMO{4}{4}{\nSx}$-MIMO
channels with constant correlation $\TxCM=\RxCM=\CC{4}{\rho}$ when
$\nSx=5$, $10$, $20$, $50$, $100$, and $\infty$ (doubly correlated
Rayleigh). This figure demonstrates that double scattering and
spatial correlation degrade the SEP performance considerably.

\subsection{MISO Double Scattering}

Finally, we consider a double-scattering MISO channel. This is a
pure transmit diversity system wherein the burden of diversity
reception at the receive terminal is moved to the transmitter.

The SEP in double-scattering MISO channels can be obtained from
\eqref{eq:MGF:SNR:2} with $\nRx=1$ as
\begin{align}    \label{eq:SEP:MPSK:MISO}
\EqF
    P_{\text{e,\,MPSK}}^\text{miso-ds}
    =
        \frac{1}{\pi}
        \int_0^{\Theta}
        \EX
            {
                \det\left(
                        \B{I}_{\nSx}
                        +\frac{g \snr}{\nSx\nTx\mathcal{R} \sin^2\theta} \,
                         \B{\Xi}_2
                         \B{\Xi}_2^\dag
                    \right)^{-1}
            }
            d\theta.
\EqE
\end{align}
Let $\SxEV{i}$, $i=1,2,\ldots,\nSx$, be the eigenvalues of $\SxCM$
in any order. Then, applying Theorem~\ref{thm:CQM:det} in
Appendix~\ref{sec:Appendix:B} to \eqref{eq:SEP:MPSK:MISO}, we
obtain
\begin{align}    
\EqF
    P_{\text{e,\,MPSK}}^\text{miso-ds}
    &=
        \frac{1}{\pi}
        \sum_{p=1}^\EVn{\SxCM}
        \sum_{q=1}^\EVn{\TxCM}
        \sum_{i=1}^\EVm{p}{\SxCM}
        \sum_{j=1}^\EVm{q}{\TxCM}
            \PFC{p}{i}{\SxCM}
            \PFC{q}{j}{\TxCM}
            \int_0^{\Theta}
                \HyperPFQ{2}{0}{
                                i,
                                j \, ;
                                -\frac{
                                        g
                                        \snr
                                        \odSxEV{p}{}
                                        \odTxEV{q}{}
                                      }
                                      {
                                        \nSx
                                        \nTx
                                        \mathcal{R}
                                        \sin^2\theta
                                      }
                                }
            d\theta
\EqE
\end{align}
where $\PFC{p}{i}{\SxCM}$ and $\PFC{q}{j}{\TxCM}$ are the
characteristic coefficients of $\SxCM$ and $\TxCM$, respectively.

\begin{figure}[t]
    \centerline{\includegraphics[width=0.7\textwidth]{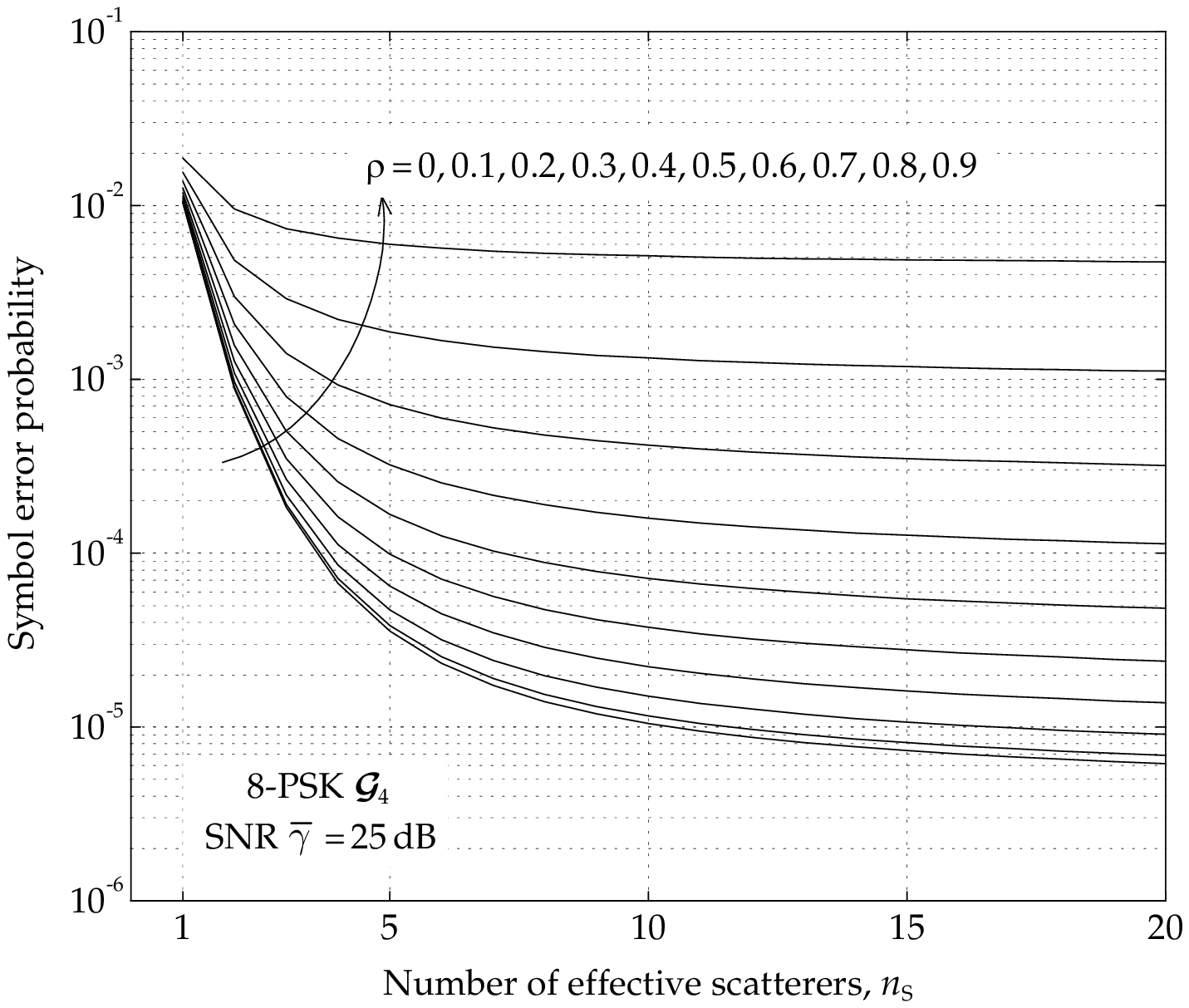}}
    \caption{
        SEP of $8$-PSK $\GF$ ($2.25$ bits/s/Hz) versus $\nSx$ in
        double-scattering $\DsMIMO{4}{1}{\nSx}$-MIMO
        channels. The transmit and scatterer
        correlations follow the constant correlation
        $\TxCM=\CC{4}{\rho}$ and $\SxCM=\CC{\nSx}{\rho}$ for $\rho=0$, $0.1$, $0.2$, $0.3$,
        $0.4$, $0.5$, $0.6$, $0.7$, $0.8$, and $0.9$. $\snr=25$ dB.
    }
    \label{fig:Fig:8}
\end{figure}

The effects of the spatial correlation and the number of effective
scatterers on the SEP performance in MISO channels can be
ascertained by referring to Fig.~\ref{fig:Fig:8}, where the SEP of
$8$-PSK $\GF$ at $\snr=25$ dB versus $\nSx$ is depicted for
double-scattering $\DsMIMO{4}{1}{\nSx}$-MIMO channels. The
transmit and scatterer correlations follow the constant
correlation $\TxCM=\CC{4}{\rho}$ and $\SxCM=\CC{\nSx}{\rho}$ where
$\rho$ varies from $0$ to $0.9$. Note that the maximum achievable
diversity order is equal to $d_\text{STBC}=4$ for $\nSx \geq 4$.
Hence, the SEP performance improves rapidly as $\nSx$ increases,
and approaches the corresponding SEP in the absence of double
scattering.

\section{Effective Fading Figure and Low-SNR Capacity}   \label{sec:Sec:4}

In this section, we access the combined effect of rank deficiency
and spatial correlation on the performance of OSTBCs in terms of
the EFF and the capacity in a low-SNR regime. It will be apparent
that these performance measures are completely characterized by
the \emph{kurtosis} of $\FN{\B{H}}$. 

\subsection{Effective Fading Figure}

One of the goals of diversity systems is to reduce the signal
fluctuation due to the stochastic nature of multipath fading.
Therefore, it is of interest to characterize the variation of the
instantaneous SNR at the output where the amount of signal
fluctuations is measured. The following measure can be used to
assess the severity of fading and the effectiveness of diversity
systems on reducing signal fluctuations.


\begin{definition}[Effective Fading Figure]     \label{def:EFF}
For the instantaneous SNR $\gamma$ at the output of interest in a
communication system subject to fading, the effective fading
figure (EFF) in dB for the output SNR $\gamma$ is defined as the
VMSR of $\gamma$, i.e.,
\begin{align}       \label{eq:EFF}
    \EFF{\gamma} \text{ (dB)}
    &\triangleq
        10\log_{10}
            \left\{
                    \frac{
                            \Var{\gamma}
                    }{
                            \left(\EX{\gamma}\right)^2
                    }
            \right\}.
    \EqE
\end{align}
%
%

\end{definition}

\mySep

It should be noted that the EFF is akin to the notions of the
normalized standard deviation (NSD) of the instantaneous combiner
output SNR \cite{WW:99:COM,WK:99:JSAC,Win:05:WCOM} and the amount
of fading (AF) \cite{Cha:79:COM,WG:03:COM}. The AF, as defined in
\cite[eq.~(2)]{Cha:79:COM}, is purely to characterize the amount
of random fluctuations in the channel itself and conveys no
information about diversity systems. In contrast, the NSD is a
measure of the signal fluctuations at the diversity combiner
output, enabling us to compare the effectiveness of diversity
combining techniques such as maximal-ratio combining (MRC),
equal-gain combining (EGC), selection combining (SC), and hybrid
section/maximal-ratio combining (H-S/MRC). If the signal
fluctuation is measured at each branch output, the EFF is
synonymous with the AF. In contrast, when the signal fluctuation
is measured at the diversity combiner output, the EFF is equal to
the square of the NSD of the instantaneous SNR at the combiner
output. The term `AF' was also confusingly used for diversity
systems in some literature with a view to bridging the philosophy
between characterizing physical channel fading and quantifying the
degree of diversity effectiveness
\cite{SL:04:VT,AS:02:COM,HO:05:WCOM}.

By definition, the efficiency of OSTBCs on reducing the severity
of fading can be assessed by
\begin{align}       \label{eq:AF:STBC}
    \AF \text{ (dB)}
    &\triangleq
        10\log_{10}
            \left\{
                    \frac{
                            \Var{\gamma_\text{STBC}}
                    }{
                            \left(\EX{\gamma_\text{STBC}}\right)^2
                    }
            \right\}
    \nonumber \\
    &=
        10\log_{10}
            \left\{
                \Kurt{\FN{\B{H}}}-1
            \right\}
    \EqE
\end{align}
where $\Kurt{\FN{\B{H}}}$ is the kurtosis of $\FN{\B{H}}$ defined
by
%
%
\begin{align}       \label{eq:kurt:FN}
    \EqF
    \Kurt{\FN{\B{H}}}
    &\triangleq
        \frac{
                \EX{
                    \left[
                        \FN{\B{H}}
                        -\EX{\FN{\B{H}}}
                    \right]^4
                }
        }{
            \left(
                \EX{
                    \left[
                        \FN{\B{H}}
                        -\EX{\FN{\B{H}}}
                    \right]^2
                }
            \right)^2
        }
    \nonumber \\
    &=
        \frac{
                \EX{
                        \FN{\B{H}}^4
                }
        }{
            \left(
                \EX{
                        \FN{\B{H}}^2
                }
            \right)^2
        }.
    \EqE
\end{align}
In \eqref{eq:kurt:FN}, the second equality follows from the fact
that the kurtosis is invariant with respect to translations of a
random variable. Note that the minimum EFF is equal to $-\infty$
dB if there is no random fluctuation in the received signal. Also,
the EFF is equal to $0$ dB for Rayleigh fading without diversity
and hence, $\AF>0$ dB means that the variation of the
instantaneous SNR in each SISO subchannel is more severe than that
in Rayleigh fading.

\subsubsection{Note on the Kurtosis of $\FN{\B{H}}$}

The kurtosis measures the peakedness or flatness of a distribution
\cite{Pea:05:Bio}. It has been revealed that this normalized form
of the fourth statistic of fading distributions plays a key role
in the low-SNR behavior of the spectral efficiency in fading
channels \cite{SV:01:IT,Ver:02:IT}. To proceed with deriving
$\Kurt{\FN{\B{H}}}$ for double-scattering
$\DsMIMO{\nTx}{\nRx}{\nSx}$-MIMO channels, we first define the
following scalar quantity related to a correlation matrix.


\begin{definition}[Correlation Figure]      \label{def:CN}
For an arbitrary $n \times n$ correlation matrix $\B{\Phi}$, the
\emph{correlation figure} of $\B{\Phi}$ is defined by
\begin{align}    \label{eq:CN}
    \CN{\B{\Phi}}
    \triangleq
        \frac{\tr\left(\B{\Phi}^2\right)}{\tr\left(\B{1}_n^2\right)}
    =
        \frac{1}{n^2}
        \tr\left(\B{\Phi}^2\right)
\end{align}
where $\B{1}_n$ denotes the $n \times n$ all-one matrix.

\end{definition}

\mySep

Note that $\frac{1}{n} \leq \CN{\B{\Phi}} \leq 1$, where the lower
and upper bounds correspond to uncorrelated and fully correlated
cases, respectively.\footnote{
        Similar to \eqref{eq:CN}, the
        \emph{correlation number} was defined as
        $\frac{1}{n}\tr\left(\B{\Phi}^2\right)$ \cite{LTV:03:IT}.
        While the correlation figure and number are the second-order statistics of
        the spectra of a correlation matrix, normalized by those of fully
        correlated and uncorrelated matrices, respectively, the
        correlation figure is bounded by $0 \leq
        \CN{\B{\Phi}} \leq 1$ for any correlation structure, as $n \to
        \infty$.
        }
The following Schur monotonicity properties hold for the
correlation figure (the proofs are given in
Appendix~\ref{sec:proof:PT:CN}).


\begin{property}     \label{ex:SC:CN}

Let $\B{\Phi}$ be an $n \times n$ correlation matrix. Then, the
correlation figure $\CN{\B{\Phi}}$ as a functional of the
eigenvalues of $\B{\Phi}$ is MIS, that is, if $\B{\Phi} \preceq
\grave{\B{\Phi}}$, then
\begin{align}
    \CN{\B{\Phi}}
    \leq
    \CNs{\grave{\B{\Phi}}}.
\end{align}

\end{property}

\mySep

\begin{property}    \label{PT:CN:2}

Let $\B{\Phi}_i$, $i=1,2,\ldots,m$, be $n_i \times n_i$
correlation matrices. Then, the product of correlation figures,
$\prod_{i=1}^m \CN{\B{\Phi}_i}$, as a functional of the
eigenvalues of $\bigotimes_{i=1}^m \B{\Phi}_i$,
is MIS, that is, if
\begin{align}
    \bigotimes_{i=1}^m
        \B{\Phi}_i
    \preceq
    \bigotimes_{i=1}^m
        \grave{\B{\Phi}}_i
    \, ,
\end{align}
then
\begin{align}
    \prod_{i=1}^m \CN{\B{\Phi}_i}
    \leq
    \prod_{i=1}^m \CNs{\grave{\B{\Phi}}_i}.
\EqE
\end{align}

\end{property}

\mySep

\begin{property}    \label{PT:CN:3}

Let $\B{\Phi}_i$, $i=1,2,\ldots,m$, be $n_i \times n_i$
correlation matrices. Then, the sum of correlation figures,
$\sum_{i=1}^m \CN{\B{\Phi}_i}$, as a functional of the eigenvalues
of $\bigoplus_{i=1}^m \frac{1}{n_i}\B{\Phi}_i$,
is MIS, that is, if
\begin{align}
    \bigoplus_{i=1}^m
        \tfrac{1}{n_i}\B{\Phi}_i
    \preceq
    \bigoplus_{i=1}^m
        \tfrac{1}{n_i}\grave{\B{\Phi}}_i
    \, ,
\end{align}
then
\begin{align}
    \sum_{i=1}^m \CN{\B{\Phi}_i}
    \leq
    \sum_{i=1}^m \CNs{\grave{\B{\Phi}}_i}.
\EqE
\end{align}

\end{property}

\mySep

The next theorem shows that $\Kurt{\FN{\B{H}}}$ depends
exclusively on the spectra of spatial correlation matrices and is
quantified solely by their correlation figures.


\begin{theorem}     \label{thm:kurt:FN}
For double-scattering $\DsMIMO{\nTx}{\nRx}{\nSx}$-MIMO channels,
the kurtosis of $\FN{\B{H}}$ is
\begin{align}    \label{eq:kurt:FN:f}
    \kappa\left(\FN{\B{H}}\right)
    =
        \CN{\TxCM}
        \CN{\RxCM}
        +
        \CN{\TxCM}
        \CN{\SxCM}
        +
        \CN{\RxCM}
        \CN{\SxCM}
        +1.
\end{align}
%
%


\begin{proof}
See Appendix~\ref{sec:proof:thm:kurt:FN}.
\end{proof}

\end{theorem}

\mySep

\begin{example}[Spatially Uncorrelated Double Scattering]     \label{ex:kurt:FN:UC-DS}

In the absence of spatial fading correlation
($\mathbb{T}_\text{uc}$), we have
\begin{align}        \label{eq:kurt:FN:UC-DS}
    \Kurt{\FN{\B{H}}}
    =
        \frac{1}
             {
                \nTx
                \nRx
             }
        +
        \frac{1}
             {
                \nTx
                \nSx
             }
        +
        \frac{1}
             {
                \nRx
                \nSx
             }
        +1.
    \EqE
\end{align}

\end{example}

\mySep

As compared with the i.i.d. case, the keyhole increases the
kurtosis of the fading distribution in SISO subchannels by twice
the reciprocal of the harmonic mean between the numbers of
transmit and receive antennas, that is,
$\frac{1}{\nTx}+\frac{1}{\nRx}$.

Next, we show the Schur monotonicity property of
$\Kurt{\FN{\B{H}}}$. 
%
%
%
%
%
%
%


\begin{corollary}     \label{cor:kurt:FN:SC}

Let
\begin{align}
    \B{\mathcal{J}}\left(\mathbb{T}\right)
    \triangleq
        \frac{\TxCM \otimes \RxCM}{\nTx\nRx}
        \oplus
        \frac{\TxCM \otimes \SxCM}{\nTx\nSx}
        \oplus
        \frac{\SxCM \otimes \RxCM}{\nSx\nRx}
\EqE
\end{align}
for a spatial correlation environment $\mathbb{T}=\left(\TxCM,
\SxCM, \RxCM\right)$. Then, the kurtosis of $\FN{\B{H}}$, as a
functional of the eigenvalues of
$\B{\mathcal{J}}\left(\mathbb{T}\right)$, is a MIS (or isotone)
function, that is, if $\B{\mathcal{J}}\left(\mathbb{T}_1\right)
\preceq \B{\mathcal{J}}\left(\mathbb{T}_2\right)$, then
\begin{align}
    \Kurt{\FN{\B{H}};\mathbb{T}_1}
    \leq
    \Kurt{\FN{\B{H}};\mathbb{T}_2}.
\end{align}
\begin{proof}
It follows immediately from Theorem~\ref{thm:kurt:FN} and
Properties~\ref{PT:CN:2} and \ref{PT:CN:3} stating the fact that
the product and sum of correlation figures preserve the
monotonicity property.
\end{proof}

\end{corollary}

\mySep

Corollary \ref{cor:kurt:FN:SC} implies that the less spatially
correlated fading results in the less peaky fading distribution of
each SISO subchannel.

\subsubsection{Note on the EFF of $\gamma_\text{STBC}$}

From Theorem~\ref{thm:kurt:FN} and \eqref{eq:AF:STBC}, it is
straightforward to see that the $\AF$ in double-scattering
$\DsMIMO{\nTx}{\nRx}{\nSx}$-MIMO channels is given by
\begin{align}
    \AF \text{ (dB)}
    =
        10\log_{10}
            \left\{
                \CN{\TxCM}
                \CN{\RxCM}
                +
                \CN{\TxCM}
                \CN{\SxCM}
                +
                \CN{\RxCM}
                \CN{\SxCM}
            \right\}
\end{align}
from which we can make the following observations on the $\AF$.

\begin{itemize}

\item

The $\AF$ as a functional of the eigenvalues of
$\B{\mathcal{J}}\left(\mathbb{T}\right)$ is MIS, that is,
\begin{align}
    \AF\left(\mathbb{T}_1\right)
    \leq
    \AF\left(\mathbb{T}_2\right)
\end{align}
whenever $\B{\mathcal{J}}\left(\mathbb{T}_1\right) \preceq
\B{\mathcal{J}}\left(\mathbb{T}_2\right)$. This reveals that the
less spatially correlated fading results in the less severe random
fluctuations in equivalent SISO subchannels induced by OSTBCs.

%

%

\item

In the absence of double scattering, $\CN{\SxCM}$ is equal to zero
and thus, the double scattering together with spatial correlation
causes  the $\AF$ to increase by the amount of
$\CN{\TxCM}\CN{\SxCM}+\CN{\RxCM}\CN{\SxCM}$. In particular, the
maximum increase in the $\AF$ is a sum of correlation figures of
the transmit and receive correlation matrices, that is,
$\CN{\TxCM}+\CN{\RxCM}$, which eventuates when $\SxCM$ goes to be
fully correlated or when the keyhole effect takes place.

%

\end{itemize}

\subsection{Low-SNR Capacity}

Recent information-theoretic studies show that the first-order
analysis of the capacity versus the SNR fails to reveal the impact
of the channel and that second-order analysis is required to
assess the wideband or low-SNR performance of communication
systems \cite{Ver:02:IT,LTV:03:IT}. In particular, it was
demonstrated that the tradeoff between the capacity in bits/s/Hz
and energy per bit required for reliable communication is the key
measure of channel capacity in a low-SNR regime.
%
In this regime, the capacity can be characterized by two
parameters, namely, i) $\SNRmin$, the \emph{minimum bit energy per
noise level} required to reliably communicate at any positive data
rate (where $E_\text{b}$ denotes the total transmitted energy per
bit), and ii) $S_0$, the \emph{low-SNR slope} (bits/s/Hz per $3$
dB) of the
capacity at the point $\SNRmin$. 

\subsubsection{General Input Signaling}        \label{sec:3:3:1}

Before proceeding to study the low-SNR capacity achieved by
OSTBCs, we first deal with the more general case of input
signaling, assuming that the fading process is ergodic and coding
is across many independent fading blocks without a delay
constraint.


\begin{theorem}     \label{thm:WSE}
Consider a general double-scattering
$\DsMIMO{\nTx}{\nRx}{\nSx}$-MIMO channel
\begin{align}
    \B{Y}
    =
        \B{HX}
        +
        \B{W}
\end{align}
where the channel matrix $\B{H}$ is given by \eqref{eq:H} at each
coherence interval and the input signal $\B{X} \in
\C^{\nTx \times \Tc}$ is subject to the power constraint
$\EX{\FN{\B{X}}^2}=\Tc \mathcal{P}$. Suppose that the receiver
knows the realization of $\B{H}$, but the transmitter has no
channel knowledge. Then, the minimum required
$\tfrac{E_\text{b}}{N_0}$ for reliable communication is
\begin{align}    \label{eq:minEbNo:DS}
    \SNRmin
    =
        \frac{\log_e 2}{\nRx}
\end{align}
and the low-SNR slope of the capacity is
\begin{align}    \label{eq:So:DS}
    S_0
    =
        \frac{
                2
             }
             {
                \CN{\TxCM}
                +
                \CN{\SxCM}
                +
                \CN{\RxCM}
                +
                \CN{\TxCM}
                \CN{\SxCM}
                \CN{\RxCM}
             }
        \quad
        \text{bits/s/Hz per $3$ dB}.
\end{align}
%
%


\begin{proof}
See Appendix~\ref{sec:proof:thm:WSE}.
\end{proof}

\end{theorem}

\mySep



From Theorem \ref{thm:WSE}, we can make the following
observations.

\begin{itemize}

\item

The $\SNRmin$ is inversely proportional to $\nRx$, whereas the
double scattering and spatial fading correlation as well as the
numbers of transmit antennas and effective scatterers do not
affect this measure. Moreover, regardless of the number of
antennas and propagation conditions, the minimum received bit
energy per noise level required for reliable communication,
$\tSNRmin$, is equal to
\begin{align}
    \tSNRmin
    =
        \nRx
        \cdot
        \SNRmin
    =
        -1.59 \text{ dB}
\end{align}
which is a fundamental feature of the channels where the additive
noise is Gaussian \cite[Theorem~1]{Ver:02:IT}.

\item

The low-SNR slope $S_0$ as a functional of the eigenvalues of
$\grave{\B{\mathcal{J}}}\left(\mathbb{T}\right)$ is MDS, that is,
if $\grave{\B{\mathcal{J}}}\left(\mathbb{T}_1\right) \preceq
\grave{\B{\mathcal{J}}}\left(\mathbb{T}_2\right)$, then
\begin{align}   \label{eq:S0:MDS}
    S_0\left(\mathbb{T}_1\right)
    \geq
    S_0\left(\mathbb{T}_2\right)
\end{align}
where $\grave{\B{\mathcal{J}}}\left(\mathbb{T}\right)$ is defined
for the environment $\mathbb{T}=\left(\TxCM, \SxCM, \RxCM\right)$
as follows:
\begin{align}
\EqF
    \grave{\B{\mathcal{J}}}\left(\mathbb{T}\right)
    \triangleq
        \frac{\TxCM}{\nTx}
        \oplus
        \frac{\SxCM}{\nSx}
        \oplus
        \frac{\RxCM}{\nRx}
        \oplus
        \frac{\TxCM \otimes \SxCM \otimes \RxCM}{\nTx\nSx\nRx}.
\EqE
\end{align}
Note that \eqref{eq:S0:MDS} follows from \eqref{eq:So:DS} and
Properties~\ref{PT:CN:2} and \ref{PT:CN:3}. This MDS property
reveals that the low-SNR slope decreases with the amount of
spatial correlation in contrast to the high-SNR capacity slope
$\min\left(\nTx,\nRx,\nSx\right)$, which is invariant with respect
to spatial correlation \cite{SL:03:IT}.

%
%
%

\end{itemize}

\mySep

\begin{example}[Dual-Antenna System]    \label{ex:dual}
Consider $\nTx=\nRx=2$. In the presence of spatially uncorrelated
double scattering, the low-SNR slope for general double-scattering
$\DsMIMO{2}{2}{\nSx}$-MIMO channels  is
\begin{align}        \label{eq:So:dual}
    S_0
    =
        2
        \cdot
        \left(
                1+
                \frac{1}{\nSx}
                \cdot
                \frac{5}{4}
        \right)^{-1}
    ~~
    \text{bits/s/Hz per $3$ dB}
\end{align}
which is bounded by $8/9 \leq S_0 \leq 2$. The lowest and highest
slopes are achieved when $\nSx=1$ (keyhole) and $\nSx=\infty$
(i.i.d.), respectively.


\end{example}


\subsubsection{OSTBC Input Signaling}

We now turn attention to the low-SNR behavior of the capacity for
double-scattering $\DsMIMO{\nTx}{\nRx}{\nSx}$-MIMO channels
employing OSTBCs.


\begin{theorem}     \label{thm:WSE:STBC}
Consider a double-scattering $\DsMIMO{\nTx}{\nRx}{\nSx}$-MIMO
channel
\begin{align*}
    \B{Y}
    =
        \B{H}
        \B{\mathcal{G}}^T
        +
        \B{W}
\end{align*}
where the channel matrix $\B{H}$ is given by \eqref{eq:H} at each
coherence interval and the OSTBC $\B{\mathcal{G}}$ is subject to
the power constraint $\E \bigl\{\FN{\B{\mathcal{G}}}^2 \bigr\}=\Tc
\mathcal{P}$. Then, the OSTBC achieves the minimum required
$\SNRmin$ same as that without the orthogonal signaling constraint
\begin{align}    \label{eq:minEbNo:STBC}
    \frac{E_\text{b}}{N_0}_\text{min}^\text{STBC}
    =
        \frac{\log_e 2}{\nRx}
\end{align}
and the low-SNR slope of the capacity 
\begin{align}       \label{eq:So:STBC}
    S_0^\text{STBC}
    =
        \frac{
                2\mathcal{R}
             }
             {
                \CN{\TxCM}
                \CN{\RxCM}
                +
                \CN{\TxCM}
                \CN{\SxCM}
                +
                \CN{\RxCM}
                \CN{\SxCM}
                +1
             }
        \quad
        \text{bits/s/Hz per $3$ dB}.
\EqE
\end{align}
%
%


\begin{proof}
See Appendix~\ref{sec:proof:thm:WSE:STBC}.
\end{proof}

\end{theorem}

\mySep

From Theorem~\ref{thm:WSE:STBC}, we can make the following
observations in parallel to \ref{sec:3:3:1}.

\begin{itemize}

\item

As compared with the general case, the use of OSTBCs does not
increase the minimum required $\tfrac{E_\text{b}}{N_0}$ for
reliable communication in MIMO channels.

\item

The low-SNR slope $S_0^\text{STBC}$ as a functional of the
eigenvalues of $\B{\mathcal{J}}\left(\mathbb{T}\right)$ is MDS,
that is, if $\B{\mathcal{J}}\left(\mathbb{T}_1\right) \preceq
\B{\mathcal{J}}\left(\mathbb{T}_2\right)$, then
\begin{align}
    S_0^\text{STBC}\left(\mathbb{T}_1\right)
    \geq
    S_0^\text{STBC}\left(\mathbb{T}_2\right).
\end{align}
In contrast, we see from \eqref{eq:C:STBC} that the high-SNR slope
of the capacity is equal to $\mathcal{R}$, which does not depend
on spatial correlation and double scattering.

\end{itemize}

\begin{example}[Alamouti's Code]    \label{ex:AC}
Consider $\nTx=\nRx=2$. In the presence of spatially uncorrelated
double scattering, the low-SNR slope for Alamouti's code with two
receive antennas is
\begin{align}        \label{eq:So:Alamouti}
    S_0^\text{STBC}
    =
        \frac{8}{5}
        \cdot
        \left(
                1+
                \frac{1}{\nSx}
                \cdot
                \frac{4}{5}
        \right)^{-1}
    ~~
    \text{bits/s/Hz per $3$ dB}
\EqE
\end{align}
which is bounded by $8/9 \leq S_0^\text{STBC} \leq 8/5$. 
\end{example}

\mySep

\mySep

\begin{figure}[t]
    \centerline{\includegraphics[width=0.7\textwidth]{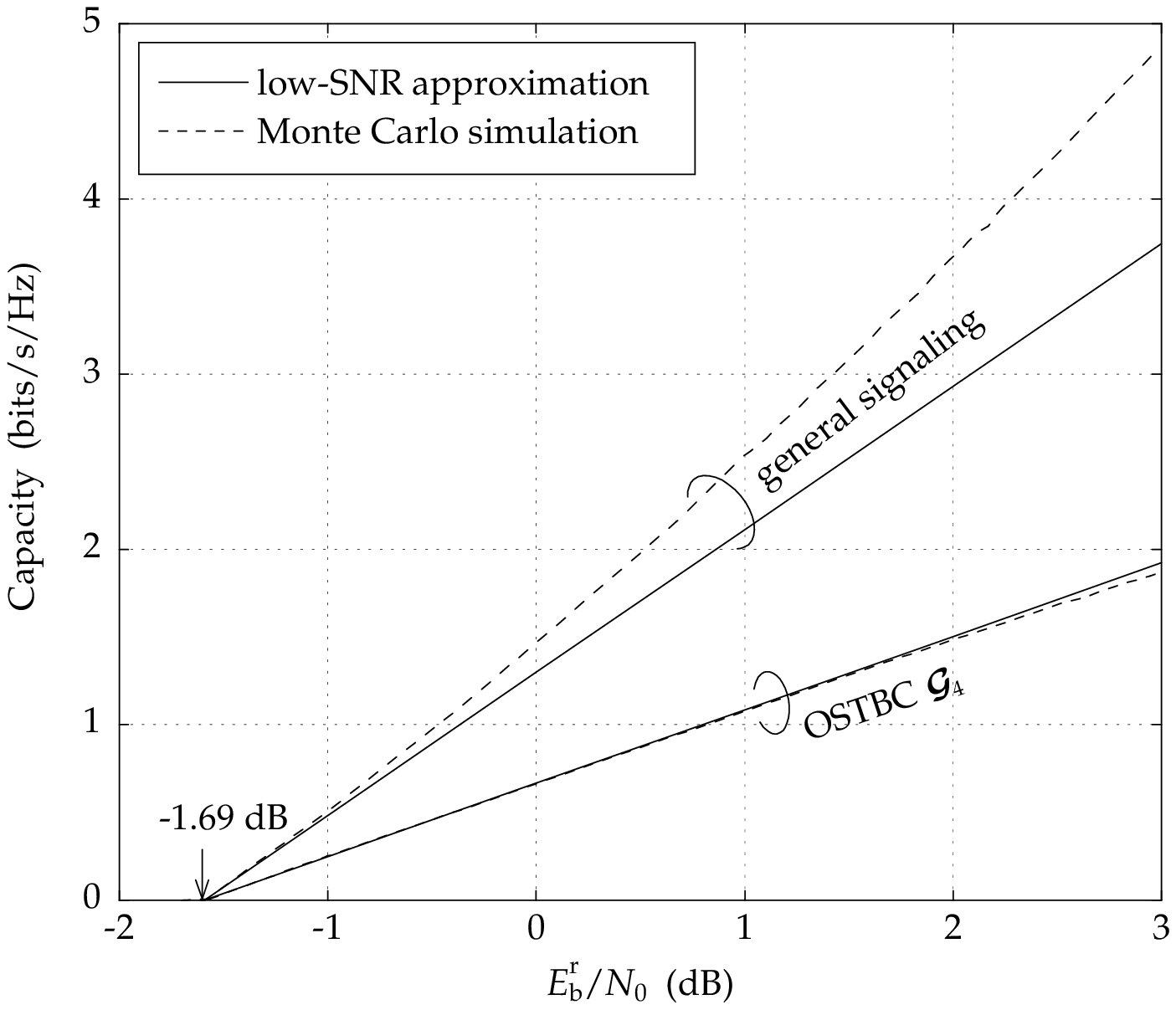}}
    \caption{
        Capacity in bits/s/Hz versus the received $\frac{E_\text{b}}{N_0}$ for the general input signaling and OSTBC
        $\GF$ in double-scattering
        $\DsMIMO{4}{4}{20}$-MIMO channels with
        exponential correlation
        $\TxCM=\RxCM=\EC{4}{0.5}$ and
        $\SxCM=\EC{20}{0.5}$.
    }
    \label{fig:Fig:3}
\end{figure}

In Fig.~\ref{fig:Fig:3}, the capacity (bits/s/Hz) versus
$\tSNRmin$ and its low-SNR approximation are depicted with and
without the signaling constraint of the OSTBC $\GF$ in
double-scattering $\DsMIMO{4}{4}{20}$-MIMO channels with
exponential correlation $\TxCM=\RxCM=\EC{4}{0.5}$ and
$\SxCM=\EC{20}{0.5}$. For the OSTBC $\GF$, the low-SNR
approximation is remarkably accurate for a fairly wide range of
$\tSNRmin$, whereas there exists some discrepancy between the
Monte Carlo simulation and the first-order approximation for the
general input signaling---approximately $11\%$ difference at
$\tSNRmin=0$ dB, for example. In this scenario, the low-SNR slopes
are $1.26$ and $2.46$ bits/s/Hz per $3$ dB with and without the
OSTBC input signaling constraint, respectively. Thus, the use of
the OSTBC $\GF$ incurs about $49\%$ reduction in the slope. 
This slope reduction is much smaller than that in a high-SNR
regime: the high-SNR slope for the OSTBC $\GF$ is
$\mathcal{R}=3/4$ and the corresponding slope for the general
signaling is equal to $\min\left(\nTx,\nRx,\nSx\right)=4$
bits/s/Hz per $3$ dB \cite{SL:03:IT}.

\section{Conclusions}  \label{sec:Sec:5}

We investigated the combined effect of rank deficiency and spatial
fading correlation on the diversity performance of MIMO systems.
In particular, we considered double-scattering MIMO channels
employing OSTBCs which use up all antennas to realize full
diversity advantage. We characterized the effects of double
scattering on the severity of fading and the low-SNR capacity by
quantifying the EFF and the capacity slope in terms of the
\emph{correlation figures} of spatial correlation matrices. The
Schur monotonicity properties were shown for these performance
measures as functionals of the eigenvalues of correlation
matrices. We also determined the required scattering richness of
the channel to achieve the full diversity order of $\nTx\nRx$.
Finally, we derived the exact SEP expressions for some classes of
double scattering, which consolidate the effects of rank
efficiency and spatial correlation on the SEP performance. On
account of the generality of channel modeling, the results of the
paper are substantial enough to encompass those for well-accepted
existing models (e.g., i.i.d./spatially correlated/keyhole MIMO
channels) as special cases of our solutions.
\appendices

\section{Majorization, Schur Monotonicity, and Correlation Matrices}   \label{sec:Appendix:A}

We use the concept of majorization
\cite{Sch:23:SBMG,She:99:ATT,And:89:LAA,HJ:91:Book,BS:85:LAA} as a
mathematical tool to characterize different spatial correlation
environments. Using the majorization theory, the analytical
framework was established in \cite{Win:05:WCOM} to assess the
performance of multiple-antenna diversity systems with different
\emph{power dispersion profiles}. In particular, monotonicity
theorems were proved for various performance measures such as the
NSD of the output SNR, the ergodic capacity, the matched-filter
bound, the inverse SEP, and the symbol error outage. The notion of
majorization has also been used in
\cite{CTKV:02:IT,WSFY:04:COM,JG:05:WCOM} as a measure of
correlation. In this appendix, we briefly discuss the basic
properties of majorization and Schur monotonicity. 

\subsection{Majorization and Correlation Matrices}

Given a real vector $\B{a}=\left(a_1,a_2,\ldots,a_n\right)^T \in
\R^n$, we rearrange its components in decreasing order as $a_{[1]}
\geq a_{[2]} \geq \cdots \geq a_{[n]}$.

\mySep

\begin{definition}      \label{def:maj}

For $\B{a}=\left(a_1,a_2,\ldots,a_n\right)^T$,
$\B{b}=\left(b_1,b_2,\ldots,b_n\right)^T \in \R^n$, we denote
$\B{a} \prec \B{b}$ and say that $\B{a}$ is \emph{weakly
majorized} (or \emph{submajorized}) by $\B{b}$ if
\begin{align}
\EqF
    \sum_{i=1}^k
        a_{[i]}
    \leq
    \sum_{i=1}^k
        b_{[i]},
    \qquad
    k=1,2,\ldots,n.
\EqE
\end{align}
If $\sum_{i=1}^n a_i = \sum_{i=1}^n b_i$ holds in addition to
$\B{a} \prec \B{b}$, then we say that $\B{a}$ is \emph{majorized}
by $\B{b}$ and denote as $\B{a} \preceq \B{b}$.

\end{definition}

\mySep

For example, if each $a_i \geq 0$ and $\sum_{i=1}^n a_i =n$, then
\begin{align}
    \left(
        1,
        1,
        \ldots,
        1
    \right)^T
    \preceq
    \left(
        a_1,
        a_2,
        \ldots,
        a_n
    \right)^T
    \preceq
    \left(
        n,
        0,
        \ldots,
        0
    \right)^T.
\end{align}
%
%
%
%
Of particular interest are the majorization relations among
Hermitian matrices in terms of their eigenvalue vectors to compare
different spatial correlation environments. A Hermitian matrix
$\B{A}$ is said to be \emph{majorized} by a Hermitian matrix
$\B{B}$, simply denoted by $\B{A} \preceq \B{B}$, if $\EV{\B{A}}
\preceq \EV{\B{B}}$ where $\EV{\cdot}$ denote the vector of
eigenvalues of a Hermitian matrix. For example, the well-known
Schur's theorem \cite[eq.~(5.5.8)]{HJ:91:Book} on the relationship
between the eigenvalues and diagonal entries of Hermitian matrices
can be written as
\begin{align}
    \B{A} \circ \B{I}_n
    \preceq
    \B{A}
    \quad
    \text{for Hermitian }
    \B{A} \in \C^{n\times n}
\EqE
\end{align}
where $\circ$ denotes a Hadamard (i.e., entrywise) product. One of
the most useful results on the eigenvalue majorization is the
following theorem.

\mySep

\begin{theorem}[{\cite[Theorem~7.1]{And:89:LAA}}]     \label{thm:eigenMaj:DSM}

A linear map $\mathcal{L}: \C^{n \times n} \to \C^{n \times n}$ is
called \emph{positive} if $\DSM{\B{A}} \geq 0$ for $\B{A}\in \C^{n
\times n}\geq 0$ and \emph{unital} if $\DSM{\B{I}_n}=\B{I}_n$. It
is said to be \emph{doubly stochastic} if $\mathcal{L}$ is a
unital positive linear map with the trace-preserving property,
i.e., $\tr \DSM{\B{A}} = \tr\left(\B{A}\right)$, $\forall \B{A}
\in \C^{n \times n}$. Let $\B{A} \in \C^{n \times n}$ be Hermitian
and $\mathcal{L}$ be a doubly stochastic map. Then,
\begin{align}
    \DSM{\B{A}}
    \preceq
    \B{A}.
\EqE
\end{align}

\end{theorem}

\mySep 

Recall that the Schur product theorem
\cite[Theorem~5.2.1]{HJ:91:Book} says that the Hadamard product of
two positive semidefinite matrices is positive semidefinite.
Therefore if $\B{\Phi}\in \C^{n \times n}$ is an arbitrary
correlation matrix and define $\DSM{\B{A}} = \B{A} \circ
\B{\Phi}$, then $\mathcal{L}$ is obviously a doubly stochastic map
on $\C^{n \times n}$.

\mySep

\begin{corollary}       \label{cor:eigenMaj:DSM}

Let $\B{A} \in \C^{n \times n}$ be Hermitian and $\B{\Phi}\in
\C^{n \times n}$ be a correlation matrix. Then,
\begin{align}
    \B{A} \circ \B{\Phi}
    \preceq
    \B{A}.
\EqE
\end{align}

\end{corollary}

\mySep

In fact, this result was first given in
\cite[Corollary~2]{BS:85:LAA} without using the notion of doubly
stochastic maps. From Corollary \ref{cor:eigenMaj:DSM}, we can
obtain the eigenvalue majorization relations for the well-known
correlation models---\emph{constant}, \emph{exponential}, and
\emph{tridiagonal correlation}---which have been widely used for
many communication problems of multiple-antenna systems (see,
e.g.,
\cite{SL:03:IT,SA:00:Book,CWZ:03:IT,SWLC:06:WCOM,LTV:03:IT,PS:60:IRE}).

\mySep

\begin{example}[Constant, Exponential, and Tridiagonal Matrices]     \label{ex:CM}

The $n$th-order constant, exponential, and tridiagonal matrices
with a coefficient $\rho$, denoted by $\CC{n}{\rho}$,
$\EC{n}{\rho}$, and $\TC{n}{\rho}$ respectively, are $n \times n$
symmetric Toeplitz matrices of the following structures:
\begin{salign}
                \label{eq:CC}
    \CC{n}{\rho}
    &=
        \begin{bmatrix}
            1   &   \rho    &   \rho    &   \cdots  &   \rho
            \\
            \rho    &   1    &   \rho    &   \cdots  &   \rho
            \\
            \vdots  &   \vdots  &   \vdots  &   \ddots  &   \vdots
            \\
            \rho    &   \rho    &   \rho    &   \cdots  &   1
        \end{bmatrix}_{n \times n}
%
%
    \\
                \label{eq:EC}
    \EC{n}{\rho}
    &=
        \begin{bmatrix}
            1   &   \rho    &   \rho^{2}    &   \cdots  &   \rho^{(n-1)}
            \\
            \rho    &   1    &   \rho   &   \cdots  &    \rho^{(n-2)}
            \\
            \vdots  &   \vdots  &   \vdots  &   \ddots  &   \vdots
            \\
            \rho^{n-1}    &   \rho^{n-2}    &   \rho^{n-3}    &   \cdots  &   1
        \end{bmatrix}_{n \times n}
%
%
%
%
   \\
    \TC{n}{\rho}
    &=
        \begin{bmatrix}
            1   &   \rho    &     &  &    &   0
            \\
            \rho    &   1    &   \rho    &    & &
            \\
                    &   \rho  &   1  &   \rho  & &
            \\
            &    &   \ddots    &   \ddots  &   \ddots  &
            \\
              &   &   & \rho & 1  &    \rho
            \\
            0   &   &   & & \rho  &    1
        \end{bmatrix}_{n \times n}.
\end{salign}
Note that $\CC{n}{\rho}$, $\EC{n}{\rho}$ with $\rho \in
\left[0,1\right]$ and $\TC{n}{\rho}$ with $\rho \in
\bigl[0,0.5/\cos\tfrac{\pi}{n+1}\bigr]$ are correlation matrices,
since they are positive semidefinite for such values of $\rho$.
Let $0 \leq \rho_1 \leq \rho_2$. Then, since
\begin{align*}
    \CC{n}{\rho_1}
    &=
        \CC{n}{\rho_2}
        \circ
        \CC{n}{\tfrac{\rho_1}{\rho_2}}
    \\
    \EC{n}{\rho_1}
    &=
        \EC{n}{\rho_2}
        \circ
        \EC{n}{\tfrac{\rho_1}{\rho_2}}
%
%
%
%
    \\
    \TC{n}{\rho_1}
    &=
        \TC{n}{\rho_2}
        \circ
        \EC{n}{\tfrac{\rho_1}{\rho_2}},
\end{align*}
it follows from Corollary \ref{cor:eigenMaj:DSM} that
\begin{align}
\EqF
    \label{eq:maj:CC}
    \CC{n}{\rho_1}
    &\preceq
    \CC{n}{\rho_2}
    \\
    \label{eq:maj:EC}
    \EC{n}{\rho_1}
    &\preceq
    \EC{n}{\rho_2}
    \\
    \label{eq:maj:TC}
    \TC{n}{\rho_1}
    &\preceq
    \TC{n}{\rho_2}.
\EqE
\end{align}
\end{example}

\mySep \mySep

\textit{Remark:} If $0 \leq \rho_1 \leq \rho_2$, then
$\CC{n}{\frac{\rho_1}{\rho_2}}$ and
$\EC{n}{\frac{\rho_1}{\rho_2}}$ are positive semidefinite. Hence,
the majorization relations \eqref{eq:maj:CC}--\eqref{eq:maj:TC}
hold, although each matrix itself is only Hermitian but may not be
positive semidefinite.

\subsection{Schur Monotonicity}

The concept of majorization is closely related to a MIS (or MDS)
function. If a function $f: \text{(a subset of) }\R^n\rightarrow
\R$ satisfies $f\left(a_1,\ldots,a_n\right) \leq
f\left(b_1,\ldots,b_n\right)$ whenever $\B{a}\preceq\B{b}$, then
$f$ is called a MIS (or isotone) function on $\text{(a subset of)
} \R^n$. The following theorem gives a necessary and sufficient
condition for $f$ to be MIS.

\mySep

\begin{theorem}[Schur 1923]      \label{thm:SC}    

Let $\mathbb{I}\subset\R$ and $f:\mathbb{I}^n\rightarrow \R$ be
continuously differentiable. Then, the function $f$ is MIS on
$\mathbb{I}^n$ if and only if
\begin{align}
    f \text{ is symmetric on } \mathbb{I}^n
\end{align}
and for all $i\neq j$,
\begin{align}    \label{eq:SC}
    \left(a_i - a_j \right)
    \left[
        \frac{\partial f}{\partial a_i}
        -
        \frac{\partial f}{\partial a_j}
    \right]
    \geq 0
    \qquad
    \forall \,
    \B{a} \in \mathbb{I}^n.
\EqE
\end{align}
\end{theorem}

\mySep \mySep

Note that Schur's condition \eqref{eq:SC} can be replaced by
\begin{align}    \label{eq:SC:1}
\EqF
    \left(a_1 - a_2 \right)
    \left[
        \frac{\partial f}{\partial a_1}
        -
        \frac{\partial f}{\partial a_2}
    \right]
    \geq 0
    \qquad
    \forall \,
    \B{a} \in \mathbb{I}^n
\EqE
\end{align}
because of the symmetry. If $f$ is MIS on $\mathbb{I}^n$, then
$-f$ is a MDS function on $\mathbb{I}^n$.

\section{Some Statistics Derived From Complex Gaussian Matrices}   \label{sec:Appendix:B}

This appendix gives useful results on some statistics derived from
complex Gaussian matrices. 
%
%
%
%
%


\subsection{Preliminary Results}      \label{sec:Sec:A:1}

\begin{lemma}   \label{le:CGM:etr}

Let $\B{X}_k \sim \CGM{m}{n}{\B{0}_{m \times
n}}{\B{\Sigma}}{\B{\Psi}_k}$, $k=1,2,\ldots,p$, be statistically
independent complex Gaussian matrices and
%
%
\begin{salign}
    \B{X}
    =
        \begin{bmatrix}
            \B{X}_1 &
            \B{X}_2 &
            \cdots  &
            \B{X}_p
        \end{bmatrix}
        \sim
        \CGM{m}{np}{\B{0}_{m \times np}}{\B{\Sigma}}{\mbox{$\bigoplus_{k=1}^p \B{\Psi}_k$}}.
\end{salign}
Then, for $\B{A} \in \C^{m \times m} \geq 0$ and
$\B{B}=\bigoplus_{k=1}^p \B{B}_k$, $\B{B}_k \in \C^{n \times n}
\geq 0$, we have
\begin{align}    \label{eq:CGM:etr}
\EqF
    \EX{
        \etr\left(
                -\B{AXBX}^\dag
            \right)
        }
    =
        \prod_{k=1}^p
        \det\left(
                \B{I}_{mn}
                + \B{A\Sigma}
                  \otimes
                  \B{\Psi}_k
                  \B{B}_k
            \right)^{-1}.
\EqE
\end{align}
%
%


\begin{proof}
Since
$
    \B{AXBX}^\dag
    =
        \sum_{k=1}^p
            \B{A}
            \B{X}_k
            \B{B}_k
            \B{X}_k^\dag
$, we have
\begin{align}      \label{eq:CGM:etr:pf:1}
\EqF
    \EX{
        \etr\left(
                -\B{AXBX}^\dag
            \right)
        }
    =
        \prod_{k=1}^p
            \EXs{\B{X}_k}{
                \etr\bigl(
                        -
                        \B{A}
                        \B{X}_k
                        \B{B}_k
                        \B{X}_k^\dag
                \bigr)
                }.
\end{align}
Therefore\footnote{
        If $\B{X}=\left(X_{ij}\right)$ is an $m \times n$ matrix of functionally independent complex variables, then
        \begin{align*}
            d\B{X}
            =
                \prod_{i=1}^m
                \prod_{j=1}^n
                d\Re X_{ij} \,
                d\Im X_{ij}.
        \end{align*}
        %
        %
        }
\begin{align}    \label{eq:CGM:etr:pf:2}
    &\EXs{\B{X}_k}{
                \etr\bigl(
                        -
                        \B{A}
                        \B{X}_k
                        \B{B}_k
                        \B{X}_k^\dag
                \bigr)
                }
    \nonumber \\
    &\qquad
    =
      c_k
      \int_{\B{X}_k}
        \etr\bigl(
                        -
                        \B{A}
                        \B{X}_k
                        \B{B}_k
                        \B{X}_k^\dag
                 -\B{\Sigma}^{-1}\B{X}_k
                 \B{\Psi}_k^{-1}\B{X}_k^\dag
            \bigr)
        d\B{X}_k
    \nonumber \\
    &\qquad
    =
      c_k
      \int_{\B{X}_k}
        \exp\left[
                -\bigl(
                    \vecOp \bigl(\B{X}_k^\dag\bigr)
                 \bigr)^\dag
                 \left\{
                    \bigl(
                        \B{\Sigma}^{T}
                        \otimes
                        \B{\Psi}_k
                    \bigr)^{-1}
                    + \B{A}^{T}
                      \otimes
                      \B{B}_k
                 \right\}
                 \vecOp \bigl(\B{X}_k^\dag\bigr)
            \right]
        d\B{X}_k
    \nonumber \\
    &\qquad
    =
      c_k
      \pi^{mn}
      \det\left\{
                \bigl(
                    \B{\Sigma}^{T}
                    \otimes
                    \B{\Psi}_k
                \bigr)^{-1}
                + \B{A}^{T}
                  \otimes
                  \B{B}_k
            \right\}^{-1}
    \nonumber \\
    &\qquad
    =
      \det\left(
                \B{I}_{mn}
                + \B{A\Sigma}
                  \otimes
                  \B{\Psi}_k
                  \B{B}_k
            \right)^{-1}
\end{align}
where $c_k=\pi^{-mn} \det\left(\B{\Sigma}\right)^{-n}
\det\left(\B{\Psi}_k\right)^{-m}$. 
%
%
Combining \eqref{eq:CGM:etr:pf:1} and \eqref{eq:CGM:etr:pf:2}
complete the proof.
\end{proof}

\end{lemma}

\mySep

\begin{lemma}   \label{le:CGM:decomp}
Let $\B{X} \sim \CGM{m}{n}{\B{0}_{m \times
n}}{\B{\Sigma}}{\B{\Psi}}$. Then, for $\B{A},\B{B} \in \C^{m
\times n}$, we have
\begin{align}    \label{eq:CGM:decomp}
    \EX{
        \etr\left(
                \B{X}^\dag\B{A}
                +\B{B}^\dag\B{X}
            \right)
        }
    = \etr\left(
                 \B{\Sigma}
                 \B{A}
                 \B{\Psi}
                 \B{B}^\dag
            \right).
\end{align}

\begin{proof}
%
%
Let $\B{M}_1$ and $\B{M}_2$ be $m \times n$ matrices such that
\begin{align}    \label{eq:CGM:decomp:pf:M}
    &\tr\bigl(
            \B{X}^\dag\B{A}
            +\B{B}^\dag\B{X}
            -\B{\Sigma}^{-1}\B{X}
             \B{\Psi}^{-1}\B{X}^\dag
        \bigr)
    \nonumber \\
    &\qquad
    =   \tr\bigl(
             \B{\Sigma}^{-1}\B{M}_1
             \B{\Psi}^{-1}\B{M}_2^\dag
        \bigr)
        + \tr\left\{
            -\B{\Sigma}^{-1}
             \left(\B{X}-\B{M}_1\right)
             \B{\Psi}^{-1}
             \left(\B{X}-\B{M}_2\right)^\dag
        \right\}.
\end{align}
Then, since
\begin{align}
\int_{\B{X}}
        \etr\bigl\{
            -\B{\Sigma}^{-1}
             \left(\B{X}-\B{M}_1\right)
             \B{\Psi}^{-1}
             \left(\B{X}-\B{M}_2\right)^\dag
            \bigr\}
        d\B{X}=\pi^{mn} \det\left(\B{\Sigma}\right)^{n}
\det\left(\B{\Psi}\right)^{m},
\end{align}
we get
\begin{align}    \label{eq:CGM:decomp:pf:2}
    \EX{
        \etr\bigl(
                \B{X}^\dag\B{A}
                +\B{B}^\dag\B{X}
            \bigr)
        }
    = \etr\bigl(
            \B{\Sigma}^{-1}\B{M}_1
             \B{\Psi}^{-1}\B{M}_2^\dag
          \bigr).
\end{align}
By comparing both the sides of \eqref{eq:CGM:decomp:pf:M}, we have
\begin{align}    \label{eq:M1}
    \B{M}_1
    &=
    \B{\Sigma A \Psi}
    \\
                \label{eq:M2}
    \B{M}_2
    &=
    \B{\Sigma B \Psi}.
\end{align}
Finally, substituting \eqref{eq:M1} and \eqref{eq:M2} into
\eqref{eq:CGM:decomp:pf:2} completes the proof.
\end{proof}
\end{lemma}

\mySep

\begin{lemma}   \label{le:CGM:CF}
Let $\B{X} \sim \CGM{m}{n}{\B{M}}{\B{\Sigma}}{\B{\Psi}}$. Then,
the characteristic function of $\B{X}$ is
\begin{align}
\EqF
    \CF{\B{X}}{\B{Z}}
    &   \triangleq
        \E\Bigl\{
            \exp\left[
               \ImUnit \,
               \Re
               \tr\left(
                        \B{XZ}^\dag
                  \right)
              \right]
        \Bigr\}
    \nonumber \\
    &   =
        \exp\left[
               \ImUnit \,
               \Re
               \tr\left(
                        \B{MZ}^\dag
                  \right)
               -
               \frac{1}{4}
               \tr\left(
                     \B{\Sigma}
                     \B{Z}
                     \B{\Psi}
                     \B{Z}^\dag
                  \right)
            \right]
\EqE
\end{align}
where $\ImUnit=\sqrt{-1}$ and $\B{Z} \in \C^{m \times n}$ is an
arbitrary matrix.

\begin{proof}
Let $\B{X}_1 \sim \CGM{m}{n}{\B{0}_{m \times
n}}{\B{\Sigma}}{\B{\Psi}}$. Then,
\begin{align}    \label{eq:CGM:CF:pf:1}
\EqF
    \CF{\B{X}}{\B{Z}}
    =
        \exp\left[
               \ImUnit \,
               \Re
               \tr\left(
                        \B{MZ}^\dag
                  \right)
            \right]
        \cdot
        \EX{
        \exp\left[
               \ImUnit \,
               \Re
               \tr\left(
                        \B{X}_1
                        \B{Z}^\dag
                  \right)
            \right]
            }.
\end{align}
Since
\begin{align}
    \Re
    \tr\left(
            \B{X}_1
            \B{Z}^\dag
       \right)
    =
    \frac{1}{2}
    \tr\bigl(
            \B{Z}^\dag
            \B{X}_1
            +
            \B{X}_1^\dag
            \B{Z}
       \bigr),
\EqE
\end{align}
it follows from Lemma \ref{le:CGM:decomp} that
\begin{align}    \label{eq:CGM:CF:pf:2}
\EqF
    \EX{
        \exp\left[
               \ImUnit \,
               \Re
               \tr\left(
                        \B{X}_1
                        \B{Z}^\dag
                  \right)
            \right]
        }
    =
        \etr\left(
                -\frac{1}{4}
                 \B{\Sigma}
                 \B{Z}
                 \B{\Psi}
                 \B{Z}^\dag
            \right).
\EqE
\end{align}
Combining \eqref{eq:CGM:CF:pf:1} and \eqref{eq:CGM:CF:pf:2}
completes the proof.
\end{proof}

\end{lemma}

\mySep

%
%
%
%

We remark that Lemma \ref{le:CGM:CF} is a counterpart result of
the real case in \cite[Theorem~2.3.2]{GN:00:Book}.

\subsection{Hypergeometric Functions of Matrix Arguments}      \label{sec:Appendix:HF}

The hypergeometric functions of matrix arguments often appear in
deriving the distributions and statistics of random matrices
\cite{Jam:60:AMS,Jam:64:AMS,Kha:66:AMS,Kha:70:SIJSSA,GN:00:Book}.
In parallel to the hypergeometric functions of a scalar argument,
the hypergeometric functions of one or two matrix arguments can be
expressed as an infinite series of zonal polynomials:\footnote{
        Zonal polynomials
        of a symmetric matrix were introduced in
        \cite{Jam:60:AMS} using group representation theory. In parallel to
        a real matrix argument, zonal polynomials of a
        Hermitian matrix were defined in \cite{Jam:64:AMS} as
        natural extension of the real case. Those polynomials are
        homogeneous symmetric functions in the eigenvalues of
        matrix argument and can be constructed in terms of homogeneous symmetric
        polynomials such as monomial symmetric functions,
        elementary symmetric functions, and Schur functions
        \cite{Tak:84:Book}.
}
\begin{align}   \label{eq:pFq:I}
\EqF
    &\matHyperPFQI{p}{q}
                 {
                    a_1,\ldots,a_p;
                    b_1,\ldots,b_q;
                    \B{A}
                 }
    =
        \sum_{k=0}^\infty
        \sum_{\kappa}
            \frac{
                    \MPS{a_1}{\kappa}
                    \cdots
                    \MPS{a_p}{\kappa}
                 }
                 {
                    \MPS{b_1}{\kappa}
                    \cdots
                    \MPS{b_q}{\kappa}
                 }
            \frac{
                    \ZP{\kappa}{\B{A}}
                 }
                 {k!}
\end{align}
\begin{align}
        \label{eq:pFq:II}
    &\matHyperPFQII{p}{q}{n}
                 {
                    a_1,\ldots,a_p;
                    b_1,\ldots,b_q;
                    \B{A},
                    \B{B}
                 }
    =
        \sum_{k=0}^\infty
        \sum_{\kappa}
            \frac{
                    \MPS{a_1}{\kappa}
                    \cdots
                    \MPS{a_p}{\kappa}
                 }
                 {
                    \MPS{b_1}{\kappa}
                    \cdots
                    \MPS{b_q}{\kappa}
                 }
            \frac{
                    \ZP{\kappa}{\B{A}}
                    \ZP{\kappa}{\B{B}}
                 }
                 {
                    k!
                    \,
                    \ZP{\kappa}{\B{I}_n}
                 }
\EqE
\end{align}
with Hermitian $\B{A} \in \C^{n \times n}$ and $\B{B} \in \C^{n
\times n}$.  In \eqref{eq:pFq:I} and \eqref{eq:pFq:II},
$\kappa=\left(k_1,k_2,\ldots,k_n\right)$ denotes a partition of
the nonnegative integer $k$ such that $k_1 \geq k_2 \geq \ldots
\geq k_n \geq 0$ and $\sum_{i=1}^n k_i =k$, $\MPS{a}{\kappa}$ is
the complex multivariate hypergeometric coefficient of the
partition $\kappa$ \cite[eq.~(84)]{Jam:64:AMS}, and
$\ZP{\kappa}{\cdot}$ is the zonal polynomial of a Hermitian matrix
\cite[eq.~(85)]{Jam:64:AMS}. Although these functions are of great
interest from an analytical point of view, the practical
difficulty lies in their numerical aspects. The determinantal
representation for the hypergeometric function of two Hermitian
matrices \cite[Lemma~3]{Kha:70:SIJSSA} settles this computational
problem and has been widely used in the literature of
multiple-antenna communication theory (see, e.g.,
\cite{CWZ:03:IT,SWLC:06:WCOM,CWZ:03:COM,CWZ:05:COM}). However,
\cite[Lemma~3]{Kha:70:SIJSSA} is valid only for the case of two
matrix arguments with the same dimension and the distinct
eigenvalues. In the following lemma, we generalize
\cite[Lemma~3]{Kha:70:SIJSSA} for the case that two matrix
arguments have the different matrix dimension and the
eigenvalues of arbitrary multiplicity.

\mySep

\begin{lemma}[Generic Determinantal Formula]     \label{le:pFq}

Let $\B{\Lambda} \in \C^{m \times m}$ and $\B{\Sigma} \in \C^{n
\times n}$, $m \leq n$, be Hermitian matrices with the ordered
eigenvalues $\lambda_1 \geq \lambda_2 \geq \ldots \geq \lambda_m$
and $\sigma_1 \geq \sigma_2 \geq \ldots \geq \sigma_n$,
respectively. Given $a_i, b_j \in \C$ where $i=1,2,\ldots,p$ and
$j=1,2,\ldots,q$, define
\begin{align}    \label{eq:pFq:g}
\EqF
    \mathcal{H}_{p,q}^{n,\nu}\left(x\right)
    \triangleq
        \HyperPFQ{p}{q}
                 {
                  a_1-n+\nu,\ldots,a_p-n+\nu;
                  b_1-n+\nu,\ldots,b_q-n+\nu;
                  x
                 }
\end{align}
%
%
%
%
\begin{align}
    \chi_{p,q}^{n,\nu}
    \triangleq
        \frac{
            \prod_{j=1}^q
                \PS{b_j-n+1}{\nu}
             }
             {
            \prod_{i=1}^p
                \PS{a_i-n+1}{\nu}
             }
\EqE
\end{align}
where $\nu$ is an arbitrary nonnegative integer,
$\PS{a}{n}=a\left(a+1\right)\cdots\left(a+n-1\right)$,
$\PS{a}{0}=1$ is the Pochhammer symbol, and
$\HyperPFQ{p}{q}{a_1,a_2,\ldots,a_p;b_1,b_2,\ldots,b_q;z}$ is the
generalized hypergeometric function of scalar argument
\cite[eq.~(9.14.1)]{GR:00:Book}. Then,
\begin{salign}    \label{eq:pFq:Det}
    &
    \matHyperPFQII{p}{q}{n}
                {a_1,\ldots,a_p;
                 b_1,\ldots,b_q;
                 \B{\Lambda},
                 \B{\Sigma}}
    \cdot
    \prod_{i<j}^m 
        \left(\lambda_j-\lambda_i\right)
    \nonumber \\
    & \hspace{1.3cm}
    =
        \frac{
                K_{p,q}^{m,n}
             }
             {
                \det\left(
                            \B{\Lambda}
                    \right)^{n-m}
             }
        \cdot
        \frac{
                \det\left(
                            \begin{bmatrix}
                                \pmb{\mathcal{Z}}_{(n-m),1}
                                &
                                \pmb{\mathcal{Z}}_{(n-m),2}
                                &
                                \cdots
                                &
                                \pmb{\mathcal{Z}}_{(n-m),\EVn{\B{\Sigma}}}
                                \\
                                \pmb{\mathcal{Y}}_{1}
                                &
                                \pmb{\mathcal{Y}}_{2}
                                &
                                \cdots
                                &
                                \pmb{\mathcal{Y}}_{\EVn{\B{\Sigma}}}
                            \end{bmatrix}
                    \right)
             }
             {
                \det\left(
                            \begin{bmatrix}
                                \pmb{\mathcal{Z}}_{(n),1}
                                &
                                \pmb{\mathcal{Z}}_{(n),2}
                                &
                                \cdots
                                &
                                \pmb{\mathcal{Z}}_{(n),\EVn{\B{\Sigma}}}
                            \end{bmatrix}
                    \right)
             }
\end{salign}
with
\begin{align}   \label{eq:Kmn}
    K_{p,q}^{m,n}
    =
        \prod_{i=1}^m
            \chi_{p,q}^{n,n-i}
            \cdot
            \left(n-i\right)!
\EqE
\end{align}
where $\pmb{\mathcal{Y}}_k=\left(\mathcal{Y}_{k,ij}\right)$ and
$\pmb{\mathcal{Z}}_{(l),k}=\left(\mathcal{Z}_{(l),k,ij}\right)$,
$l \leq n$, $k=1,2,\ldots,\EVn{\B{\Sigma}}$, are $m \times
\EVm{k}{\B{\Sigma}}$ and $l \times \EVm{k}{\B{\Sigma}}$ matrices,
 whose $\left(i,j\right)$th entries are given respectively by
\begin{gather}
\EqF
        \label{eq:pFq:Det:S}
    \mathcal{Y}_{k,ij}
    =
        \frac{
                \lambda_i^{j-1}
             }
             {
                \chi_{p,q}^{n,j-1}
             }
        \cdot
        \mathcal{H}_{p,q}^{n,j}
            \left(
                \lambda_i
                \odEV{\sigma}{k}{}
            \right)
%
%
%
%
    \\
        \label{eq:pFq:Det:R}
    \mathcal{Z}_{(l),k,ij}
    =
        \PS{i-j+1}{j-1}
        \odEV{\sigma}{k}{i-j}.
\EqE
\end{gather}
In particular, for
$\matHyperPFQII{0}{0}{n}{\B{\Lambda},\B{\Sigma}}$, $K_{p,q}^{m,n}$
in \eqref{eq:Kmn} and the $\left(i,j\right)$th entry of
$\pmb{\mathcal{Y}}_k$ in \eqref{eq:pFq:Det:S} reduce to
\begin{align}   \label{eq:0F0:Kmn}
\EqF
    K_{0,0}^{m,n}
    &=
        \prod_{i=1}^m
            \left(n-i\right)!
%
%
%
%
    \\
        \label{eq:0F0:S}
    \mathcal{Y}_{k,ij}
    &=
        \lambda_i^{j-1}
        e^{
            \lambda_i
            \odEV{\sigma}{k}{}
          }.
\EqE
\end{align}

\begin{proof}
Let us dilate the $m \times m$ matrix $\B{\Lambda}$ to the $n
\times n$ matrix $\B{\Lambda} \oplus \B{0}_{n-m}$ by affixing zero
elements. 
%
%
Then, this augmented matrix $\B{\Lambda} \oplus \B{0}_{n-m}$ has
the eigenvalues $\lambda_1,\lambda_2,\ldots,\lambda_m$ and $(n-m)$
additional zero eigenvalues. Note that zonal polynomials depend on
its Hermitian matrix arguments through Schur functions in the
eigenvalues of matrix arguments
\cite{Jam:64:AMS,Kha:66:AMS,Kha:70:SIJSSA,Tak:84:Book}. Since
Schur functions are invariant to augmenting zero elements
\cite{Mac:99:Book}, it is easy to show that
\begin{align}       \label{eq:ZP:PT}
\EqF
    \ZP{\kappa}{\B{\Lambda} \oplus \B{0}_{n-m}}
    =
        \ZP{\kappa}{\B{\Lambda}}.
\EqE
\end{align}
%
%
%
%
Let $\lambda_{m+1},\lambda_{m+2},\ldots,\lambda_{n}$ be $(n-m)$
additional zero eigenvalues and denote the left-hand side of
\eqref{eq:pFq:Det} by $\mathrm{LHS}_{\eqref{eq:pFq:Det}}$ for
convenience.
Then, it follows from \eqref{eq:ZP:PT} and
\cite[Lemma~3]{Kha:70:SIJSSA} that
\begin{align}    \label{eq:pFq:Det:Khatri}
\EqF
    \mathrm{LHS}_{\eqref{eq:pFq:Det}}
    =
        K_{p,q}^{n,n}
        \,
        \frac{
                \det\limits_{1\leq i,j \leq n}\left(
                            \mathcal{H}_{p,q}^{n,1}
                                    \left(
                                            \lambda_i
                                            \sigma_j
                                    \right)
                    \right)
             }
             {
                \prod_{i<j}^n
                    \left(\lambda_j-\lambda_i\right)
                    \left(\sigma_j-\sigma_i\right)
             }
        \cdot
        \prod_{i<j}^m
            \left(\lambda_j-\lambda_i\right).
\EqE
\end{align}

From a computational point of view, \eqref{eq:pFq:Det:Khatri}
presents numerical difficulty since the Vandermonde determinant
$\prod_{i<j}^n \left( \lambda_j -\lambda_i \right)$ or
$\prod_{i<j}^n \left( \sigma_j -\sigma_i \right)$ becomes zero
when some of the $\lambda_i$'s or $\sigma_i$'s are
equal. 
%
%
This can be alleviated by using Cauchy's mean value theorem (or
L'H\^{o}spital's rule): 
%
%
\begin{align}       \label{eq:pFq:limit:1}
\EqF
    \mathrm{LHS}_{\eqref{eq:pFq:Det}}
    &=
        K_{p,q}^{n,n}
        \,
        \lim_{
                \B{\sigma}
                \to
                \tilde{\B{\sigma}}
             }
        \,
        \lim_{
                \left\{
                \lambda_{k}
                \right\}_{k=m+1}^n
                \to 0
             }
        \,
        \frac{
                \det\limits_{1\leq i,j \leq n}\left(
                            \mathcal{H}_{p,q}^{n,1}
                                    \left(
                                            \lambda_i
                                            \sigma_j
                                    \right)
                    \right)
             }
             {
                \prod_{i<j}^n
                    \left(\lambda_j-\lambda_i\right)
                    \left(\sigma_j-\sigma_i\right)
             }
        \cdot
        \prod_{i<j}^m
            \left(\lambda_j-\lambda_i\right)
\EqE
\end{align}
where $\B{\sigma} \to \tilde{\B{\sigma}}$ means that
\begin{align*}
\EqF
    &
    \bigl\{
        \sigma_i
    \bigr\}_{i=1}^{\EVm{1}{\B{\Sigma}}}
    \to
        \odEV{\sigma}{1}{},
    \\
    &
    \bigl\{
        \sigma_i
    \bigr\}_{i=\EVm{1}{\B{\Sigma}}+1}^{\EVm{1}{\B{\Sigma}}+\EVm{2}{\B{\Sigma}}}
    \to
        \odEV{\sigma}{2}{},
    \\
    & \quad
    \vdots
    \\
    &
    \bigl\{
        \sigma_i
    \bigr\}_{i=n-\EVm{\EVn{\B{\Sigma}}}{\B{\Sigma}}+1}^{n}
    \to
        \odEV{\sigma}{\EVn{\B{\Sigma}}}{}.
\EqE
\end{align*}

Let $n$-dimensional vectors $\B{u}\left(z\right)$ and
$\B{v}\left(z\right)$ be
\begin{align}
    \label{eq:pFq:Det:s}
    \B{u}\left(z\right)
    &=
        \left(
                \mathcal{H}_{p,q}^{n,1}\left(\sigma_1 z\right),
                \mathcal{H}_{p,q}^{n,1}\left(\sigma_2 z\right),
                \ldots,
                \mathcal{H}_{p,q}^{n,1}\left(\sigma_n z\right)
        \right)
    \\
    \label{eq:pFq:Det:r}
    \B{v}\left(z\right)
    &=
        \left(
                1,z,\ldots,z^{n-1}
        \right)
\EqE
\end{align}
and let $\B{u}^{\left(k\right)}\left(z\right)$ and
$\B{v}^{\left(k\right)}\left(z\right)$ be the $k$th derivatives of
$\B{u}\left(z\right)$ and $\B{v}\left(z\right)$ with respect to
$z$, respectively. Note that the $j$th components
$u_j^{\left(k\right)}\left(z\right)$ and
$v_j^{\left(k\right)}\left(z\right)$, $j=1,2,\ldots,n$, of
$\B{u}^{\left(k\right)}\left(z\right)$ and
$\B{v}^{\left(k\right)}\left(z\right)$ are given respectively by
\begin{align}
\EqF
        \label{eq:pFq:Det:s:1}
    u_j^{\left(k\right)}\left(z\right)
    &=
        \frac{
                \sigma_j^k
             }
             {
                \chi_{p,q}^{n,k}
             }
        \cdot
        \mathcal{H}_{p,q}^{n,k+1}\left(\sigma_j z\right)
    \\
        \label{eq:pFq:Det:r:1}
    v_j^{\left(k\right)}\left(z\right)
    &=
        \PS{j-k}{k}
        z^{j-k-1}
\EqE
\end{align}
where \eqref{eq:pFq:Det:s:1} follows from the differentiation
identity of \cite[eq.~(7.2.3.47)]{PBM:90:Book:v3}. Then, taking
the limits on $\lambda_{k}$'s, we get
\begin{align}    \label{eq:pFq:Lambda:1}
\EqF
        \lim_{
                \left\{
                \lambda_{k}
                \right\}_{k=m+1}^n
                \to 0
             }
        \,
        \frac{
                \det\limits_{1\leq i,j \leq n}\left(
                            \mathcal{H}_{p,q}^{n,1}
                                    \left(
                                            \lambda_i
                                            \sigma_j
                                    \right)
                    \right)
             }
             {
                \prod_{i<j}^n
                    \left(\lambda_j-\lambda_i\right)
             }
        =
            \frac{
                    \det\left(
                                \left[
                                    \begin{smallmatrix}
                                        \B{U}_\text{A}
                                        \\
                                        \B{U}_\text{B}
                                    \end{smallmatrix}
                                \right]
                        \right)
                 }
                 {
                    \det\left(
                                \left[
                                    \begin{smallmatrix}
                                        \B{V}_\text{A}
                                        \\
                                        \B{V}_\text{B}
                                    \end{smallmatrix}
                                \right]
                        \right)
                 }
\EqE
\end{align}
with the $\left(n-m\right) \times n$ matrices
%
%
\begin{salign}
    \B{U}_\text{A}
    &=
        \left(U_{\text{A},ij}\right)
    =
        \begin{bmatrix}
            \B{u}^{\left(0\right)}\left(0\right)
            \\
            \B{u}^{\left(1\right)}\left(0\right)
            \\
            \vdots
            \\
            \B{u}^{\left(n-m-1\right)}\left(0\right)
        \end{bmatrix}
    \\[5pt]
    \B{V}_\text{A}
    &=
        \left(V_{\text{A},ij}\right)
    =
        \begin{bmatrix}
            \B{v}^{\left(0\right)}\left(0\right)
            \\
            \B{v}^{\left(1\right)}\left(0\right)
            \\
            \vdots
            \\
            \B{v}^{\left(n-m-1\right)}\left(0\right)
        \end{bmatrix},
\end{salign}
and the $m \times n$ matrices
$\B{U}_\text{B}=\left(\mathcal{H}_{p,q}^{n,1}\left(\lambda_i\sigma_j\right)\right)$
and $\B{V}_\text{B}=\left(\lambda_i^{j-1}\right)$. From
\eqref{eq:pFq:Det:s:1} and \eqref{eq:pFq:Det:r:1}, it is easy to
see that the $\left(i,j\right)$th entries of $\B{U}_\text{A}$ and
$\B{V}_\text{A}$ are given respectively by
\begin{salign}
    U_{\text{A},ij}
    &=
        u_j^{(i-1)}\left(0\right)
    =
        \frac{
                \sigma_j^{i-1}
             }
             {
                \chi_{p,q}^{n,i-1}
             }
    \\[5pt]
    V_{\text{A},ij}
    &=
        v_j^{(i-1)}\left(0\right)
    =
        \begin{cases}
            \left(i-1\right)! \, ,
            &
            \text{if $i=j$}
            \\[5pt]
            0 \, ,
            &
            \text{otherwise}.
        \end{cases}
\end{salign}
Now, using the result on the determinant of a partitioned matrix
\begin{salign}      \label{eq:Det:partition}
    \det\left(
                \begin{bmatrix}
                    \B{A}   &   \B{B}
                    \\
                    \B{C}   &   \B{D}
                \end{bmatrix}
        \right)
    =
        \det\left(\B{A}\right)
        \det\left(
                    \B{D}
                    -
                    \B{C}
                    \B{A}^{-1}
                    \B{B}
            \right),
    \quad
    \text{if $\B{A}$ is invertible},
\end{salign}
we have
\begin{salign}      \label{eq:pFq:Lambda:2}
    \det\left(
                \begin{bmatrix}
                    \B{V}_\text{A}
                    \\
                    \B{V}_\text{B}
                \end{bmatrix}
        \right)
    &=
        \prod_{l=1}^{n-m}
            \left(l-1\right)!
        \cdot
        \det\left(
                    \begin{bmatrix}
                        \lambda_1^{n-m}     &
                        \lambda_1^{n-m+1}   &
                        \cdots              &
                        \lambda_1^{n-1}
                        \\
                        \lambda_2^{n-m}     &
                        \lambda_2^{n-m+1}   &
                        \cdots              &
                        \lambda_2^{n-1}
                        \\
                        \vdots              &
                        \vdots              &
                        \ddots              &
                        \vdots
                        \\
                        \lambda_m^{n-m}     &
                        \lambda_m^{n-m+1}   &
                        \cdots              &
                        \lambda_m^{n-1}
                    \end{bmatrix}
            \right)
    \nonumber \\[3pt]
    &=
        \prod_{l=1}^{n-m}
            \left(l-1\right)!
        \,
        \prod_{k=1}^m
            \lambda_k^{n-m}
        \prod_{i<j}^m
            \left(
                    \lambda_j
                    -\lambda_i
            \right).
\end{salign}
Hence, combining \eqref{eq:pFq:limit:1}, \eqref{eq:pFq:Lambda:1},
and \eqref{eq:pFq:Lambda:2} gives
\begin{align}       \label{eq:pFq:limit:2}
\EqF
    \mathrm{LHS}_{\eqref{eq:pFq:Det}}
    &=
        \frac{
                K_{p,q}^{m,n}
             }
             {
                \det\left(
                            \B{\Lambda}
                    \right)^{n-m}
             }
        \,
        \lim_{
                \B{\sigma}
                \to
                \tilde{\B{\sigma}}
             }
        \,
        \frac{
                    \det\left(
                                \left[
                                    \begin{smallmatrix}
                                        \tilde{\B{U}}_\text{A}
                                        \\
                                        \B{U}_\text{B}
                                    \end{smallmatrix}
                                \right]
                        \right)
             }
             {
                \prod_{i<j}^n
                    \left(\sigma_j-\sigma_i\right)
             }
\EqE
\end{align}
where $\tilde{\B{U}}_\text{A}=\left(\sigma_j^{i-1}\right)$ is the
$\left(n-m\right) \times n$ submatrix of the Vandermonde matrix of
$\sigma_1,\sigma_2,\ldots,\sigma_n$.

Using similar steps leading to \eqref{eq:pFq:Lambda:1}, we obtain
\begin{salign}    \label{eq:pFq:Det:1}
    \lim_{\B{\sigma} \to \tilde{\B{\sigma}}}
        \frac{
                    \det\left(
                                \left[
                                    \begin{smallmatrix}
                                        \tilde{\B{U}}_\text{A}
                                        \\
                                        \B{U}_\text{B}
                                    \end{smallmatrix}
                                \right]
                        \right)
             }
             {
                \prod_{i<j}^n
                    \left(\sigma_j-\sigma_i\right)
             }
    =
        \frac{
                \det\left(
                            \begin{bmatrix}
                                \pmb{\mathcal{Z}}_{(n-m),1}
                                &
                                \pmb{\mathcal{Z}}_{(n-m),2}
                                &
                                \cdots
                                &
                                \pmb{\mathcal{Z}}_{(n-m),\EVn{\B{\Sigma}}}
                                \\
                                \pmb{\mathcal{Y}}_{1}
                                &
                                \pmb{\mathcal{Y}}_{2}
                                &
                                \cdots
                                &
                                \pmb{\mathcal{Y}}_{\EVn{\B{\Sigma}}}
                            \end{bmatrix}
                    \right)
             }
             {
                \det\left(
                            \begin{bmatrix}
                                \pmb{\mathcal{Z}}_{(n),1}
                                &
                                \pmb{\mathcal{Z}}_{(n),2}
                                &
                                \cdots
                                &
                                \pmb{\mathcal{Z}}_{(n),\EVn{\B{\Sigma}}}
                            \end{bmatrix}
                    \right)
             }
\end{salign}
where the $\left(i,j\right)$th entries of $m \times
\EVm{k}{\B{\Sigma}}$ matrices $\pmb{\mathcal{Y}}_k$ and $l \times
\EVm{k}{\B{\Sigma}}$ matrices $\pmb{\mathcal{Z}}_{(l),k}$, $l \leq
n$, $k=1,2,\ldots,\EVn{\B{\Sigma}}$, are given by
\eqref{eq:pFq:Det:S} and \eqref{eq:pFq:Det:R}, respectively.
Finally, substituting \eqref{eq:pFq:Det:1} into
\eqref{eq:pFq:limit:2} completes the proof of the lemma.
\end{proof}

\end{lemma}

\mySep

As a by-product of Lemma \ref{le:pFq}, we obtain the following
determinantal formula for the hypergeometric function of one
matrix argument.

\mySep

\begin{corollary}   \label{cor:pFq}

If $\B{\Sigma}=\B{I}_n$ in Lemma \ref{le:pFq}, then we have
\begin{align}    \label{eq:pFq:Det:one}
\EqF
    {}_{p}\tilde{F}_{q}
        \left(
                a_1,\ldots,a_p;
                b_1,\ldots,b_q;
                \B{\Lambda}
        \right)
    \cdot
    \prod_{i<j}^m
        \left(
                \lambda_j
                -\lambda_i
        \right)
    =
        \det_{1\leq i,j \leq m}
            \left(
                    \lambda_i^{j-1}
                    \mathcal{H}_{p,q}^{n,n-m+j}
                        \left(
                                \lambda_i
                        \right)
            \right).
\EqE
\end{align}

\begin{proof}
The result follows immediately from  \eqref{eq:Det:partition} and
Lemma~\ref{le:pFq} with $\EVn{\B{\Sigma}}=1,
\EVm{1}{\B{\Sigma}}=n$, and $\odEV{\sigma}{1}{}=1$.
\end{proof}

\end{corollary}


\subsection{Some Statistics}      \label{sec:Sec:A:3}

\begin{lemma}   \label{le:CGM:CumTr}
Let $\B{X} \sim \CGM{m}{n}{\B{0}_{m \times
n}}{\B{\Sigma}}{\B{\Psi}}$. Then, for $\B{A}\in \C^{m \times m}
\geq 0$ and $\B{B}\in \C^{n \times n} \geq 0$, the $k$th-order
cumulant of $\tr\bigl(\B{AXBX}^\dag\bigr)$ is
\begin{align}    \label{eq:CGM:CumTr}
\EqF
    \mathbb{C}\mathrm{um}_k\Bigl\{
                    \tr\left(
                            \B{AXBX}^\dag
                       \right)
                \Bigr\}
    &\triangleq
        \left(-1\right)^k
        \left.
            \frac{d^k}{ds^k}
            \ln \MGF{\tr(\B{AXBX}^\dag)}{s}
        \right|_{s=0}
    \nonumber \\
    &=
        \left(k-1\right)! \,
        \tr\bigl\{
                \left(
                        \B{A\Sigma}
                \right)^k
           \bigr\}
        \tr\bigl\{
                \left(
                        \B{\Psi B}
                \right)^k
           \bigr\}
\EqE
\end{align}
where $\MGF{\tr(\B{AXBX}^\dag)}{s}\triangleq
\E\bigl\{\etr\bigl(-s\B{AXBX}^\dag\bigr)\bigr\}$ is the MGF of
$\tr\bigl(\B{AXBX}^\dag\bigr)$.

\mySep

\begin{proof}
Since
\begin{align*}
    \tr\bigl(
            \B{AXBX}^\dag
       \bigr)
    =
        \bigl(
            \vecOp\bigl(
                        \B{X}^\dag
                  \bigr)
        \bigr)^\dag
        \bigl(
            \B{A}^T
            \otimes
            \B{B}
        \bigr)
        \vecOp\bigl(
                    \B{X}^\dag
              \bigr)
\EqE
\end{align*}
is a quadratic form in complex Gaussian variables, whose
characteristic function has been reported in \cite{Tur:60:Bio}, it
can be readily shown that
\begin{align}
\EqF
    \MGF{\tr(\B{AXBX}^\dag)}{s}
    &=
       \det\left\{
                \B{I}_{mn}
                +
                s
                \bigl(
                    \B{\Sigma}^T
                    \otimes
                    \B{\Psi}
                \bigr)
                \bigl(
                    \B{A}^T
                    \otimes
                    \B{B}
                \bigr)
           \right\}^{-1}
    \nonumber \\
    &=
       \det\left(
                \B{I}_{mn}
                +
                s
                \B{A\Sigma}
                \otimes
                \B{\Psi B}
           \right)^{-1}.
\end{align}
Therefore,
\begin{align}    \label{eq:CGM:DtrCF}
    \frac{d^k}{ds^k}
    \ln \MGF{\tr(\B{AXBX}^\dag)}{s}
    =
        \left(-1\right)^k
        \left(k-1\right)!
        \tr\left\{
                \Bigl[
                    \left(
                        \B{I}_{mn}
                        +
                        s
                        \B{A\Sigma}
                        \otimes
                        \B{\Psi B}
                    \right)^{-1}
                    \left(
                        \B{A\Sigma}
                        \otimes
                        \B{\Psi B}
                    \right)
                \Bigr]^k
            \right\}.
\EqE
\end{align}
Hence, we obtain the result \eqref{eq:CGM:CumTr} from
\eqref{eq:CGM:DtrCF} with $s=0$.
\end{proof}

\end{lemma}

\mySep

We remark that the cumulants, except for the first-order cumulant,
are invariant with respect to translations of a random variable.
The first and second order cumulants are the mean and variance of
the underlying random variable, respectively, and other
higher-order statistics can also be obtained from general
relationships between the cumulants and moments.
Lemma~\ref{le:CGM:CumTr} reveals that all cumulants of
$\tr\bigl(\B{AXBX}^\dag\bigr)$ as functionals of the eigenvalues
of $\B{A\Sigma}$ and $\B{\Psi B}$ are MIS.

\mySep

\begin{lemma}   \label{le:CGM:TrSq}
Let $\B{X} \sim \CGM{m}{n}{\B{0}_{m \times
n}}{\B{\Sigma}}{\B{\Psi}}$. Then, for $\B{A}\in \C^{m \times m}
\geq 0$ and $\B{B}\in \C^{n \times n} \geq 0$,  we have
\begin{align}    \label{eq:CGM:TrSq}
    \EX{
        \tr\left[
                \bigl(\B{AXBX}^\dag\bigr)^2
           \right]
       }
    =
        \tr^2\left(
                    \B{A\Sigma}
             \right)
        \tr\bigl\{
                \left(
                    \B{\Psi B}
                \right)^2
            \bigr\}
       +\tr^2\left(
                    \B{\Psi B}
             \right)
        \tr\bigl\{
                \left(
                    \B{A\Sigma}
                \right)^2
            \bigr\}.
\EqE
\end{align}

\begin{proof}
We first start with the characteristic function of
$\B{S}=\left(S_{ij}\right)=\B{A}^{1/2}\B{X}\B{B}^{1/2}$. Let
$\tilde{\B{\Sigma}}=\bigl(\tilde{\Sigma}_{ij}\bigr)=\B{A}^{1/2}\B{\Sigma}\B{A}^{1/2}$
and
$\tilde{\B{\Psi}}=\bigl(\tilde{\Psi}_{ij}\bigr)=\B{B}^{1/2}\B{\Psi}\B{B}^{1/2}$.
Then, 
\begin{align}   \label{eq:CGM:CF}
\EqF
    \CF{\B{S}}{\B{Z}}
    &   =
        \EX{\exp\left[
               \ImUnit \,
               \Re
               \tr\bigl(
                        \B{A}^{1/2}
                        \B{X}
                        \B{B}^{1/2}
                        \B{Z}^\dag
                  \bigr)
              \right]
            }
    \nonumber \\
    &   =
        \CF{\B{X}}{\B{A}^{1/2}\B{Z}\B{B}^{1/2}}
    \nonumber \\
    &
    \stackrel{\left(a\right)}{=}
        \etr\left(
                -\frac{1}{4}
                 \tilde{\B{\Sigma}}
                 \B{Z}
                 \tilde{\B{\Psi}}
                 \B{Z}^\dag
            \right)
    \nonumber \\
    &   =
        e^{\varphi\left(\B{Z}\right)}
\EqE
\end{align}
where (a) follows from Lemma \ref{le:CGM:CF} and
\begin{align}    \label{eq:CGM:CF:varphi}
\EqF
    \varphi\left(\B{Z}\right)
    =   -\frac{1}{4}
         \sum_{i=1}^m
         \sum_{p=1}^m
         \sum_{q=1}^n
         \sum_{j=1}^n
            \tilde{\Sigma}_{ip}
            Z_{pj}
            \tilde{\Psi}_{jq}
            Z_{iq}^\ast.
\EqE
\end{align}
It follows from the characteristic function $\CF{\B{S}}{\B{Z}}$ in
\eqref{eq:CGM:CF} that
\begin{align}   \label{eq:CGM:TrSq:e}
\EqF
    \EX{S_{i_1 j_1}
        S_{i_2 j_2}^\ast
        S_{i_3 j_3}
        S_{i_4 j_4}^\ast
        }
    &   =
        \frac{1}{\ImUnit^4}
        \left.
        \frac{\partial \CF{\B{S}}{\B{Z}}}
             {\partial
              Z_{i_1 j_1}
              \partial
              Z_{i_2 j_2}^\ast
              \partial
              Z_{i_3 j_3}
              \partial
              Z_{i_4 j_4}^\ast
             }
        \right|_{\B{Z}=\B{0}}
    \nonumber \\
    &   =
        \frac{1}{\ImUnit^4}
        \left.
            \left[
                \frac{\partial \varphi_3\left(\B{Z}\right)}
                     {\partial
                      \Re
                      Z_{i_4 j_4}
                     }
                - \ImUnit \,
                \frac{\partial \varphi_3\left(\B{Z}\right)}
                     {\partial
                      \Im
                      Z_{i_4 j_4}
                     }
            \right]
        \right|_{\B{Z}=\B{0}}
    \nonumber \\
    &   =
        \tilde{\Sigma}_{i_1 i_2}
        \tilde{\Psi}_{j_1 j_2}^\ast
        \tilde{\Sigma}_{i_3 i_4}
        \tilde{\Psi}_{j_3 j_4}^\ast
        +
        \tilde{\Sigma}_{i_1 i_4}
        \tilde{\Psi}_{j_1 j_4}^\ast
        \tilde{\Sigma}_{i_3 i_2}
        \tilde{\Psi}_{j_3 j_2}^\ast
\end{align}
with
\begin{align}
    \label{eq:CGM:TrSq:varphi:1}
    \varphi_1\left(\B{Z}\right)
    &=  e^{\varphi\left(\B{Z}\right)}
        \left[
        \frac{\partial \varphi\left(\B{Z}\right)}
             {\partial
              \Re
              Z_{i_1 j_1}
             }
             +
             \ImUnit \,
             \frac{\partial \varphi\left(\B{Z}\right)}
                  {\partial
                   \Im
                   Z_{i_1 j_1}
                  }
        \right]
    \\
    \label{eq:CGM:TrSq:varphi:2}
    \varphi_2\left(\B{Z}\right)
    &=   \frac{\partial \varphi_1\left(\B{Z}\right)}
             {\partial
              \Re
              Z_{i_2 j_2}
             }
             -
             \ImUnit \,
             \frac{\partial \varphi_1\left(\B{Z}\right)}
                  {\partial
                   \Im
                   Z_{i_2 j_2}
                  }
    \\
    \label{eq:CGM:TrSq:varphi:3}
    \varphi_3\left(\B{Z}\right)
    &=   \frac{\partial \varphi_2\left(\B{Z}\right)}
             {\partial
              \Re
              Z_{i_3 j_3}
             }
             +
             \ImUnit \,
             \frac{\partial \varphi_2\left(\B{Z}\right)}
                  {\partial
                   \Im
                   Z_{i_3 j_3}
                  }.
\EqE
\end{align}
%
%


Using \eqref{eq:CGM:TrSq:e}, we obtain
\begin{align}    \label{eq:CGM:TrSq:pf}
\EqF
    \EXs{\B{X}}
        {
        \tr\left[
                \bigl(\B{AXBX}^\dag\bigr)^2
            \right]
       }
    &=
    \EXs{\B{S}}
        {
        \tr\left[
                \bigl(\B{SS}^\dag\bigr)^2
            \right]
       }
    \nonumber \\
    &=
         \sum_{i=1}^m
         \sum_{p=1}^n
         \sum_{q=1}^n
         \sum_{j=1}^m
            \EX{
                S_{ip}
                S_{jp}^\ast
                S_{jq}
                S_{iq}^\ast
               }
    \nonumber \\
    &=
         \sum_{i=1}^m
         \sum_{p=1}^n
         \sum_{q=1}^n
         \sum_{j=1}^m
            \left(
            \tilde{\Sigma}_{i j}
            \tilde{\Psi}_{p p}
            \tilde{\Sigma}_{j i}
            \tilde{\Psi}_{q q}
            +
            \tilde{\Sigma}_{i i}
            \tilde{\Psi}_{p q}
            \tilde{\Sigma}_{j j}
            \tilde{\Psi}_{q p}
            \right)
    \nonumber \\
    &=
        \tr^2\bigl(
                    \tilde{\B{\Sigma}}
             \bigr)
        \tr\bigl(
                    \tilde{\B{\Psi}}^2
           \bigr)
       +\tr^2\bigl(
                    \tilde{\B{\Psi}}
             \bigr)
        \tr\bigl(
                    \tilde{\B{\Sigma}}^2
           \bigr)
\EqE
\end{align}
from which \eqref{eq:CGM:TrSq} follows readily.
\end{proof}

\end{lemma}

\mySep

\begin{theorem}     \label{thm:CGM:Tr2Sq}
Let $\B{X}_1 \sim \CGM{m}{p}{\B{0}_{m \times
p}}{\B{\Sigma}_1}{\B{\Psi}_1}$ and $\B{X}_2 \sim
\CGM{p}{n}{\B{0}_{p \times n}}{\B{\Sigma}_2}{\B{\Psi}_2}$ be
statistically independent complex Gaussian matrices. Then,
\begin{align}    \label{eq:CGM:Tr2Sq:1}
\EqF
    &\EXs{\B{X}_1,\B{X}_2}
        {
            \tr^2\bigl(
                        \B{X}_1
                        \B{X}_2
                        \B{X}_2^\dag
                        \B{X}_1^\dag
                 \bigr)
        }
    \nonumber \\
    &\quad
    =
        \tr\left(
                \B{\Sigma}_1^2
           \right)
        \tr^2\left(
                \B{\Psi}_1
                \B{\Sigma}_2
             \right)
        \tr\left(
                \B{\Psi}_2^2
           \right)
        +
        \tr\left(
                \B{\Sigma}_1^2
           \right)
        \tr^2\left(
                \B{\Psi}_2
           \right)
        \tr\bigl\{
                \left(
                        \B{\Psi}_1
                        \B{\Sigma}_2
                \right)^2
           \bigr\}
    \nonumber \\
    &\qquad
        +
        \tr^2\left(
                \B{\Sigma}_1
           \right)
        \tr\bigl\{
                \left(
                        \B{\Psi}_1
                        \B{\Sigma}_2
                \right)^2
           \bigr\}
        \tr\left(
                \B{\Psi}_2^2
           \right)
        +
        \tr^2\left(
                \B{\Sigma}_1
           \right)
        \tr^2\left(
                \B{\Psi}_1
                \B{\Sigma}_2
           \right)
        \tr^2\left(
                \B{\Psi}_2
           \right)
\end{align}
and
\begin{align}    \label{eq:CGM:Tr2Sq:2}
    &\EXs{\B{X}_1,\B{X}_2}
        {
            \tr\left[
                    \bigl(
                        \B{X}_1
                        \B{X}_2
                        \B{X}_2^\dag
                        \B{X}_1^\dag
                    \bigr)^2
                \right]
        }
    \nonumber \\
    &\quad
    =
        \tr^2\left(
                \B{\Sigma}_1
           \right)
        \tr^2\left(
                \B{\Psi}_1
                \B{\Sigma}_2
             \right)
        \tr\left(
                \B{\Psi}_2^2
           \right)
        +
        \tr^2\left(
                \B{\Sigma}_1
           \right)
        \tr^2\left(
                \B{\Psi}_2
           \right)
        \tr\bigl\{
                \left(
                        \B{\Psi}_1
                        \B{\Sigma}_2
                \right)^2
           \bigr\}
    \nonumber \\
    &\qquad
        +
        \tr\bigl\{
                \left(
                        \B{\Psi}_1
                        \B{\Sigma}_2
                \right)^2
           \bigr\}
        \tr\left(
                \B{\Psi}_2^2
           \right)
        \tr\left(
                \B{\Sigma}_1^2
           \right)
        +
        \tr^2\left(
                \B{\Psi}_1
                \B{\Sigma}_2
           \right)
        \tr^2\left(
                \B{\Psi}_2
           \right)
        \tr\left(
                \B{\Sigma}_1^2
           \right).
\EqE
\end{align}

\begin{proof}
Using the first two cumulants from Lemma \ref{le:CGM:CumTr}, 
we get
\begin{align}       \label{eq:CGM:Tr2Sq:pf:1}
\EqF
    &\EXs{\B{X}_1,\B{X}_2}
        {
            \tr^2\bigl(
                        \B{X}_1
                        \B{X}_2
                        \B{X}_2^\dag
                        \B{X}_1^\dag
                 \bigr)
        }
    \nonumber \\
    & \quad
    =
        \EXs{\B{X}_2}
            {
                \tr\left(
                        \B{\Sigma}_1^2
                   \right)
                \tr\left[
                        \bigl(
                            \B{X}_2
                            \B{X}_2^\dag
                            \B{\Psi}_1
                        \bigr)^2
                   \right]
                +
                \tr^2\left(
                        \B{\Sigma}_1
                   \right)
                \tr^2\bigl(
                            \B{X}_2
                            \B{X}_2^\dag
                            \B{\Psi}_1
                     \bigr)
           }
\EqE
\end{align}
where it follows from Lemma \ref{le:CGM:TrSq} that
\begin{align}        \label{eq:CGM:Tr2Sq:pf:2}
\EqF
    \EXs{\B{X}_2}
        {
            \tr\left[
                    \bigl(
                            \B{X}_2
                            \B{X}_2^\dag
                            \B{\Psi}_1
                        \bigr)^2
                \right]
        }
    =
        \tr^2\left(
                    \B{\Psi}_1
                    \B{\Sigma}_2
             \right)
        \tr\left(
                    \B{\Psi}_2^2
            \right)
        +
        \tr^2\left(
                    \B{\Psi}_2
             \right)
        \tr\bigl\{
                \left(
                    \B{\Psi}_1
                    \B{\Sigma}_2
                \right)^2
            \bigr\}
\end{align}
and from Lemma \ref{le:CGM:CumTr} that
\begin{align}        \label{eq:CGM:Tr2Sq:pf:3}
    \EXs{\B{X}_2}
        {
            \tr^2
                    \bigl(
                            \B{X}_2
                            \B{X}_2^\dag
                            \B{\Psi}_1
                    \bigr)
        }
    =
        \tr\bigl\{
                \left(
                    \B{\Psi}_1
                    \B{\Sigma}_2
                \right)^2
            \bigr\}
        \tr\left(
                    \B{\Psi}_2^2
            \right)
        +
        \tr^2\left(
                    \B{\Psi}_1
                    \B{\Sigma}_2
             \right)
        \tr^2\left(
                    \B{\Psi}_2
             \right).
\EqE
\end{align}
Combining \eqref{eq:CGM:Tr2Sq:pf:1}--\eqref{eq:CGM:Tr2Sq:pf:3}
yields the desired result \eqref{eq:CGM:Tr2Sq:1}.

Similar to \eqref{eq:CGM:Tr2Sq:pf:1}, we have
\begin{align}       \label{eq:CGM:Tr2Sq:pf:4}
\EqF
    &\EXs{\B{X}_1,\B{X}_2}
        {
            \tr\left[
                    \bigl(
                        \B{X}_1
                        \B{X}_2
                        \B{X}_2^\dag
                        \B{X}_1^\dag
                    \bigr)^2
               \right]
        }
    \nonumber \\
    & \quad
    =
        \EXs{\B{X}_2}
            {
                \tr^2\left(
                        \B{\Sigma}_1
                   \right)
                \tr\left[
                        \bigl(
                            \B{X}_2
                            \B{X}_2^\dag
                            \B{\Psi}_1
                        \bigr)^2
                   \right]
                +
                \tr^2\bigl(
                            \B{X}_2
                            \B{X}_2^\dag
                            \B{\Psi}_1
                     \bigr)
                \tr\left(
                        \B{\Sigma}_1^2
                   \right)
           }.
\EqE
\end{align}
From \eqref{eq:CGM:Tr2Sq:pf:2}--\eqref{eq:CGM:Tr2Sq:pf:4}, we
obtain the desired result \eqref{eq:CGM:Tr2Sq:2}.
\end{proof}

\end{theorem}

\mySep

%
%

\begin{theorem}  \label{thm:CWM:eigenPDF}
Let $\B{X} \sim \CGM{m}{n}{\B{0}_{m \times
n}}{\B{\Sigma}}{\B{I}_n}$, $m \leq n$, and
$\sigma_1,\sigma_2,\ldots,\sigma_m$ be the eigenvalues of
$\B{\Sigma}$ in any order. Then, the joint pdf of the ordered
eigenvalues $\lambda_1 \geq \lambda_2 \geq \ldots \geq \lambda_m
>0$ of a central complex Wishart matrix $\B{XX}^\dag \sim
\CWM{m}{n}{\B{\Sigma}}$ is given by
\begin{salign}    \label{eq:CWM:eigenPDF}
    p_{\B{\lambda}}\left(
                            \lambda_1,
                            \lambda_2,
                            \ldots,
                            \lambda_m
                   \right)
    =
        \mathcal{A}^{-1}
        \det\left(
                    \begin{bmatrix}
                        \B{G}_1
                        &
                        \B{G}_2
                        &
                        \cdots
                        &
                        \B{G}_\EVn{\B{\Sigma}}
                    \end{bmatrix}
            \right)
        \det_{1 \leq i,j \leq m}
            \left(
                    \lambda_j^{i-1}
            \right)
        \prod_{k=1}^m
            \lambda_k^{n-m}
\end{salign}
where
\begin{salign}    \label{eq:CWM:eigenPDF:K}
    \mathcal{A}
    =
        K_{0,0}^{m,n}
        \cdot
        \det\left(
                    \begin{bmatrix}
                        \B{\mathcal{B}}_1
                        &
                        \B{\mathcal{B}}_2
                        &
                        \cdots
                        &
                        \B{\mathcal{B}}_\EVn{\B{\Sigma}}
                    \end{bmatrix}
             \right)
\end{salign}
and $\B{G}_k=\bigl(G_{k,ij}\bigr)$ and
$\B{\mathcal{B}}_k=\bigl(\mathcal{B}_{k,ij}\bigr)$,
$k=1,2,\ldots,\EVn{\B{\Sigma}}$, are $m \times
\EVm{k}{\B{\Sigma}}$ matrices, whose $\left(i,j\right)$th entries
are given respectively by
\begin{align}    \label{eq:CWM:eigenPDF:V}
\EqF
    G_{k,ij}
    &=
        \oEV{\lambda}{i}{j-1}
        e^{-\oEV{\lambda}{i}{}/\odEV{\sigma}{k}{}}
%
%
    \\
        \label{eq:CWM:eigenPDF:U}
    \mathcal{B}_{k,ij}
    &=
        \left(-1\right)^{i-j}
        \PS{i-j+1}{j-1}
        \odEV{\sigma}{k}{n-i+j}.
\EqE
\end{align}

\begin{proof}
The joint eigenvalue density
$p_{\B{\lambda}}\left(\lambda_1,\lambda_2,\ldots,\lambda_m\right)$
is given by \cite[eq.~(95)]{Jam:64:AMS} in terms of the
hypergeometric
function of matrix arguments. 
%
To render this joint pdf more amenable to further analysis and
computationally tractable, we apply Lemma~\ref{le:pFq} to
\cite[eq.~(95)]{Jam:64:AMS}, which results in
\eqref{eq:CWM:eigenPDF} after some algebra.
\end{proof}

\end{theorem}

\mySep

Note that \eqref{eq:CWM:eigenPDF} is valid for any covariance
matrix $\B{\Sigma}$ with the eigenvalues of arbitrary multiplicity
and hence, generalizes the previous determinantal representation
for the joint eigenvalue pdf of Wishart matrices. If
$\B{\Sigma}=\B{I}_m$ in Theorem \ref{thm:CWM:eigenPDF}, all of the
eigenvalues are identically equal to one and hence, with
$\EVn{\B{\Sigma}}=1$, $\EVm{1}{\B{\Sigma}}=m$, and
$\odEV{\sigma}{1}{}=1$, \eqref{eq:CWM:eigenPDF} reduces to
\cite[eq.~(6)]{CWZ:03:IT}. Furthermore, if all the eigenvalues of
$\B{\Sigma}$ are distinct, then, with $\EVn{\B{\Sigma}}=m$ and
$\EVm{1}{\B{\Sigma}}=\EVm{2}{\B{\Sigma}}=\ldots=\EVm{m}{\B{\Sigma}}=1$,
\eqref{eq:CWM:eigenPDF} reduces to \cite[eq.~(18)]{CWZ:03:IT}.

\mySep\mySep

\begin{theorem}  \label{thm:CQM:eigenPDF}
Let $\B{X} \sim \CGM{m}{n}{\B{0}_{m \times
n}}{\B{I}_m}{\B{\Psi}}$, $m \leq n$, $\B{A} \in \C^{n \times n}$
be Hermitian positive definite, and
$\beta_1,\beta_2,\ldots,\beta_n$ be the eigenvalues of
$\B{A}^{1/2}\B{\Psi}\B{A}^{1/2}$ in any order. Then, the joint pdf
of the ordered eigenvalues $\lambda_1 \geq \lambda_2 \geq \ldots
\geq \lambda_m
>0$ of a matrix quadratic form $\B{XAX}^\dag$ is given by
\begin{salign}    \label{eq:CQM:eigenPDF}
    &p_{\B{\lambda}}\left(
                            \lambda_1,
                            \lambda_2,
                            \ldots,
                            \lambda_m
                   \right)
    \nonumber \\
    &\quad
    =
        \frac{
            \det\left(
                    \begin{bmatrix}
                        \B{\mathcal{V}}_{(n-m),1}
                        &
                        \B{\mathcal{V}}_{(n-m),2}
                        &
                        \cdots
                        &
                        \B{\mathcal{V}}_{(n-m),\EVn{\B{A}^{1/2}\B{\Psi}\B{A}^{1/2}}}
                        \\
                        \B{Q}_{1}
                        &
                        \B{Q}_{2}
                        &
                        \cdots
                        &
                        \B{Q}_{\EVn{\B{A}^{1/2}\B{\Psi}\B{A}^{1/2}}}
                    \end{bmatrix}
                \right)
            }{
                K_{0,0}^{m,m}
                \det\left(
                            \B{A}
                            \B{\Psi}
                    \right)^m
                \det\left(
                    \begin{bmatrix}
                        \B{\mathcal{V}}_{(n),1}
                        &
                        \B{\mathcal{V}}_{(n),2}
                        &
                        \cdots
                        &
                        \B{\mathcal{V}}_{(n),\EVn{\B{A}^{1/2}\B{\Psi}\B{A}^{1/2}}}
                    \end{bmatrix}
                \right)
            }
        \det_{1 \leq i,j \leq m}
            \left(
                    \lambda_j^{i-1}
            \right)
\end{salign}
where $\B{Q}_k=\bigl(Q_{k,ij}\bigr)$ and
$\B{\mathcal{V}}_{(l),k}=\bigl(\mathcal{V}_{(l),k,ij}\bigr)$, $l
\leq n$,
$k=1,2,\ldots,\varrho\bigl(\B{A}^{1/2}\B{\Psi}\B{A}^{1/2}\bigr)$,
are $m \times \tau_k\bigl(\B{A}^{1/2}\B{\Psi}\B{A}^{1/2}\bigr)$
and $l \times \tau_k\bigl(\B{A}^{1/2}\B{\Psi}\B{A}^{1/2}\bigr)$
matrices, whose $\left(i,j\right)$th entries are given
respectively by
\begin{align}    \label{eq:CQM:eigenPDF:V}
\EqF
    Q_{k,ij}
    &=
        \oEV{\lambda}{i}{j-1}
        e^{-\oEV{\lambda}{i}{}/\odEV{\beta}{k}{}}
%
%
    \\
        \label{eq:CQM:eigenPDF:U}
    \mathcal{V}_{(l),k,ij}
    &=
        \left(-1\right)^{i-j}
        \PS{i-j+1}{j-1}
        \odEV{\beta}{k}{-i+j}.
\EqE
\end{align}

\begin{proof}
Let $\B{S}=\B{XAX}^\dag$, then $\B{S} \sim
\CQM{m}{n}{\B{A}}{\B{I}_m}{\B{\Psi}}$ is a positive-definite
quadratic form in the complex Gaussian matrix
\cite[Definition~II.3]{SL:03:IT}. 
%
%
%
Using the pdf \cite[(2)]{SWLC:06:WCOM}, we can write the joint
eigenvalue pdf of $\B{S}$ in the form
\begin{align}   \label{eq:CQM:eigenPDF:pf:2}
\EqF
    p_{\B{\lambda}}\left(
                            \lambda_1,
                            \lambda_2,
                            \ldots,
                            \lambda_m
                   \right)
    &=
        \frac{
                \pi^{m\left(m-1\right)}
             }
             {
                \tilde{\Gamma}_m\left(m\right)
             }
        \int_{\B{U} \in \mathcal{U}\left(m\right)}
            \PDF{\B{S}}{\B{UDU}^\dag}
            \prod_{i<j}^m
                \left(
                    \lambda_i
                    -\lambda_j
                \right)^2
            \left[d\B{U}\right]
    \nonumber \\
    &=
        \frac{
                \pi^{m\left(m-1\right)}
                \det\left(\B{A\Psi}\right)^{-m}
            }{
                \tilde{\Gamma}_m\left(n\right)
                \tilde{\Gamma}_m\left(m\right)
            }
        \,
        \matHyperPFQII{0}{0}{n}{
                \B{D},
                -\B{\Psi}^{-1}
                \B{A}^{-1}
            }
        \prod_{k=1}^m
            \lambda_k^{n-m}
        \prod_{i<j}^m
            \left(
                    \lambda_i
                    -\lambda_j
            \right)^2
\EqE
\end{align}
where
$\B{D}=\mathrm{diag}\left(\lambda_1,\lambda_2,\ldots,\lambda_m\right)$,
$\tilde{\Gamma}_m\left(\alpha\right)=\pi^{m\left(m-1\right)/
2}\prod_{i=0}^{m-1}\Gamma\left(\alpha-i\right)$ with
$\Re\left(\alpha\right)>m-1$ is the complex multivariate gamma
function and $\Gamma\left(\cdot\right)$ is the gamma function. In
\eqref{eq:CQM:eigenPDF:pf:2},
$\mathcal{U}\left(m\right)=\left\{\B{U}:\B{UU}^\dag=\B{I}_m\right\}$
is the unitary group of order $m$ and $\left[d\B{U}\right]$ is the
unitary invariant Haar measure on the unitary group
$\mathcal{U}\left(m\right)$ normalized to make the total volume
unity. Similar to Theorem~\ref{thm:CWM:eigenPDF}, we obtain the
desired result \eqref{eq:CQM:eigenPDF} applying Lemma~\ref{le:pFq}
to \eqref{eq:CQM:eigenPDF:pf:2}.
\end{proof}

\end{theorem}

\mySep


\begin{definition}[Characteristic Coefficient]      \label{def:PFC}

Let $\B{A}$ be an $n \times n$ Hermitian matrix with the
eigenvalues $\alpha_1,\alpha_2,\ldots,\alpha_n$ in any order.
Then, the $\left(i,j\right)$th \emph{characteristic coefficient}
$\PFC{i}{j}{\B{A}}$, $i=1,2,\ldots,\EVn{\B{A}}$,
$j=1,2,\ldots,\EVm{i}{\B{A}}$, is defined as a partial fraction
expansion coefficient of $\det\left(\B{I}_n + \xi
\B{A}\right)^{-1}$ such that
\begin{align}       \label{eq:PF:CP}
    \det\left(
                \B{I}_n
                +
                \xi
                \B{A}
        \right)^{-1}
    &=
        \prod_{i=1}^\EVn{\B{A}}
            \bigl(
                    1+
                    \xi
                    \odEV{\alpha}{i}{}
            \bigr)^{-\EVm{i}{\B{A}}}
    \nonumber \\
    &=
        \sum_{i=1}^\EVn{\B{A}}
        \sum_{j=1}^{\EVm{i}{\B{A}}}
            \PFC{i}{j}{\B{A}}
            \bigl(
                    1+
                    \xi
                    \odEV{\alpha}{i}{}
            \bigr)^{-j}
\EqE
\end{align}
where $\xi$ is a scalar constant such that  $\B{I}_n + \xi \B{A}$
is nonsingular. The $\left(i,j\right)$th characteristic
coefficient $\PFC{i}{j}{\B{A}}$ can be determined by
\begin{align}        \label{eq:ChC}
\EqF
    \PFC{i}{j}{\B{A}}
    &=
        \frac{
                1
             }
             {
                \varpi_{i,j}! \,
                \odEV{\alpha}{i}{\varpi_{i,j}}
             }
        \cdot
        \Biggl[
        \frac{d^{\varpi_{i,j}}}{d\upsilon^{\varpi_{i,j}}}
            \bigl(
                    1+
                    \upsilon
                    \odEV{\alpha}{i}{}
            \bigr)^{\EVm{i}{\B{A}}}
            \det\left(
                        \B{I}_n
                        +
                        \upsilon
                        \B{A}
                \right)^{-1}
        \Biggr]
        \Biggr|_{\upsilon=-1/\odEV{\alpha}{i}{}}
    \nonumber \\
    &=
        \frac{
                \left(-1\right)^{\varpi_{i,j}}
             }
             {
                \odEV{\alpha}{i}{\varpi_{i,j}}
             }
        \sum_{
                \begin{smallmatrix}
                    k_1+k_2+\ldots+k_\EVn{\B{A}}
                    =
                        \varpi_{i,j}
                    \\
                    k_l \in \left\{0,\mathbb{N}\right\}
                    \text{ for }
                    \forall l \neq i
                    \\
                    k_i=0
                \end{smallmatrix}
            }
            \,
            \prod_{
                    \substack{
                                l=1
                                \\
                                l \neq i
                             }
                  }^\EVn{\B{A}}
                \binom{
                        \EVm{l}{\B{A}}+k_l-1
                      }
                      {k_l}
                \frac{
                        \odEV{\alpha}{l}{k_l}
                     }
                     {
                        \left(
                                1-
                                \frac{\odEV{\alpha}{l}{}}
                                     {\odEV{\alpha}{i}{}}
                        \right)^{
                                    \EVm{l}{\B{A}}+k_l
                                }
                     }
\EqE
\end{align}
where $\varpi_{i,j}=\EVm{i}{\B{A}}-j$. 

\end{definition}


\mySep

Note that the characteristic coefficients are invariant with
respect to the constant $\xi$ and only a function of the spectra
of $\B{A}$. In addition, it can be seen from \eqref{eq:PF:CP} with
$\xi=0$ that the sum of all the characteristic coefficients is
equal to one. By definition, we have
\begin{salign}
    \PFC{1}{j}{\B{I}_n}
    =
        \begin{cases}
            \,
            0,
            &
            j=1,2,\ldots,n-1
            \\[5pt]
            \,
            1,
            &
            j=n.
        \end{cases}
\end{salign}

\mySep

\begin{example}[Constant Correlation Matrix]    \label{ex:CC:PFC}
Consider a constant correlation matrix $\CC{n}{\rho}$. Since the
eigenvalues of $\CC{n}{\rho}$ are $1+\left(n-1\right)\rho$ and
$1-\rho$ with $n-1$ multiplicity, it is easy to show that the
characteristic coefficients of $\CC{n}{\rho}$, $\rho \in
\left(0,1\right)$, are
\begin{salign}      \label{eq:CC:CC:1}
    \mathcal{X}_{1,1}\bigl(\CC{n}{\rho}\bigr)
    &=
        \left(
                \frac{
                        n\rho
                     }
                     {
                        1
                        -\rho
                        +n\rho
                     }
        \right)^{-n+1}
%
%
%
%
    \\
            \label{eq:CC:CC:2}
    \mathcal{X}_{2,j}\bigl(\CC{n}{\rho}\bigr)
    &=
        -\frac{
                1-\rho
             }
             {
                1
                -\rho
                +n\rho
             }
        \cdot
        \left(
                \frac{
                        n\rho
                     }
                     {
                        1
                        -\rho
                        +n\rho
                     }
        \right)^{-n+j}
\end{salign}
where $j=1,2,\ldots,n-1$.

\end{example}

\mySep\mySep

\begin{theorem}   \label{thm:CGM:detVec}
Let $\B{X} \sim \CGM{m}{n}{\B{0}_{m \times
n}}{\B{\Sigma}}{\B{I}_n}$, $m \leq n$, and
$\sigma_1,\sigma_2,\ldots,\sigma_m$ be the eigenvalues of
$\B{\Sigma}$. Let $\B{A}$ be a $\nu \times \nu$
positive-semidefinite matrix with the eigenvalues
$\alpha_1,\alpha_2,\ldots,\alpha_\nu$. Then, for $\xi \geq 0$, we
have
\begin{salign}    \label{eq:CGM:detVec}
    \EX{
        \det\left(
                \B{I}_{m \nu}
                +
                \xi \,
                \B{A}
                \otimes
                \B{XX}^\dag
            \right)^{-1}
        }
    =
        \mathcal{A}^{-1}
        \det\left(
                    \begin{bmatrix}
                        \B{\Omega}_1
                        &
                        \B{\Omega}_2
                        &
                        \cdots
                        &
                        \B{\Omega}_\EVn{\B{\Sigma}}
                    \end{bmatrix}
            \right)
\end{salign}
where
$\mathcal{A}$ is given in \eqref{eq:CWM:eigenPDF:K} and
$\B{\Omega}_k=\left(\Omega_{k,ij}\right)$,
$k=1,2,\ldots,\EVn{\B{\Sigma}}$, are $m \times
\EVm{k}{\B{\Sigma}}$ matrices whose $\left(i,j\right)$th entry is
given by
\begin{align}    \label{eq:CGM:detVec:O}
\EqF
    \Omega_{k,ij}
    =
        \sum_{p=1}^\EVn{\B{A}}
        \sum_{q=1}^\EVm{p}{\B{A}}
            \Bigl\{
            &
                \PFC{p}{q}{\B{A}} \,
                \odEV{\sigma}{k}{n-m+i+j-1}
                \left(
                    n-m+i+j-2
                \right)!
    \nonumber \\
    & \quad \times
                \HyperPFQ{2}{0}{
                                n-m+i+j-1,
                                q \, ;
                                -\xi
                                \odEV{\alpha}{p}{}
                                \odEV{\sigma}{k}{}
                                }
            \Bigr\}
\EqE
\end{align}
where $\PFC{p}{q}{\B{A}}$ is the $\left(p,q\right)$th
characteristic coefficient of $\B{A}$.


\begin{proof}
%
From Theorem~\ref{thm:CWM:eigenPDF}, we have
\begin{salign}    \label{eq:CGM:detVec:pf}
    &\EX{
        \det\left(
                \B{I}_{m \nu}
                +
                \xi \,
                \B{A}
                \otimes
                \B{XX}^\dag
            \right)^{-1}
        }
    \nonumber \\
    & \qquad
    =
    \EX{
        \prod_{p=1}^\EVn{\B{A}}
            \det\left(
                    \B{I}_{m}
                    +
                    \xi
                    \odEV{\alpha}{p}{} \,
                    \B{XX}^\dag
                \right)^{-\EVm{p}{\B{A}}}
        }
    \nonumber \\
    & \qquad
    =
        \idotsint\limits_{
                0 < \lambda_m \leq
                \ldots \leq \lambda_1 < \infty
             }
             \,
             \prod_{k=1}^{m}
             \prod_{p=1}^\EVn{\B{A}}
                \bigl(
                    1+\xi
                      \odEV{\alpha}{p}{}
                      \lambda_k
                \bigr)^{-\EVm{p}{\B{A}}}
             p_{\B{\lambda}}\bigl(
                                    \lambda_1,
                                    \lambda_2,
                                    \ldots,
                                    \lambda_m
                            \bigr)
             d\lambda_1
             d\lambda_2
             \cdots
             d\lambda_m
    \nonumber \\
    & \qquad
    \stackrel{\left(a\right)}{=}
        \frac{1}{m! \, \mathcal{A}}
        \,
        \underbrace{
            \int_{0}^{\infty}
            \cdots
            \int_{0}^{\infty}
        }_{m \text{-fold}}
            ~
             \prod_{k=1}^{m}
                \left\{
                     \lambda_k^{n-m}
                     \prod_{p=1}^\EVn{\B{A}}
                        \bigl(
                                1+\xi
                                \odEV{\alpha}{p}{}
                                \lambda_k
                        \bigr)^{-\EVm{p}{\B{A}}}
                \right\}
    \nonumber \\
    & \hspace{4.7cm}
    \times
             \det\left(
                        \begin{bmatrix}
                            \B{G}_1
                            &
                            \B{G}_2
                            &
                            \cdots
                            &
                            \B{G}_\EVn{\B{\Sigma}}
                        \end{bmatrix}
                  \right)
             \det_{1 \leq i,j \leq m}
                \left(
                        \lambda_j^{i-1}
                \right)
             \,
             d\lambda_1
             d\lambda_2
             \cdots
             d\lambda_m
    \nonumber \\
    & \qquad
    \stackrel{\left(b\right)}{=}
        \mathcal{A}^{-1}
        \det\left(
                    \begin{bmatrix}
                        \B{\Omega}_1
                        &
                        \B{\Omega}_2
                        &
                        \cdots
                        &
                        \B{\Omega}_\EVn{\B{\Sigma}}
                    \end{bmatrix}
            \right)
\end{salign}
where $\left(a\right)$ follows from the fact that the integrand is
symmetric in $\lambda_1,\lambda_2,\ldots,\lambda_m$ and
$\left(b\right)$ follows from the generalized Cauchy--Binet
formula \cite[Appendix]{CWZ:03:IT}, \cite[Lemma~2]{SWLC:06:WCOM},
yielding the $\left(i,j\right)$th entry of $m \times
\EVm{k}{\B{\Sigma}}$ matrices $\B{\Omega}_k$,
$k=1,2,\ldots,\EVn{\B{\Sigma}}$, as
\begin{align}    \label{eq:CGM:detVec:O:1}
\EqF
    \Omega_{k,ij}
    =
        \int_0^\infty
            \prod_{p=1}^\EVn{\B{A}}
                \bigl(
                    1+\xi
                      \odEV{\alpha}{p}{}
                      \lambda
                \bigr)^{-\EVm{p}{\B{A}}}
            \lambda^{n-m+i+j-2}
            e^{-\lambda/\odEV{\sigma}{k}{}}
        d\lambda.
\EqE
\end{align}
Using a partial fraction decomposition, \eqref{eq:CGM:detVec:O:1}
can be written as
\begin{align}    \label{eq:CGM:detVec:O:2}
\EqF
    \Omega_{k,ij}
    =
        \sum_{p=1}^\EVn{\B{A}}
        \sum_{q=1}^\EVm{p}{\B{A}}
            \PFC{p}{q}{\B{A}}
            \int_0^\infty
                \bigl(
                    1+\xi
                      \odEV{\alpha}{p}{}
                      \lambda
                \bigr)^{-q}
                \lambda^{n-m+i+j-2}
                e^{-\lambda/\odEV{\sigma}{k}{}}
            d\lambda
\EqE
\end{align}
where the characteristic coefficients $\PFC{p}{q}{\B{A}}$ is given
by \eqref{eq:ChC}. We complete the proof of the theorem by
evaluating the integral in \eqref{eq:CGM:detVec:O:2} with the help
of the following integral identity:
\begin{align}    \label{eq:II:1}
    \int_0^\infty
        \left(
            1+ax
        \right)^{\mu-1}
        x^{n-1}
        e^{-x/b}
        dx
    =
        b^n
        \left(n-1\right)! \
        \HyperPFQ{2}{0}{
                        n,
                        -\mu+1 \, ;
                        -ab
                        }
\EqE
\end{align}
where $a,b>0$, $n \in \mathbb{N}$, and $\mu \in \C$.
\end{proof}

\end{theorem}

\mySep\mySep

\begin{corollary}   \label{cor:CGM:detVec:UC}
Let $\B{X} \sim \CGM{m}{n}{\B{0}_{m \times
n}}{\B{\Sigma}}{\B{I}_n}$, $m \leq n$. 
Then, for $\nu \in \mathbb{N}$, we have
\begin{align}    \label{eq:CGM:detVec:UC}
\EqF
    \EX{
        \det\bigl(
                \B{I}_{m}
                +
                \xi \,
                \B{XX}^\dag
            \bigr)^{-\nu}
        }
    &=
        \frac{
                \det\left(
                            \B{\Omega}
                    \right)
             }
             {
                \prod_{i=1}^m
                    \left(n-i\right)!
                    \left(i-1\right)!
             }
\EqE
\end{align}
where
$\B{\Omega}=\left(\Omega_{ij}\right)$ is the $m \times m$ Hankel
matrix whose $\left(i,j\right)$th entry is given by
\begin{align}    \label{eq:CGM:detVec:UC:G}
\EqF
    \Omega_{ij}
    =
        \left(
                n-m+i+j-2
        \right)! \
        \HyperPFQ{2}{0}{
                        n-m+i+j-1,
                        \nu \,;
                        -\xi
                        }.
\EqE
\end{align}

\begin{proof}
It follows immediately from Theorem \ref{thm:CGM:detVec} with
$\B{\Sigma}=\B{I}_m$, $\B{A}=\B{I}_\nu$, $\EVn{\B{\Sigma}}=1$,
$\EVm{1}{\B{\Sigma}}=m$, $\odEV{\sigma}{1}{}=1$, $\EVn{\B{A}}=1$,
$\EVm{1}{\B{A}}=\nu$, and $\odEV{\alpha}{1}{}=1$.
\end{proof}

\end{corollary}

\mySep\mySep

\begin{theorem}   \label{thm:CQM:det}
Let $\B{X} \sim \CGM{m}{n}{\B{0}_{m \times
n}}{\B{\Sigma}}{\B{\Psi}}$, $\sigma_{i}$, $i=1,2,\ldots,m$, and
$\psi_{j}$, $j=1,2,\ldots,n$, be the eigenvalues of $\B{\Sigma}$
and $\B{\Psi}$, respectively. Then, for $\xi \geq 0$, we have
\begin{align}    \label{eq:CQM:det}
\EqF
    \EX{
        \det\left(
                \B{I}_{m}
                +
                \xi \,
                \B{XX}^\dag
            \right)^{-1}
        }
    =
        \sum_{p=1}^\EVn{\B{\Sigma}}
        \sum_{q=1}^\EVn{\B{\Psi}}
        \sum_{i=1}^\EVm{p}{\B{\Sigma}}
        \sum_{j=1}^\EVm{q}{\B{\Psi}}
            \PFC{p}{i}{\B{\Sigma}}
            \PFC{q}{j}{\B{\Psi}} \,
            \HyperPFQ{2}{0}{
                            i,
                            j \, ;
                            -\xi \,
                            \odEV{\sigma}{p}{}
                            \odEV{\psi}{q}{}
                            }
\EqE
\end{align}
where $\PFC{p}{i}{\B{\Sigma}}$ and $\PFC{q}{j}{\B{\Psi}}$ are the
$\left(p,i\right)$th and $\left(q,j\right)$th characteristic
coefficients of $\B{\Sigma}$ and $\B{\Psi}$, respectively.

\begin{proof}
It follows from Lemmas \ref{le:CGM:etr} and \ref{le:CGM:decomp}
that
\begin{align}   \label{eq:CQM:det:pf:1}
\EqF
    \det\left(
            \B{I}_{m}
            +
            \xi \,
            \B{XX}^\dag
        \right)^{-1}
    &=
        \EXs{\B{y}_1}{
                        \etr\left(
                                -\xi
                                 \B{X}^\dag
                                 \B{y}_1
                                 \B{y}_1^\dag
                                 \B{X}
                            \right)
                      }
    \nonumber \\
    &=
        \EXs{\B{y}_1,\B{y}_2}{
                        \etr\left(
                                 \xi
                                 \B{y}_1^\dag
                                 \B{X}
                                 \B{y}_2
                                 -
                                 \B{y}_2^\dag
                                 \B{X}^\dag
                                 \B{y}_1
                            \right)
                      }
\EqE
\end{align}
%
%
where $\B{y}_1 \sim \CGM{m}{1}{\B{0}_{m \times 1}}{\B{I}_m}{1}$
and $\B{y}_2 \sim \CGM{n}{1}{\B{0}_{n \times 1}}{\B{I}_n}{1}$.
Denoting the left-hand side of \eqref{eq:CQM:det} by
$\mathrm{LHS}_{\eqref{eq:CQM:det}}$ and using
\eqref{eq:CQM:det:pf:1}, we have
\begin{align}   \label{eq:CQM:det:pf:2}
\EqF
    \mathrm{LHS}_{\eqref{eq:CQM:det}}
    &=
        \EXs{\B{y}_1,\B{y}_2}{
                        \EXs{\B{X}}{
                               \etr\left(
                                         \xi
                                         \B{y}_2
                                         \B{y}_1^\dag
                                         \B{X}
                                         -
                                         \B{X}^\dag
                                         \B{y}_1
                                         \B{y}_2^\dag
                                    \right)
                                    }
                              }
    \nonumber \\
    &=
        \EXs{\B{y}_1,\B{y}_2}{
                        \exp\left(
                                -\xi
                                 \B{y}_1^\dag
                                 \B{\Sigma}
                                 \B{y}_1
                                 \B{y}_2^\dag
                                 \B{\Psi}
                                 \B{y}_2
                            \right)
                            }.
\EqE
\end{align}

Now, introducing a delta function to decouple the expectations for
$\B{y}_1$ and $\B{y}_2$ in \eqref{eq:CQM:det:pf:2} yields
\begin{align}    \label{eq:CQM:det:pf:3}
\EqF
    \mathrm{LHS}_{\eqref{eq:CQM:det}}
    &=
        \EXs{
                \B{y}_1,
                \B{y}_2
            }
            {
                \int_{-\infty}^{\infty}
                    e^{
                        -\xi
                         z
                         \B{y}_2^\dag
                         \B{\Psi}
                         \B{y}_2
                      }
                    \delta\bigl(
                                 z
                                 -
                                 \B{y}_1^\dag
                                 \B{\Sigma}
                                 \B{y}_1
                          \bigr)
                    dz
            }
    \nonumber \\
    &
    \stackrel{\left(a\right)}{=}
        \frac{1}{2\pi}
        \EXs{
                \B{y}_1,
                \B{y}_2
            }
            {
                \int_{-\infty}^{\infty}
                \int_{-\infty}^{\infty}
                    e^{
                        -\xi
                         z
                         \B{y}_2^\dag
                         \B{\Psi}
                         \B{y}_2
                      }
                    e^{
                         \ImUnit
                         \left(
                                 z
                                 -
                                 \B{y}_1^\dag
                                 \B{\Sigma}
                                 \B{y}_1
                         \right)
                         \omega
                      }
                    d\omega
                    dz
            }
    \nonumber \\
    &=
        \frac{1}{2\pi}
        \int_{-\infty}^{\infty}
        \int_{-\infty}^{\infty}
            e^{\ImUnit \omega z}
            \EXs{\B{y}_1}{
                    \etr\bigl(
                        -\ImUnit
                         \omega
                         \B{\Sigma}
                         \B{y}_1
                         \B{y}_1^\dag
                        \bigr)
                }
            \EXs{\B{y}_2}{
                    \etr\bigl(
                        -\xi
                         z
                         \B{\Psi}
                         \B{y}_2
                         \B{y}_2^\dag
                        \bigr)
                }
            d\omega
            dz
    \nonumber \\
    &
    \stackrel{\left(b\right)}{=}
        \frac{1}{2\pi}
        \int_{-\infty}^{\infty}
        \int_{-\infty}^{\infty}
            e^{\ImUnit \omega z}
            \det\left(
                    \B{I}_m
                    +
                    \ImUnit
                    \omega
                    \B{\Sigma}
                \right)^{-1}
            \det\left(
                    \B{I}_n
                    +
                    \xi
                    z
                    \B{\Psi}
                \right)^{-1}
            d\omega
            dz
    \nonumber \\
    &
    \stackrel{\left(c\right)}{=}
        \frac{1}{2\pi}
        \sum_{p=1}^\EVn{\B{\Sigma}}
        \sum_{q=1}^\EVn{\B{\Psi}}
        \sum_{i=1}^\EVm{p}{\B{\Sigma}}
        \sum_{j=1}^\EVm{q}{\B{\Psi}}
        \Biggl\{
            \PFC{p}{i}{\B{\Sigma}}
            \PFC{q}{j}{\B{\Psi}}
    \nonumber \\
    & \hspace{3cm}
    \times
            \int_{-\infty}^{\infty}
            \int_{-\infty}^{\infty}
                e^{\ImUnit \omega z}
                \bigl(
                    1+\ImUnit
                      \odEV{\sigma}{p}{}
                      \omega
                \bigr)^{-i}
                \bigl(
                    1+\xi
                      \odEV{\psi}{q}{}
                      z
                \bigr)^{-j}
            d\omega
            dz
        \Biggr\}
\EqE
\end{align}
where $\left(a\right)$ is obtained by replacing the delta function
with its Fourier representation, $\left(b\right)$ follows from
Lemma \ref{le:CGM:etr}, and $\left(c\right)$ is obtained from
Definition~\ref{def:PFC}. Using the integral identity, for $a>0$,
$\ell \in \mathbb{N}$, and $z \in \mathbb{R}$,
\begin{align}    \label{eq:II:2}
    \int_{-\infty}^\infty
        e^{\ImUnit \omega z}
        \left(
            1+\ImUnit
              a
              \omega
        \right)^{-\ell}
        d\omega
    =
        \frac{
                \pi
                z^{\ell-1}
                e^{-\sqrt{z^2}/a}
             }
             {
                a^\ell
                \left(\ell-1\right)!
             }
        \left(
            1+\mathrm{sign}\left(z\right)
        \right),
\EqE
\end{align}
\eqref{eq:CQM:det:pf:3} can be written as
\begin{align}    \label{eq:CQM:det:pf:4}
\EqF
    \mathrm{LHS}_{\eqref{eq:CQM:det}}
    =
        \sum_{p=1}^\EVn{\B{\Sigma}}
        \sum_{q=1}^\EVn{\B{\Psi}}
        \sum_{i=1}^\EVm{p}{\B{\Sigma}}
        \sum_{j=1}^\EVm{q}{\B{\Psi}}
            \frac{
                    \PFC{p}{i}{\B{\Sigma}}
                    \PFC{q}{j}{\B{\Psi}}
                 }
                 {
                    \odEV{\sigma}{p}{i}
                    \left(i-1\right)!
                 }
            \int_{0}^{\infty}
                \bigl(
                    1+\xi
                      \odEV{\psi}{q}{}
                      z
                \bigr)^{-j}
                z^{i-1}
                e^{-z/\odEV{\sigma}{p}{}}
            dz.
\EqE
\end{align}
Finally, we obtain the desired result \eqref{eq:CQM:det} by
evaluating the integral in \eqref{eq:CQM:det:pf:4} with the help
of \eqref{eq:II:1}.
\end{proof}

\end{theorem}

\section{Proofs}   \label{sec:Appendix:C}

\subsection{Proof of Theorem~\ref{thm:DO}}  \label{sec:proof:thm:DO}

We first prove Theorem~\ref{thm:DO} for $M$-ary phase shift keying
($M$-PSK) signaling. The SEP of the OSTBC with $M$-PSK
constellation can be expressed as \cite{SL:03:CL,SL:04:VT}
\begin{align}    \label{eq:SEP:MPSK}
\EqF
    P_{\text{e,\,MPSK}}
    =
        \frac{1}{\pi}
        \int_0^{\Theta}
            \MGF{\gamma_\text{STBC}}
                {
                    \frac{g}{\sin^2\theta};
                    \snr
                }
            d\theta
\EqE
\end{align}
where $\Theta=\pi-\pi/M$ and $g=\sin^2\left(\pi/M\right)$.
From \eqref{eq:SEP:MPSK}, we can obtain the upper bound as
\begin{align}
\EqF
    P_{\text{e,\,MPSK}}
    \leq
        \left(
            1-\frac{1}{M}
        \right)
        \MGF{\gamma_\text{STBC}}
            {
                g;
                \snr
            }
\EqE
\end{align}
which becomes tighter as $\snr$ increases \cite{SA:00:Book}, and
hence yields
\begin{align}
\EqF
    d_\text{STBC}
    =
        \lim_{\snr \to \infty}
            \frac{
                -\log \MGF{\gamma_\text{STBC}}
                         {
                            g;
                            \snr
                         }
                 }
                 {\log \snr}.
\EqE
\end{align}
Therefore, the asymptotic behavior of the MGF
$\MGF{\gamma_\text{STBC}}{s;\snr}$ at large $\snr$ reveals a
high-SNR slope of the SEP curve.

Suppose that $\snr$ is sufficiently large. For $\nTx \leq \nRx$,
it follows from \eqref{eq:MGF:SNR:1} that
\begin{align}
\EqF
    \log \MGF{\gamma_\text{STBC}}{g;\snr}
    &\approx
        -
        \underbrace{\mathrm{rank}\left(
                         \B{\Xi}_1^\dag
                         \B{\Xi}_1
                         \SxCM
                         \otimes
                         \TxCM
                      \right)}_{
                                \nTx
                                \cdot
                                \min\left(\nRx,\nSx\right)
                               }
        \cdot
        \log 
                         \snr
        +
        \text{constant}.
\EqE
\end{align}
Similarly, using \eqref{eq:MGF:SNR:2}, we have for $\nTx > \nRx$,
\begin{align}
\EqF
    \log \MGF{\gamma_\text{STBC}}{g;\snr}
    &\approx
        -
        \underbrace{\mathrm{rank}\left(
                            \RxCM
                            \otimes
                            \B{\Xi}_2
                            \B{\Xi}_2^\dag
                      \right)}_{
                                \nRx
                                \cdot
                                \min\left(\nTx,\nSx\right)
                               }
        \cdot
        \log 
                \snr
        +
        \text{constant}'.
\end{align}
Hence,
\begin{align}
    d_\text{STBC}
    =
        \min\left(\nTx,\nRx\right)
        \cdot
        \min\left\{
                    \max\left(\nTx,\nRx\right),
                    \nSx
            \right\}
\EqE
\end{align}
from which \eqref{eq:DO:STBC} follows immediately. For a general
case of arbitrary two-dimensional signaling constellation with
polygonal decision boundaries, the SEP can be written as a convex
combination of terms akin to \eqref{eq:SEP:MPSK}
\cite{DBW:99:COM}. Hence, we can easily generalize the proof to
the case of any two-dimensional signaling constellation.

\subsection{Proofs of Property~\ref{ex:SC:CN}--\ref{PT:CN:3}}  \label{sec:proof:PT:CN}

\subsubsection{Proof of Property~\ref{ex:SC:CN}}

Let $\lambda_1,\lambda_2,\ldots,\lambda_n$ be the eigenvalues of
$\B{\Phi}$. Then, the correlation figure $\CN{\B{\Phi}}$ defined
in Definition~\ref{def:CN} can be written as
\begin{align}
    \CN{\B{\Phi}}
    =
        \frac{1}{n^2}
        \sum_{k=1}^n
            \lambda_k^2
\end{align}
which is symmetric in $\lambda_1,\lambda_2,\ldots,\lambda_n$ and
holds Schur's condition \eqref{eq:SC:1}. Hence, we complete the
proof.

\subsubsection{Proof of Property~\ref{PT:CN:2}}

Since $\prod_{i=1}^m \CN{\B{\Phi}_i}=\CN{\bigotimes_{i=1}^m
\B{\Phi}_i}$, it follows immediately from Property~\ref{ex:SC:CN}.

\subsubsection{Proof of Property~\ref{PT:CN:3}}

Let $\lambda_1^{(i)},\lambda_2^{(i)},\ldots,\lambda_n^{(i)}$ be
the eigenvalues of $\B{\Phi}_i$ ($i=1,2,\ldots,m$). Then,
$\sum_{i=1}^m \CN{\B{\Phi}_i}$ can be written as
\begin{align}
\EqF
    \sum_{i=1}^m
        \CN{\B{\Phi}_i}
    =
        \sum_{i=1}^m
        \sum_{k=1}^{n_i}
            \left(\frac{\lambda_k^{(i)}}{n_i}\right)^2
\EqE
\end{align}
which is symmetric in
$\bigl\{\frac{1}{n_i}\lambda_1^{(i)},\frac{1}{n_i}\lambda_2^{(i)},\ldots,\frac{1}{n_i}\lambda_{n_i}^{(i)}\bigr\}_{i=1}^m$
and holds Schur's condition \eqref{eq:SC:1}. Since
$\bigl\{\frac{1}{n_i}\lambda_1^{(i)},\frac{1}{n_i}\lambda_2^{(i)},\ldots,\frac{1}{n_i}\lambda_{n_i}^{(i)}\bigr\}_{i=1}^m$
are the eigenvalues of $\bigoplus_{i=1}^m
\frac{1}{n_i}\B{\Phi}_i$, we complete the proof.

\subsection{Proof of Theorem~\ref{thm:kurt:FN}}  \label{sec:proof:thm:kurt:FN}

Using Theorem~\ref{thm:CGM:Tr2Sq} in
Appendix~\ref{sec:Appendix:B}, we get
\begin{align}       \label{eq:H:FN:4}
    \EqF
    \E \bigl\{\FN{\B{H}}^4\bigr\}
    &=
        \EXs{\, \B{\Xi}_1,\B{\Xi}_2}
            {
                \tr^2\left(
                         \frac{1}{\nSx}
                         \B{\Xi}_1
                         \B{\Xi}_2
                         \B{\Xi}_2^\dag
                         \B{\Xi}_1^\dag
                    \right)
            }
    \nonumber \\
    &=
        \left(\frac{\nRx}{\nSx}\right)^2
        \tr\left(\TxCM^2\right)
        \tr\left(\SxCM^2\right)
        +
        \tr\left(\TxCM^2\right)
        \tr\left(\RxCM^2\right)
        +
        \left(\frac{\nTx}{\nSx}\right)^2
        \tr\left(\RxCM^2\right)
        \tr\left(\SxCM^2\right)
        +
        \left(\nTx\nRx\right)^2.
    \EqE
\end{align}
Combining \eqref{eq:kurt:FN}, \eqref{eq:CN}, and
\eqref{eq:H:FN:4}, together with the fact that $\E
\bigl\{\FN{\B{H}}^2 \bigr\}=\nTx \nRx$, yields
\eqref{eq:kurt:FN:f}.

\subsection{Proof of Theorem~\ref{thm:WSE}}  \label{sec:proof:thm:WSE}
In this case, the ergodic capacity (or Shannon-sense mean
capacity) is given by the well-known expression
\cite{Fos:96:BLTJ,FG:98:WPC,Tel:99:ETT}
\begin{align}
\EqF
    C\left(\snr\right)
    =
        \EX{
            \log_2
            \det\left(
                        \B{I}_\nRx
                        +
                        \frac{\snr}{\nTx}
                        \B{HH}^\dag
                \right)
            }
    \qquad
    \text{bits/s/Hz}
\EqE
\end{align}
which is achieved by the complex Gaussian input $\B{X} \sim
\tilde{\mathcal{N}}_{\nTx,\Tc}\bigl(\B{0}_{\nTx \times
\Tc},\frac{\mathcal{P}}{\nTx}\B{I}_\nTx,\B{I}_{\Tc}\bigr)$.

From \cite[(35)]{Ver:02:IT} and \cite[Theorem~9]{Ver:02:IT}, we
get
\begin{align}
\EqF
    \SNRmin
    =
        \frac{
                \nTx
                \log_e 2
             }
             {
                \EX{
                        \FN{\B{H}}^2
                   }
             }
    =
        \frac{\log_e 2}{\nRx}
\EqE
\end{align}
and
\begin{align}       \label{eq:So:DS:pf}
    S_0
    &=
        \frac{
                2
                \left(\EX{\FN{\B{H}}^2}
                  \right)^2
             }
             {
                \EX{
                    \tr\left[
                            \left(
                                    \B{HH}^\dag
                            \right)^2
                       \right]
                   }
             }
    \nonumber \\
    &=
        \frac{
                2
                \left(
                    \nTx
                    \nRx
                    \nSx
                \right)^2
             }
             {
                \EXs{\B{\Xi}_1,\B{\Xi}_2}
                    {
                    \tr\left[
                            \left(
                                    \B{\Xi}_1
                                    \B{\Xi}_2
                                    \B{\Xi}_2^\dag
                                    \B{\Xi}_1^\dag
                            \right)^2
                       \right]
                   }
             }.
\EqE
\end{align}
Using Definition~\ref{def:CN} and Theorem~\ref{thm:CGM:Tr2Sq} in
Appendix~\ref{sec:Appendix:B}, \eqref{eq:So:DS:pf} can be
expressed in terms of the correlation figures of $\TxCM$, $\RxCM$,
and $\SxCM$ as in \eqref{eq:So:DS}.

\subsection{Proof of Theorem~\ref{thm:WSE:STBC}}  \label{sec:proof:thm:WSE:STBC}

Due to the channel decoupling property of OSTBCs, the Shannon
capacity of OSTBC MIMO channels can be written as
\begin{align}        \label{eq:C:STBC}
\EqF
    C_\text{STBC}\left(\snr\right)
    =
        \mathcal{R}
        \cdot
        \EX{
            \log_2\left(
                        1+
                        \frac{
                                \snr
                                \FN{\B{H}}^2
                             }
                             {
                                \nTx\mathcal{R}
                             }
                  \right)
           }
    \qquad
    \text{bits/s/Hz}
\EqE
\end{align}
which is achieved by complex Gaussian inputs $x_k \sim
\mathcal{CN}\bigl(0,\frac{\mathcal{P}}{\nTx\mathcal{R}}\bigr)$.
From \cite[(35)]{Ver:02:IT}, \cite[Theorem~9]{Ver:02:IT} and the
first two derivatives of \eqref{eq:C:STBC} at $\snr=0$, it is easy
to show \eqref{eq:minEbNo:STBC} and
\begin{align}
\EqF
    S_0^\text{STBC}
    =
        \frac{
                2\mathcal{R}
             }
             {
                \Kurt{\FN{\B{H}}}
             }
\EqE
\end{align}
from which and Theorem \ref{thm:kurt:FN}, \eqref{eq:So:STBC}
follows readily.

\section*{Acknowledgment}

The authors would like to thank the anonymous reviewers for their
helpful feedback on the paper. They also wish to thank J.~S.~Kwak
and I.~Keliher for their comments and careful reading of the
manuscript.


\clearpage

\clearpage

\clearpage

\clearpage

\begin{figure}[t]
    \centerline{\includegraphics[width=0.7\textwidth]{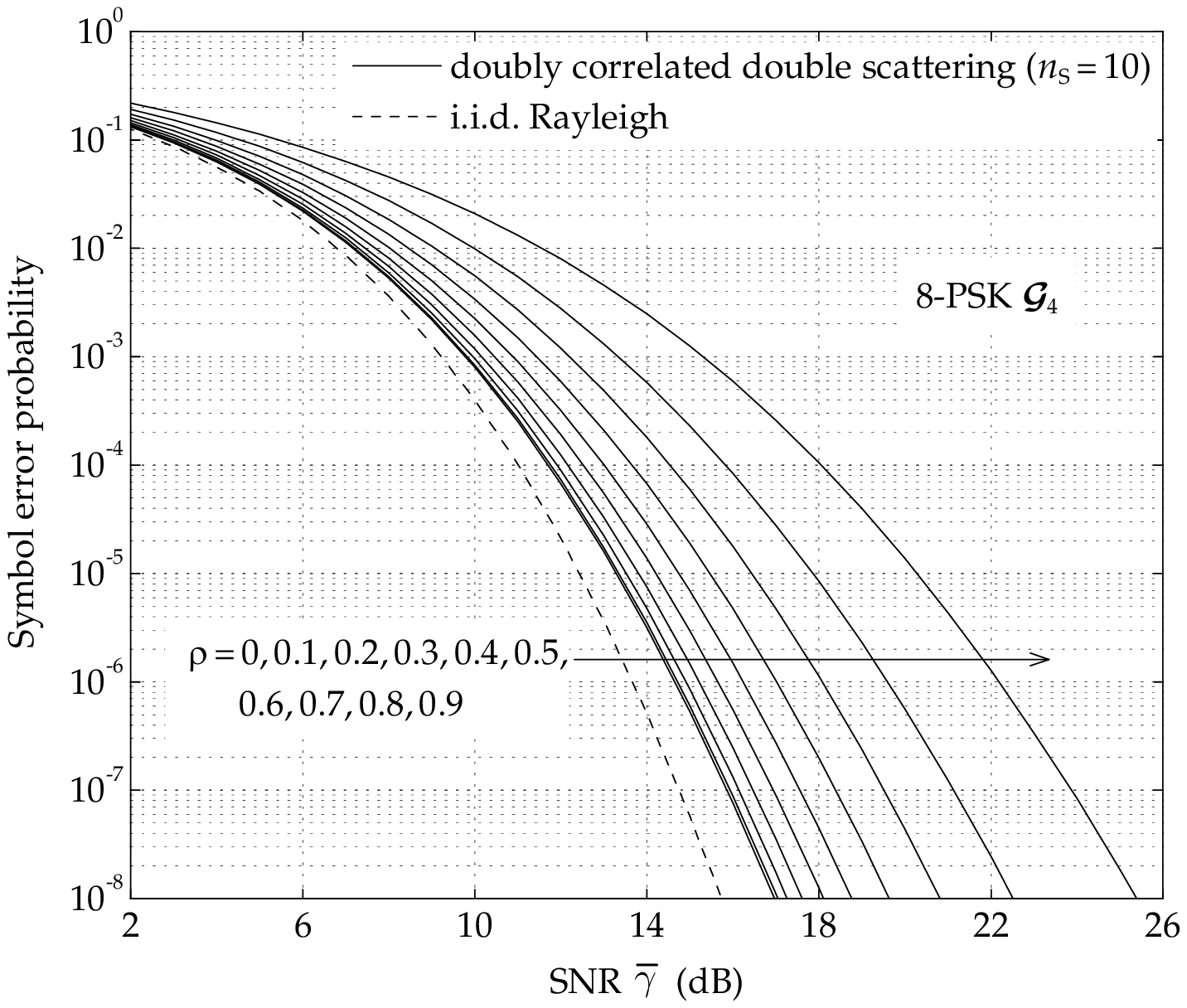}}
    \caption{
        SEP of $8$-PSK $\GF$ ($2.25$ bits/s/Hz) versus $\snr$ in doubly correlated
        double-scattering $\DsMIMO{4}{4}{10}$-MIMO
        channels. The transmit and receive
        correlations follow the constant correlation
        $\TxCM=\RxCM=\CC{4}{\rho}$ for $\rho=0$ (spatially uncorrelated double-scattering),
        $0.1$, $0.2$, $0.3$, $0.4$, $0.5$, $0.6$, $0.7$, $0.8$, and $0.9$. For comparison, the
        SEP for i.i.d.\ Rayleigh-fading MIMO channels is also
        plotted.
    }
    \label{fig:Fig:6}
\end{figure}

\clearpage

\clearpage

%
%

\clearpage

\end{document}